\documentclass[aps,prb,amsmath,amssymb,twocolumn,longbibliography,showpacs,superscriptaddress,10pt]{revtex4-1}

\usepackage{graphicx}
\usepackage{dcolumn}
\usepackage{bm}
\usepackage{tikz,pgfplots}
\usetikzlibrary{arrows}
\usepackage{color}
\usepackage{xcolor}
\usepackage[unicode=true, bookmarks=false,breaklinks=false,pdfborder={0 0 1},backref=false,colorlinks=false]{hyperref}
\usepackage{makeidx}
\hypersetup{colorlinks=true,citecolor=blue,linkcolor=blue,filecolor=blue,urlcolor=blue}
\usepackage{subfigure}

\newcommand{\ud}{\mathrm{d}}    

\newcommand{\s}{\sigma}
\newcommand{\om}{\omega}
\newcommand{\p}{\prime}
\newcommand{\ii}{\imath}
\newcommand{\al}{\alpha}
\newcommand{\z}{\mbox{\boldmath$\zeta$}}

\newcommand*{\temp}{\multicolumn{1}{|c}}

\usepackage{mathrsfs}

\bibpunct{[}{]}{,}{n}{}{}	

\begin{document}

\title{Phononic heat transport in nanomechanical structures: steady-state and pumping}

\author{Marcone I. Sena-Junior}
\affiliation{Instituto de F\'{i}sica, Universidade Federal Fluminense, 24210-346 Niter\'oi - RJ, Brazil }
\affiliation{Instituto de F\'{i}sica, Universidade Federal de Alagoas, 57072-900 Macei\'o - AL, Brazil}
\affiliation{Escola Polit\'{e}cnica de Pernambuco, Universidade de Pernambuco, 50720-001 Recife - PE, Brazil }
\author{Leandro R. F. Lima}
\author{Caio H. Lewenkopf}
\affiliation{Instituto de F\'{i}sica, Universidade Federal Fluminense, 24210-346 Niter\'oi - RJ, Brazil }

\date{\today}

\begin{abstract}
We study the heat transport due to phonons in nanomechanical structures using a phase space representation 
of non-equilibrium Green's functions. This representation accounts for the atomic degrees of freedom making it 
particularly suited for the description of small (molecular) junctions systems. We {rigorously} show that for the
steady state limit our formalism correctly recovers the heuristic Landauer-like heat conductance for a quantum 
coherent molecular system coupled to thermal reservoirs.
We find general expressions for the non-stationary heat current due to an external periodic drive. In both cases we 
discuss the quantum thermodynamic properties of the systems. 
We apply our formalism to the case of a diatomic molecular junction.
\end{abstract}

\pacs{65.80.-g,05.60.Gg,44.10.+i,66.70.-f}
\maketitle

\section{Introduction}
\label{sec:introduction}

Significant progress has been recently achieved on the understanding of phononic heat 
transfer at the molecular level \cite{Dubi2011,Li2012,Chen2005}. 
In addition to the investigation of fundamental aspects of the problem \cite{Dubi2011,Dhar2008}, 
several authors have realized that phonons, usually regarded as an energy waste, can be manipulated 
and controlled to carry and process information.
Exploring analogies with electrons and photons, theoretical proposals have been put forward aiming the 
fabrication of devices such as thermal diodes \cite{Li2004}, thermal transistors \cite{Li2006,Joulain2016}, and 
thermal logic gates \cite{Wang-Li2007}, some of them already experimentally verified \cite{Chang2006, 
Narayana2012, Martinez-Perez2015}. 
These ideas have given rise to the emerging field of phononics \cite{Li2012,Madovan2013}. 

The presence of an external time-dependent drive, such as an external force or time-varying 
thermal bath temperature, gives another interesting twist to the problem, making possible to 
explore non-equilibrium phenomena such as directed heat pumping and cooling 
\cite{Galperin2009,Santandrea2011, Ren2010, Li2012, Arrachea2012, Arrachea2013,Li2015,Beraha2016}.

{
Early reports on the measurement of quantized 
thermal conductance in suspended nanostructures \cite{Tighe1997, Schwab2000} attracted attention 
to the field. More recently, ballistc thermal conductance has been experimentally studied in carbon 
nanotubes \cite{Yang2002,Prasher2009,Marconnet2013}, silicon nanowires \cite{Bourgeois2007, Maire2017}, as well 
as molecular and atomic contacts \cite{Wang2007exp,Cui2017}. The experimental advances in these 
studies are remarkable and pose important challenges 
to the quantum theory of thermal conductance \cite{Marconnet2013,Cui2017}.
}

One of the fundamental tools for the theoretical study of non-equilibrium properties of quantum 
systems is the non-equilibrium Green's functions (NEGF) theory \cite{Rammer1986,Kamenev2011}. 
This approach, originally developed for fermionic systems \cite{Caroli1971,Meir1992}, 
has been nicely adapted to describe the 
heat transfer in small junctions systems \cite{Ozpineci2001, Segal2003, Yamamoto2006, Dhar2008, 
Wang2007, Wang2008, Wang2014}. 
Despite its success, the implementation of the NEGF to calculate phonon heat currents 
driven by a temperature difference between source and drain still has some caveats, like the need 
to symmetrize the heat current {to obtain the standard Landauer-like transmission formula} 
\cite{Wang2007, Wang2008, Wang2014}. 
The relevance of NEGF for phononics calls for a deeper {and careful} analysis of the formalism.

The purpose of the paper is twofold. First, we present a rigorous method for the description of quantum 
thermal transport properties due to phonon or atomic degrees of freedom using nonequilibrium Green's 
function in phase space. 
We show that our formal developments solve the problems of the previous works \cite{Wang2007,Wang2008, 
Wang2014} and recover the well known Landauer-like formula for the stationary heat current in the ballistic 
regime \cite{Pendry1983,Rego1998,Angelescu1998,Blencowe1999,Mingo2003,Chalopin2013}. 
Second, we extend the formalism to address systems under the influence of a time dependent drive. 
As an example, we derive general expressions for the heat current pumped by an external time-dependent 
periodic potential for a system coupled to two thermal reservoirs at the same temperature.

We {show how to apply our method by analyzing} the steady-state heat transport properties of a diatomic molecule coupled 
to thermal reservoirs by semi-infinite linear harmonic chains.
Next, we study the heat current pumped through the system due to a time-dependent driving force and discuss 
its thermodynamic properties. 

The paper is organized as follows: In Sec.~\ref{sec:GF} we introduce the phase space representation of the 
Green's functions on which our derivations are built. 
We begin Sec.~\ref{sec:model} by presenting the model Hamiltonian addressed in this study. 
We then use the Green's function formalism to derive expressions for the thermal current due 
to a source-drain temperature difference and the heat current pumped by an external periodical 
drive of the system atomic degrees of freedom. In Sec.~\ref{sec:application}, we apply our results 
to the simple model of a diatomic molecular junction. Finally, we present our conclusions in Sec.~\ref{sec:conclusion}.

\section{Green's functions in phase space}
\label{sec:GF}

In this the section we {use} a phase space representation of non-equilibrium Green's functions
\cite{Dhar2006,Dhar2012}. 
We show that this representation is very convenient for a canonical quantization of the displacements 
$\vec{u}\equiv (u_{1}, \ldots, u_{n})$ and their canonical conjugated momenta $\vec{p}\equiv (p_{1},
 \ldots, p_{n})$ in a $2n$-dimensional phase space.

Let us consider a quadratic Hamiltonian expressed in terms of space phase variables 
$(\vec{u}, \vec{p})$ representing a system of coupled oscillators. 
The model Hamiltonian reads
\begin{equation}
\label{hamil}
H(t)=\frac{1}{2}\,\vec{p}^{\;\text{T}}\cdot\vec{p} + \frac{1}{2}\,\vec{u}^{\;\text{T}}\cdot\hat{K}(t)\cdot\vec{u} 
\equiv \frac{1}{2}\,\mbox{\boldmath$\zeta$}^{\,\text{T}}\cdot\check{\mathcal{M}}(t)\cdot\mbox{\boldmath$\zeta$},
\end{equation}
where, for the sake of compactness, we assume that the masses are identical and have unit value. $\hat{K}(t)$ 
is the force constant matrix that represents the couplings of the oscillators network. The dynamic variable 
$\mbox{\boldmath$\zeta$}$ and the matrix $\check{\mathcal{M}}$ have the symplectic structure 
\begin{equation}
\mbox{\boldmath$\zeta$}=
\begin{pmatrix}
\vec{u}\\
\vec{p}
\end{pmatrix} \qquad\text{and}\qquad
\check{\mathcal{M}}(t)=
\begin{pmatrix}
\hat{K}(t) & \hat{0}\\
\hat{0} & \hat{I}
\end{pmatrix},
\end{equation}
where $\hat{I}$ is the identity matrix.

The equation of motion for ${\bm\zeta}$ reads
\begin{subequations}
\begin{equation}\label{derivada}
\frac{\ud}{\ud t}\z=\check{\mathcal{Q}}\cdot\frac{\partial}{\partial\z}H = -\check{\mathcal{K}}(t)\cdot\z, 
\end{equation}
where
\begin{equation}\label{eq_4b}
\check{\mathcal{Q}}=
\begin{pmatrix}
\hat{0} & \hat{I}\\
-\hat{I} & \hat{0}
\end{pmatrix}\quad\text{and}\quad
\check{\mathcal{K}}(t)\equiv-\check{\mathcal{Q}}\cdot\check{\mathcal{M}}(t)=
\begin{pmatrix}
\hat{0} & -\hat{I}\\
\hat{K}(t) & \hat{0}
\end{pmatrix}.
\end{equation}
\end{subequations}

We define the phase space correlation functions $\hat{C}(\tau,\tau^{\prime})$ on the Keldysh contour \cite{Rammer1986}  as
\begin{equation}
\label{corr_keldy}
\check{C}(\tau,\tau^{\p})\equiv\frac{1}{\ii\,\hbar}\left\langle\mathbb{T}_{\mathcal{C}}\,\z(\tau)\otimes\z(\tau^{\p})\right\rangle\equiv
\begin{pmatrix}
\hat{C}^{(uu)} & \hat{C}^{(up)}\\
\hat{C}^{(pu)} & \hat{C}^{(pp)}
\end{pmatrix}(\tau,\tau^{\p}),
\end{equation}
where 
$\ii\hbar\,\hat{C}^{(\alpha\beta)}\equiv\langle\mathbb{T}_{\mathcal{C}}\,\vec{\alpha}(\tau)
\otimes\vec{\beta}(\tau^{\p})\rangle$. 
The correlation functions $\hat{C}^{(\alpha\beta)}(\tau,\tau^{\prime})$ are a straightforward phase space
ge\-ne\-ra\-li\-za\-tion of standard Green's functions \cite{Rammer1986,Kamenev2011}, as we discuss below.

As standard \cite{Rammer1986},  the \textit{greater}, \textit{lesser}, \textit{time-ordered}, 
and \textit{anti-time-ordered} correlations functions read
\begin{subequations}\label{set_5}
\begin{align}
\check{C}^{>}(t,t^{\p}) &= (\ii\,\hbar\,)^{-1}\big\langle\z(t)\otimes\z(t^{\p})\big\rangle,\\
\check{C}^{<}(t,t^{\p})  &= \left[\check{C}^{>}(t,t^{\p})\right]^{\text{T}},\\
\check{C}^{\mathbb{T}}(t,t^{\p}) &= \theta(t-t^{\p})\,\check{C}^{>}(t,t^{\p}) + \theta(t^{\p}-t)\,\check{C}^{<}(t,t^{\p}),\\
\check{C}^{\overline{\mathbb{T}}}(t,t^{\p}) &= \theta(t^{\p}-t)\,\check{C}^{>}(t,t^{\p}) + \theta(t-t^{\p})\,\check{C}^{<}(t,t^{\p}),
\end{align}
\end{subequations}
where $(\check{C}^{\,\mathbb{T}} + \check{C}^{\,\overline{\mathbb{T}}} - \check{C}^{>} - \check{C}^{<}\,)(t,t^{\p}) = 0$.

Alternatively, the correlation functions can be represented by their \textit{retarded} $\check{C}^{r}$, 
\textit{advanced} $\check{C}^{a}$, and \textit{Keldysh} $\check{C}^{K}$ components, namely
\begin{subequations}
\label{set_6}
\begin{align}
\check{C}^{r}(t,t^{\p}) & =\frac{1}{2}(\,\check{C}^{\,\mathbb{T}}+\check{C}^{>}-\check{C}^{<}-\check{C}^{\,\overline{\mathbb{T}}}\,)(t,t^{\p})\nonumber\label{eq_6a}\\
& = \theta(t-t^{\p})\,\left(\,\check{C}^{>}-\check{C}^{<}\,\right)(t,t^{\p}),\\
\check{C}^{a}(t,t^{\p}) & =\frac{1}{2}(\,\check{C}^{\,\mathbb{T}}-\check{C}^{>}+\check{C}^{<}-\check{C}^{\,\overline{\mathbb{T}}}\,)(t,t^{\p})\nonumber\label{eq_6b}\\
& = \theta(t^{\p}-t)\,\left(\,\check{C}^{<}-\check{C}^{>}\,\right)(t,t^{\p}),\\
\check{C}^{K}(t,t^{\p}) & =\frac{1}{2}(\,\check{C}^{\,\mathbb{T}}+\check{C}^{>}+\check{C}^{<}+\check{C}^{\,\overline{\mathbb{T}}}\,)(t,t^{\p})\nonumber\\
& = \left(\,\check{C}^{>}+\check{C}^{<}\,\right)(t,t^{\p}).
\label{eq_6c}
\end{align}
\end{subequations}

Using Eqs.~\eqref{derivada} and \eqref{set_5} we obtain the equations of motion for  
$\check{C}^{\gtrless}(t,t^{\p})$ and $\check{C}^{\mathbb{T},\overline{\mathbb{T}}}(t,t^{\p})$,
namely
\begin{subequations}\label{set_7}
\begin{align}
&\left(\check{\mathcal{I}}\,\frac{\partial}{\partial t} + \check{\mathcal{K}}(t)\right)\cdot\check{C}^{\gtrless}(t,t^{\p})=0,\label{Cgreater}\\
&\left(\check{\mathcal{I}}\,\frac{\partial}{\partial t} + \check{\mathcal{K}}(t)\right)\cdot\check{C}^{\mathbb{T},\overline{\mathbb{T}}}(t,t^{\p})=\pm\delta(t-t^{\p})\,\check{\mathcal{Q}}.
\end{align}
\end{subequations}
Similarly, using Eqs.~\eqref{derivada} and \eqref{set_7}, we show that $\check{C}^{\text{K}}(t,t^{\p})$ and 
$\check{C}^{r,a}(t,t^{\p})$  satisfy
\begin{subequations}\label{set_8}
\begin{align}
&\left(\check{\mathcal{I}}\,\frac{\partial}{\partial t}+\check{\mathcal{K}}(t)\right)\cdot 
\check{C}^{K}(t,t^{\p})=0,\label{keldysh}\\
&\left(\check{\mathcal{I}}\,\frac{\partial}{\partial t} +\check{\mathcal{K}}(t)\right)\cdot 
\check{C}^{r,a}(t,t^{\p})=\delta(t-t^{\p})\,\check{\mathcal{Q}}\label{eq_green},
\end{align}
\end{subequations}
where $\check{\mathcal{I}}$ is the $2n\times 2n$ identity matrix. To obtain Eq.~\eqref{keldysh}, 
we use the identity $\check{C}^{>}(t,t)-\check{C}^{<}(t,t)=\check{\mathcal{Q}}$, that follows from 
the canonical commutations relations. 

To make the notation compact, we write the correlation function in a block structure as 
(\textit{Keldysh space})$\,\otimes\,$(\textit{symplectic space}) in its irreducible representation, 
namely
\begin{align}\label{structure_KS}
\breve{\mathcal{C}}(t,t^{\p})=&
\begin{pmatrix}
\check{C}^{K} & \check{C}^{r}\\
\check{C}^{a} & \check{0} 
\end{pmatrix}(t,t^{\p})
\nonumber\\
\equiv &\;\sigma_{1}\otimes\check{\mathcal{G}}(t,t^{\p})
 + \text{homogeneous solution},
\end{align}
where $\sigma_1$ is the first Pauli matrix. Note that $\check{\mathcal{G}}(t,t^{\p})$ has also a 
symplectic structure and satisfies (by inspection) the equation of motion 
\begin{align}
\left(\check{\mathcal{I}}\,\frac{\partial}{\partial t}+\check{\mathcal{K}}(t)\right)\cdot \check{\mathcal{G}}(t,t^{\p}) = \delta(t-t^{\p})\, \check{\mathcal{Q}},
\label{diff}
\end{align}
with a self-adjoint equation
\begin{equation}\label{diff_self}
\check{\mathcal{G}}(t,t^{\p})\cdot\left(\check{\mathcal{I}}\,\overleftarrow{\frac{\partial}{\partial t^{\p}}} + 
\check{\mathcal{K}}^{\text{T}}(t^{\p})\right) =- \delta(t-t^{\p})\,\check{\mathcal{Q}}.
\end{equation}
Using Eqs.~\eqref{diff} and \eqref{diff_self} we obtain the following identity
\begin{align}
\frac{\ud}{\ud t}\check{\mathcal{G}}(t,t) &\equiv\left(\frac{\partial}{\partial t}+
\frac{\partial}{\partial t^{\p}}\right)\check{\mathcal{G}}(t,t^{\p})\Bigg\vert_{t=t^{\prime}}\nonumber\\
&=-\check{\mathcal{K}}(t)\cdot \check{\mathcal{G}}(t,t) - 
\check{\mathcal{G}}(t,t)\cdot\check{\mathcal{K}}^{\text{T}}(t).
\end{align}

Performing the Keldysh rotation \cite{Kamenev2011} in Eq.~\eqref{structure_KS}, we obtain 
a reducible representation of the correlation function in terms of the quantities defined in 
Eq.~\eqref{set_6} as
\begin{align}
\breve{\mathcal{P}}\cdot\breve{\mathcal{C}}(t,t^{\p})\cdot\breve{\mathcal{P}}^{\text{T}}=&
\begin{pmatrix}
\check{C}^{\mathbb{T}} & \check{C}^{<}\\
\check{C}^{>} & \check{C}^{\overline{\mathbb{T}}} 
\end{pmatrix}(t,t^{\p})\nonumber\\
\equiv &\;\sigma_{3}\otimes\check{\mathcal{G}}(t,t^{\p}) + \text{homog. solution},
\end{align}
where $\breve{\mathcal{P}}=\frac{1}{\sqrt{2}}\left(I_{2} + \ii\, \sigma_{2}\right)\otimes\check{\mathcal{I}}$ 
and $\sigma_{2}$ is the second matrix of Pauli.

Let us now introduce the frequency representation of the correlation functions.
Assuming time translational invariance, {\it i.e.}, that the matrix $\check{\mathcal{K}}$ does not 
depend on time, one defines $\underline{\check{\mathcal{G}}}[\omega]$ in terms of the 
Fourier transform
\begin{equation}\label{fourier}
\underline{\check{\mathcal{G}}}[\omega] = \int_{-\infty}^{\infty}\!\ud(t-t^\prime)
\,\text{e}^{\ii\omega (t-t^{\p})}\,\underline{\check{\mathcal{G}}}(t-t^\p),
\end{equation}
for $\underline{\check{\mathcal{G}}}(t-t^\p)=\check{\mathcal{G}}(t,t^\p)$. We study the time-dependent 
problem in Sec.~\ref{sec_pumping}.

By inserting Eq.~\eqref{fourier} in \eqref{diff} [or in \eqref{diff_self}], we write
\begin{align}
\label{12}
\underline{\check{\mathcal{G}}}[\omega]=&\left(-\ii\,\omega\,\check{\mathcal{I}} + 
\check{\mathcal{K}}\right)^{-1}\cdot\check{\mathcal{Q}} 
\nonumber\\ =&  -\check{\mathcal{Q}}\cdot\left(\ii\,\omega\,\check{\mathcal{I}} +
 \check{\mathcal{K}}^{\text{T}}\right)^{-1},
\end{align}
where
\begin{equation}\label{wang}
\underline{\check{\mathcal{G}}}[\omega] \equiv 
\begin{pmatrix}
\check{\mathcal{G}}^{(uu)}[\omega]  & \check{\mathcal{G}}^{(up)}[\omega] \\
\check{\mathcal{G}}^{(pu)}[\omega]  & \check{\mathcal{G}}^{(pp)}[\omega]
\end{pmatrix}
=
\begin{pmatrix}
\hat{G}[\omega]  & \ii\,\omega\,\hat{G}[\omega]\\
-\ii\,\omega\,\hat{G}[\omega]   & \hat{G}[\omega]\cdot\hat{K}
\end{pmatrix},
\end{equation}
with 
\begin{equation}\label{G_wang}
\hat{G}[\omega]=(\omega^{2}\,\hat{I} - \hat{K})^{-1}.
\end{equation}

Equation~\eqref{wang} has been obtained in Ref.~\onlinecite{Wang2007} by directly taking the 
Fourier transform of the displacement $\lbrace u_{i}\rbrace$ and the canonically conjugate 
momentum operators $\lbrace p_{i}\rbrace$.
We note that despite being very appealing, this straightforward  procedure is {formally} problematic, 
since the canonical commutation relations  $\left[u_{i}(t), p_{j}(t)\right]$ can not be consistently 
defined in the frequency domain {(see Appendix~\ref{commutation} for more details)}. 
This problem can be circumvented \cite{Wang2014} by performing the Fourier transform 
of the phase space correlation functions, as described above.

The Green's function $\hat{G}[\omega]$ can be represented as
\begin{equation}
\label{xx}
\hat{G}[\omega] =\frac{1}{2}\int_{-\infty}^{\infty}\dfrac{\ud \bar{\omega}}{2\pi}\,\hat{J}(\bar{\omega})
\,\left(\frac{1}{\omega-\bar{\omega}}-\frac{1}{\omega+\bar{\omega}}\right),
\end{equation}
where the spectral operator $\hat{J}(\bar{\omega})$ is
\begin{equation}
\hat{J}(\omega) = 
2\pi \sum_{j}\frac{1}{\omega_{j}}\,\delta(\omega-\omega_{j})\, \vert j\rangle\langle j\vert.
\end{equation}
Here we have used that $\hat{K}$ is a positive-semidefinite matrix \cite{Bollobas2013},
which sa\-tis\-fies $\hat{K}\vert j\rangle = \omega_{j}^2\,\vert j\rangle$ with $\omega_{j}\geqslant 0$ 
(recall that $\langle j\vert j^{\p}\rangle=\delta_{j, j^{\p}}$ and $\sum_{j}\vert j\rangle\langle j\vert =\hat{I}$). 

The general expression \eqref{xx} does not distinguish the retarded, advanced, ordered, 
and anti-ordered components of $\hat{G}[\omega]$. 
A proper representation of the components requires a re\-gu\-la\-ri\-za\-tion around the 
poles $\omega =\pm\bar{\omega}$ of Eq.~\eqref{xx}, namely
\begin{subequations}
\label{green_14}
\begin{align}
\hat{G}^{r,a}[\omega] & = \frac{1}{2}\int\limits_{-\infty}^{\infty}\dfrac{\ud \bar{\omega}}{2\pi}\,
\hat{J}(\bar{\omega})\left(\frac{1}{\omega-\bar{\omega}\pm \ii 0^{+}}-\frac{1}{\omega+\bar{\omega}\pm \ii 0^{+}}\right)
\nonumber\\
& =\left[(\omega\pm\ii 0^{+})^{2}\,\hat{I}-\hat{K}\right]^{-1},
\label{15a}\\
\hat{G}^{\mathbb{T},\overline{\mathbb{T}}}[\omega] & = \frac{1}{2}\int\limits_{-\infty}^{\infty}\dfrac{\ud \bar{\omega}}{2\pi}\,\hat{J}(\bar{\omega})\left(\frac{1}{\omega-\bar{\omega}\pm \ii 0^{+}}-\frac{1}{\omega+\bar{\omega}\mp \ii 0^{+}}\right)\nonumber \\
& = \left[\omega^{2}\,\hat{I} - (\sqrt{\hat{K}}\mp\ii 0^{+}\,\hat{I})^{2}\right]^{-1}. \label{15b}
\end{align}
\end{subequations}
The Green's functions $\hat{G}^{r,a}(t,t^{\p})$ and $\hat{G}^{\mathbb{T},\overline{\mathbb{T}}}(t,t^{\p})$ 
are obtained by the inverse Fourier transform of Eqs.~\eqref{green_14} and are consistent with 
Eqs.~\eqref{set_5} and \eqref{set_6}, as they should.

Substituting Eqs.~\eqref{15a} and \eqref{wang} in the inverse Fourier transform Eq.~\eqref{fourier}, we 
write the retarded component of $\underline{\mathcal{G}}(t-t^{\p})$  as
\begin{subequations}
\begin{multline}
\underline{\check{\mathcal{G}}}^{r}(t-t^{\p})=\theta(t-t^{\prime})\\
\times\begin{pmatrix}
-\frac{\sin\left[\sqrt{\hat{K}}(t-t^{\p})\right]}{\sqrt{\hat{K}}} & \cos\left[\sqrt{\hat{K}}\,(t-t^{\p})\right]\\
-\cos\left[\sqrt{\hat{K}}\,(t-t^{\p})\right] & -\sqrt{\hat{K}}\,\sin\left[\sqrt{\hat{K}}(t-t^{\p})\right]
\end{pmatrix}\\
+\;\text{solution of homogeneous equation},
\end{multline}
and $\underline{\check{\mathcal{G}}}^{a}(t-t^{\p}) = -\,\underline{\check{\mathcal{G}}}^{r}(t^{\p}-t)$, 
where $\check{\mathcal{G}}^{r,a}(0^{\pm})=\mathcal{\check{Q}}$. 

Similarly, the ordered and anti-ordered components read
\begin{multline}
\underline{\check{\mathcal{G}}}^{\mathbb{T},\overline{\mathbb{T}}}(t-t^{\p})=\\
\begin{pmatrix}
\frac{1}{2\ii\,\sqrt{\hat{K}}}\text{e}^{\mp\ii\sqrt{\hat{K}}\,\vert t- t^{\p}\vert}   &  \pm\frac{1}{2}\text{sgn}(t-t^{\prime})\,\text{e}^{\mp\ii\sqrt{\hat{K}}\,\vert t- t^{\p}\vert}\\
\pm\frac{1}{2}\text{sgn}(t^{\prime}-t)\,\text{e}^{\mp\ii\sqrt{\hat{K}}\,\vert t- t^{\p}\vert} &  \frac{1}{2\ii}\sqrt{\hat{K}}\cdot \text{e}^{\mp\ii\sqrt{\hat{K}}\,\vert t- t^{\p}\vert}
\end{pmatrix}
\\
+\;\text{solution of homogeneous equation},
\end{multline}
\end{subequations}
which satisfy $\check{\mathcal{G}}^{\mathbb{T}}(0^{\pm}) - \check{\mathcal{G}}^{\overline{\mathbb{T}}}(0^{\pm})=\pm\check{\mathcal{Q}}$.

The Keldysh component of the correlation function is, in general, more demanding to obtain. 
As standard, the exception is the equilibrium case. In this limit, the fluctuation-dissi\-pa\-tion 
theo\-rem \cite{Haug2008} relates the Keldysh component of the correlation function of a 
bosonic system to its retarded and advanced components as
\begin{align}\label{thermal}
\hat{G}^{K}_{\rm eq}[\omega] & =\big(\hat{G}^{r}[\omega]-\hat{G}^{a}[\omega]\big)\left(2 f(\omega)+1\right),
\end{align}
where $f(\omega)=\left(\text{e}^{\beta\hbar\omega} -1\right)^{-1}$ is the Bose-Einstein distribution function. One can also write
\begin{equation}\label{20_0}
\hat{G}^{>}_{\rm eq}[\omega] + \sigma\,\hat{G}^{<}_{\rm eq}[\omega] = \ii\,\hat{A}[\omega]\,\big( 2 f(\omega)\,\delta_{\sigma,+} + 1\big),
\end{equation}
where $\s=\pm 1$ and 
\begin{equation}
\ii\,\hat{A}(\omega)=\hat{G}^{r}[\om] - \hat{G}^{a}[\om]=\frac{1}{2\ii}\left[\hat{J}(\om)-\hat{J}(-\om)\right].
\end{equation}

As a result, the equilibrium lesser and greater Green's functions are given by
\begin{subequations}\label{21}
\begin{align}
& \hat{G}^{<}_{\rm eq}[\omega] = \ii\,\hat{A}(\omega) \,f(\omega),\\
& \hat{G}^{>}_{\rm eq}[\omega] = \ii\,\hat{A}(\omega) \,\left(f(\omega)+1\right).
\end{align}
\end{subequations}

\section{Model Hamiltonian}
\label{sec:model}

In this section, we describe the heat transport pro\-per\-ties of a molecular junction 
modeled by a central region $C$ representing a nanostructure coupled by multiple 
leads connected to reservoirs in thermal equilibrium \cite{Wang2007,Wang2008}. 
We recall that we only consider thermal transport due vibrational degrees of freedom, 
which is the dominant mechanism in insulator systems.

This partition scheme allows one to write the general Hamiltonian of Eq.~\eqref{hamil} as
\begin{equation}\label{eq:general_Hamiltonian}
H(t)=\sum_{\alpha}H_{\alpha}(t)\;  + H_{C}(t) + H_{T}(t),
\end{equation}
where 
\begin{subequations}\label{set_eq27}
\begin{align}
&H_{\alpha}(t)= H_{\alpha}^{0} +  U_{\alpha\alpha}(t),\\
&H_{C}(t) = H_{C}^{0} +  U_{CC}(t),\\
&H_{T}(t) = \sum_{\alpha}\Big[\,U_{C\alpha}(t) + U_{\alpha C}(t)\,\Big], \label{4bb}
\end{align}
\end{subequations}
correspond to the Hamiltonian of the $\alpha$-lead, central region and tunneling, respectively.
We define the decoupled Hamiltonian $H_{a}^{0}$ corresponding to the $a$-partition as
\begin{subequations}
\begin{equation}
H_{a}^{0}\equiv\frac{1}{2}\,\vec{p}^{\,\,\text{T}}_{a}\cdot\vec{p}_{a}\,+\,\frac{1}{2}\,\vec{u}^{\,\text{T}}_{a}
\cdot K_{a a}^{0}\cdot\vec{u}_{a}
\end{equation}
and the coupling Hamiltonian $U_{ab}(t)$ bet\-ween $a$ and $b$-partitions as
\begin{equation}
U_{ab}(t)\equiv\frac{1}{2}\,\vec{u}^{\,\text{T}}_{a}\cdot V_{a b}(t)\cdot\vec{u}_{b}.
\end{equation}
\end{subequations}
The force constant matrix in Eq.~\eqref{hamil} is decomposed as $\hat{K}(t)=\hat{K}^{0}+\hat{V}(t)$, 
where $\hat{K}^{0}$ gives the dynamical matrix of the decoupled partitions 
\begin{align}
\hat{K}^{0}&=\left[\bigoplus_{\alpha} K_{\alpha}^{0}\right]\oplus K^{0}_{C},\label{24a}
\end{align}
and $\hat{V}(t)$ corresponds to the coupling between different partitions, namely
\begin{align}
\hat{V}(t)&=\left[\bigoplus_{\alpha} V_{\alpha\alpha}(t)\right]\oplus V_{CC}(t)\;\; +\;\;\hat{V}_{\text{mixed}}(t).\label{24b}
\end{align}
These definitions allow us to write the tunneling Hamiltonian $H_{T}(t)$  as  
\begin{align}
H_{T}(t) =\frac{1}{2}\, \vec{u}^{\,\text{T}}\cdot\hat{V}_{\text{mixed}}(t)\cdot\vec{u},\label{4b}
\end{align}
where $\vec{u}\equiv\left[\bigoplus_{\alpha}\vec{u}_{\alpha}\right]\oplus\vec{u}_{C}$. Note 
that $\hat{V}=\hat{V}^{\text{T}}$ and therefore $V_{\alpha C}=V_{C\alpha}^{\text{T}}$ for 
all $\alpha$ terminals.

The model Hamiltonian in Eq.~\eqref{eq:general_Hamiltonian} includes $V_{aa}$ ($a=\alpha, C$) terms 
that have not been explicitly accounted for by previous works \cite{Wang2007, Wang2008, Wang2014}. 
{Neglecting $V_{aa}$ can be problematic for the consistency of NEGF. This can be seen using the adiabatic switch-on picture, the standard implementation of NEGF in the steady-state regime (A discussion of different implementation schemes can be found, for instance, in Ref.~\cite{Odashima2017}).} 
The absorption  of $V_{aa}$ into $K^{0}_{\alpha\alpha}$ modifies the free Green's functions making their calculation troublesome. This issue becomes clear in the formal development below as well as in the applications {discussed} in Sec.~\ref{sec:application}.

To discuss the thermodynamic properties of the system it is convenient to describe the molecular junction 
as formed by reservoirs coupled to an extended central region, which we refer to as ``molecule". Accordingly, we write Eq.~\eqref{eq:general_Hamiltonian} as
\begin{equation}
H(t) = \sum_{\alpha}H_{\alpha}(t) + H_{M}(t),
\end{equation}
where the molecule Hamiltonian reads
\begin{equation}
H_{M}(t)\equiv H_{C}(t) + H_{T}(t).
\end{equation}

The energy of the extended molecule is defined as $E_{M}(t)\equiv\left\langle H_{M}(t)\right\rangle$, 
namely
\begin{align}
\label{EM}
&E_{M}(t)=\frac{\ii\hbar}{2}\,{\rm Tr} \Big\lbrace C_{CC}^{<(pp)}(t,t) + K_{CC}(t)\cdot C_{CC}^{<(uu)}(t,t)
\nonumber\\
&+\sum_{\alpha}\left[V_{C\alpha}(t)\cdot C_{\alpha C}^{<(uu)}(t,t) \, +  C_{C\alpha}^{<(uu)}(t,t)\cdot V_{\alpha C}(t)\right]\Big\rbrace,
\end{align}
where the components of lesser functions are explicit given by 
\begin{subequations}
\begin{align}
\ii\hbar\,\big[C^{<(pp)}_{a b}(t,t^{\prime})\big]_{k,k^{\prime}} &= \big\langle \left[\vec{p}_{b}(t^{\prime})\right]_{k^{\prime}}\; \left[\vec{p}_{a}(t)\right]_{k} \big\rangle;\\
\ii\hbar\,\big[C^{<(uu)}_{a b}(t,t^{\prime})\big]_{n,n^{\p}} &= \big\langle \left[\vec{u}_{b}(t^{\prime})\right]_{n^{\p}}\; \left[\vec{u}_{a}(t)\right]_{n} \big\rangle ; \\
\ii\hbar\,\big[C^{<(up)}_{a b}(t,t^{\prime})\big]_{n,k} &= \big\langle \left[\vec{p}_{b}(t^{\prime})\right]_{k}\; \left[\vec{u}_{a}(t)\right]_{n} \big\rangle;\\
\ii\hbar\,\big[C^{<(pu)}_{a b}(t,t^{\prime})\big]_{k,n} &= \big\langle \left[\vec{u}_{b}(t^{\prime})\right]_{n}\; \left[\vec{p}_{a}(t)\right]_{k} \big\rangle.
\end{align}
\end{subequations}
with $a, b = \lbrace C, \alpha\rbrace$, in line with Eq.~\eqref{corr_keldy}.

One can define the  thermal current flowing through an open molecule connected to multiple 
reservoirs by comparing its energy variation 
\begin{align}\label{displacement_I}
\frac{\ud E_{M}(t)}{\ud t}=& \left\langle\frac{\ud H_{M}(t)}{\ud t}\right\rangle 
\nonumber \\ =&
 \frac{\ii}{\hbar} \left\langle\left[H(t), H_{M}(t)\right]\right\rangle +
 \left\langle\frac{\partial H_{M}(t)}{\partial t}\right\rangle
\end{align}
with the energy continuity equation, expressed as
\begin{equation}\label{displacement_II}
\frac{\ud E_{M}(t)}{\ud t} = \sum_{\alpha} J_{\alpha}(t) + \Phi(t),
\end{equation}
where one associates $J_{\alpha}(t)$ to the thermal current from $\alpha$-reservoir into 
the molecule and $\Phi(t)$ is power developed by the ac sources (or drives) in the molecule. 
Hence, by inspection one infers that
\begin{align}
J_{\alpha}(t) = -\frac{i}{\hbar}\left\langle\left[H(t), H_{\alpha}(t)\right]\right\rangle  \label{IL}
\end{align}
and
\begin{align}\label{PHII}
& \Phi(t) = \left\langle\frac{\partial H_{M}(t)}{\partial t}\right\rangle. 
\end{align}

{Using the equation-of-motion method \cite{Haug2008}}, we write the thermal current from $\alpha$-reservoir into the molecule in terms of the 
correlation functions as 
\begin{align}
J_{\alpha}(t) 
=&\,\text{Re}\left[\text{Tr}\left\lbrace V_{C\alpha}(t)\cdot\ii\hbar\,C^{<(pu)}_{\alpha C}(t,t)\right\rbrace\right],\label{eq_ILL}
\end{align}
while the power developed by the external time-dependent drives reads
\begin{align}
\Phi(t)  
= & \text{Re}\left[\text{Tr}\left\lbrace \frac{1}{2}\,\dot{V}_{CC}(t)\cdot \ii\hbar\,C_{CC}^{<(uu)}(t,t)\right.\right.\nonumber\\
&\left.\left.  \qquad\qquad + \sum_{\alpha}\dot{V}_{C\alpha}(t)\cdot \ii\hbar\,C_{\alpha C}^{<(uu)}(t,t) \right\rbrace\right].\label{eq_Phii}
\end{align}

In the following subsections we study separately the steady-state transport ($\dot{V}_{ab}=0$) and the heat transport due to pumping by an external drive ($\dot{V}_{ab}\neq 0$) for $a,b = \lbrace \alpha, C\rbrace$.

\subsection{Steady-state transport}\label{Sec_Sst}

Let us now calculate the steady-state thermal current flowing from the $\alpha$-lead due to a 
temperature difference in the reservoirs. Here, we consider the heat current expression \eqref{IL} 
for a time-independent coupling matrix $\hat{V}$. 

Since the Hamiltonian does not explicitly depends on time, it is convenient to work in the 
frequency representation. The Fourier transform of $C^{<(up)}_{C\alpha}(t,t^{\p})$ is 
\begin{align}\label{eq_29}
C_{C\alpha}^{<(up)}(t,t^{\p}) &=\int_{-\infty}^{\infty}
\frac{\ud\omega}{2\pi}\,\text{e}^{-\ii\omega (t-t^{\p})}\,C_{C\alpha}^{<(up)}[\omega],
\end{align}
where $C_{C\alpha}^{<(up)}[\omega]=\ii\,\omega\,G^{<}_{C\alpha}[\omega]$. 
Substituting Eq.~\eqref{eq_29} into Eq.~\eqref{IL}, we cast the steady-state heat current as 
\begin{equation}
\label{40}
J_{\alpha}^{(S)}=\int_{-\infty}^{\infty}\,\frac{\ud\omega}{4\pi}\,\hbar\omega\,
{\rm Tr}\!\left\{ V_{C\alpha}\cdot G^{<}_{\alpha C}[\omega] - G^{<}_{C\alpha}[\omega]\cdot V_{\alpha C} \right\} .
\end{equation}

The system Green's function $\hat{G}[\omega]=\big(\omega^2\,\hat{I} - \hat{K}\big)^{-1}$  satisfies the Dyson equation
\begin{align}
\label{dyson_I}
\hat{G}[\omega] & = \hat{g}[\omega]+\hat{g}[\omega]\cdot\hat{V}\cdot\hat{G}[\omega] 
\nonumber \\ & = 
 \hat{g}[\omega]+\hat{G}[\omega]\cdot\hat{V}\cdot\hat{g}[\omega],
\end{align}
where $\hat{K}=\hat{K}^{0}+\hat{V}$ and $\hat{g}[\omega]=\big(\omega^2\,\hat{I} - \hat{K}^{0}\big)^{-1}$. 
Note that the free Green's function $ \hat{g}[\omega]$ is block diagonal in the partitions. 

From Eq.~\eqref{dyson_I} we obtain 
\begin{subequations}\label{43}
\begin{align}
& G_{C\alpha}[\om] =G_{CC}[\om]\cdot V_{C\alpha}\cdot \tilde{g}_{\alpha}[\om],\label{34a}\\
& G_{\alpha C}[\om] =\tilde{g}_{\alpha}[\om]\cdot V_{\alpha C}\cdot G_{CC}[\om],\label{34b}\\
& G_{CC}[\om] = \left(\tilde{g}_{C}[\om]^{-1}-\tilde{\Sigma}[\om]\right)^{-1},\label{34c}\\
& G_{\alpha\beta}[\om] = \tilde{g}_{\alpha}[\om]\cdot V_{\alpha C}\cdot G_{CC}[\om]\cdot V_{C\beta}\cdot\tilde{g}_{\beta}[\om]\nonumber\\
&\qquad\quad\quad + \delta_{\alpha\beta}\;\tilde{g}_{\alpha}[\om],\label{34d}
\end{align}
\end{subequations}
where, for notational convenience, we introduce an \textit{effective embedding self-energy} 
\begin{equation}\label{eq:self-energy_VgV}
\tilde{\Sigma}[\om]=\sum_{\alpha}\tilde{\Sigma}_{\alpha}[\om] =
\sum_\alpha V_{C\alpha}\cdot\tilde{g}_{\alpha}[\om]\cdot V_{\alpha C},
\end{equation}
and an \textit{effective free Green's function}
\begin{align}\label{set_32}
&\tilde{g}_{a}[\om]^{-1}=g_{a}[\om]^{-1} - V_{aa}\quad\text{with}\quad a=\lbrace \alpha,C\rbrace,
\end{align}
where $g_{a}[\om]=(\om^2 I_{a}- K^{0}_{a})^{-1}$. 
{ 
In Sec.~\ref{sec_application_ss} and in Appendix~\ref{sec:freegf} we discuss the importance of 
including $V_{aa}$ in the surface Green's function.
}
For $a=\alpha$, it corresponds to Green's function in thermal equilibrium with the $\alpha$-reservoir 
at a temperature $T_{\alpha}$.
Hence, using Eqs.~\eqref{15a} and \eqref{21} we write obtain 
\begin{subequations}\label{set_33}
\begin{align}
& g^{<}_{\alpha}[\omega] =\ii A_{\alpha}(\omega)\,f_{\alpha}(\om),\\
& g^{>}_{\alpha}[\omega] = \ii A_{\alpha}(\omega)\,\Big(1+f_{\alpha}(\om)\Big),\\
& g^{r,a}_{\alpha}[\omega] =\left[\left(\omega \pm \ii 0^{+}\right)^{2}\,I_{\alpha}-K^{0}_{\alpha}\right]^{-1},\label{set_33_c}
\end{align}
\end{subequations}
where $\ii A_{\alpha}(\omega) \equiv g^{r}_{\alpha}[\om]-g^{a}_{\alpha}[\om]$ is the 
\textit{$\alpha$-lead ``free" spectral function} and $f_{\alpha}(\om) =
\left(\text{e}^{\beta_{\al}\hbar\om}-1\right)^{-1}$ with $\beta_{\alpha}=1/k_{B}T_{\alpha}$.
{
In general, the retarded and advanced surface Green's functions ${g}_{\alpha}^{r,a}[\om]$ are
computed by de\-ci\-ma\-tion techniques \cite{Sancho1985,Wang2007}.
}

The lesser components of $G_{\alpha C}$ and $G_{C\alpha}$ are obtained by applying the Langreth rules \cite{Rammer1986,Haug2008} to Eq.~\eqref{43}. By in\-ser\-ting the result in Eq.~\eqref{40}, we obtain
\begin{align}
\label{20}
J_{\alpha}^{(S)} = \int_{-\infty}^{\infty} &\frac{\ud\om}{4\pi}\,\hbar\omega\,\text{Tr} \left\lbrace  
G^{<}_{C C}[\om]\cdot\big(\tilde{\Sigma}_{\alpha}^{r}[\om]-\tilde{\Sigma}_{\alpha}^{a}[\om]\big)\right.
\nonumber \\
&\ \ \ \ \ \ \ \left.  -\big(G^{r}_{C C}[\om] - G^{a}_{C C}[\om]\big)\cdot\tilde{\Sigma}_{\alpha}^{<}[\om]
\right\rbrace.
\end{align}

The self-energies are given in terms of
\begin{subequations}
\begin{align}
& \tilde{g}^{<}_{\alpha}[\omega] =\ii \tilde{A}_{\alpha}(\omega)\,f_{\alpha}(\om),\label{37a}\\
& \tilde{g}^{>}_{\alpha}[\omega] = \ii\, \tilde{A}_{\alpha}(\omega)\,\bigl(1+f_{\alpha}(\om)\bigl),\label{37b}\\
& \tilde{g}^{r,a}_{\alpha}[\omega] =\left[\left(\omega \pm \ii 0^{+}\right)^{2}\,I_{\alpha}-K_{\alpha}\right]^{-1},\label{37c}
\end{align}
\end{subequations}
where $K_{\alpha}=K_{\alpha}^{0}\,+\,V_{\alpha\alpha}$ and $\ii\,\tilde{A}_{\alpha}(\omega)=\tilde{g}^{r}_{\alpha}[\om]-\tilde{g}^{a}_{\alpha}[\om]$.
Hence, 
\begin{subequations}\label{set:gamma}
\begin{equation}\label{width}
\tilde{\Sigma}_{\alpha}^{r}[\om]-\tilde{\Sigma}_{\alpha}^{a}[\om] = 
\ii V_{C\alpha}\cdot\tilde{A}_{\alpha}[\om]\cdot V_{\alpha C}
\equiv-\ii \,\tilde{\Gamma}_{\alpha}[\om],
\end{equation}
where $\tilde{\Gamma}_{\alpha}[\om]$ is the $\alpha$-contact line width function. Similarly, 
\begin{equation}\label{Eq_48b}
\tilde{\Sigma}_{\alpha}^{<}[\om] = V_{C\alpha}\cdot\tilde{g}^{<}_{\al}[\om]\cdot V_{\al C} = -\ii\, f_{\alpha}(\om)\,\tilde{\Gamma}_{\al}[\om].
\end{equation}
\end{subequations}

By expressing the self-energies in terms of the line width functions, we write the heat current as
\begin{multline}\label{51}
J_{\alpha}^{(S)} =\int_{-\infty}^{\infty}\frac{\ud\om}{4\pi\ii}\,\hbar\omega\,\text{Tr}\left\lbrace\tilde{\Gamma}_{\alpha}[\omega]\cdot\Big[G^{<}_{CC}[\omega]\right. \\
\left. -\, f_{\alpha}(\omega)\Big(G^{r}_{CC}[\omega]-G^{a}_{CC}[\omega]\Big)\Big]\right\rbrace.
\end{multline}

Applying the Langreth rules to Eq.~\eqref{34c} and using Eq.~\eqref{set:gamma}, we obtain 
\begin{subequations}\label{52}
\begin{align}
& G^{<}_{CC}[\om]=-\sum_{\alpha}G^{r}_{CC}[\om]\cdot\ii\,\tilde{\Gamma}_{\alpha}[\om]\cdot G^{a}_{CC}[\om]\;f_{\alpha}(\om),\\
& G^{r}_{CC}[\om]-G^{a}_{CC}[\om]=-\sum_{\alpha} G^{r}_{CC}[\om]\cdot\ii\,\tilde{\Gamma}_{\alpha}[\om]\cdot G^{a}_{CC}[\om],
\end{align}
\end{subequations}
that are inserted  in Eq.~\eqref{51} to finally arrive at the steady-state heat current
\begin{equation}
\label{53}
J_{\alpha}^{(S)}=\sum_{\beta} \int_{0}^{\infty}\frac{\ud\omega}{2\pi}\hbar\omega\, \mathcal{T}_{\alpha\beta}(\omega)\, \Big[f_{\alpha}(\omega)-f_{\beta}(\omega)\Big],
\end{equation}
where
\begin{equation}
\label{transmission}
\mathcal{T}_{\alpha\beta}(\omega)\equiv \text{Tr}\left\{\tilde{\Gamma}_{\alpha}[\omega]\cdot G^{r}_{CC}[\omega]
\cdot\tilde{\Gamma}_{\beta}[\omega]\cdot G^{a}_{CC}[\omega]\right\},
\end{equation}
rigorously obtaining the Landauer heat conductance that has been phenomenologically put forward \cite{Rego1998} 
and adopted by several authors, see for instance, Refs.~\onlinecite{Mingo2003,Chalopin2013, Zhang2007}. 
As a consequence, the numerical implementation of the heat current $J_{\alpha}^{(S)}$ given by Eq.~\eqref{53}, 
is obviously the same as the one using the 
scattering matrix \cite{Mingo2003,Zhang2007}. 

{
The explicitly symmetric tunneling Hamiltonian $H_T(t)$, Eq.~\eqref{4bb}, leads to an 
expression for the heat current $J_\alpha(t)$ with terms depending on both $V_{C\alpha}$ 
and $V_{\alpha C}$.  This ensures that  $J_\alpha(t)$ accounts for processes corresponding 
to the heat flow from the central region $C$ to the $\alpha$-lead as well as from $\alpha$ to $C$.
Our result differs from the heat current derived by Wang and collaborators \cite{Wang2007,Wang2008,Wang2014}. 
These authors derive the heat current using the Hamiltonian without explicitly taking into account processes
associated to $V_{L C}$ (corresponding to $\alpha=L$). 
The obtained expression for heat current depends only on the hybrid Green's function $G^{<}_{CL}$.
Furthermore, the absence of $V_{\alpha C}$ (or $V_{C\alpha}$) in their Hamiltonian implies that the
self-energy $\Sigma_{L}=V_{CL}\cdot g_{L}\cdot V_{LC}$ has to be introduced in a somewhat arbitrary manner. 
Moreover, Refs.~\cite{Wang2007,Wang2008,Wang2014} need the {\it ad hoc} symmetrization, $J = (J_L + J_L^* - J_R - J_R^*)/4$, to obtain the well known Caroli formula for the transmission since the integrand of Eq.~(\ref{40}) is not purely real in the absence of $V_{\alpha C}$ (or $V_{C\alpha}$).
}



The transmission coefficient $\mathcal{T}_{\alpha\beta}(\om)$ is interpreted as the probability of 
an energy  $\hbar\omega$ to be transmitted from the reservoir $\alpha$ to the reservoir $\beta$ 
and has the same structure of the Meir-Wingreen formula \cite{Meir1992} that describes the 
electronic conductance of fully coherent systems of non-interacting electrons.

It is straightforward  to verify that $\mathcal{T}_{\alpha\beta}(\omega)=\mathcal{T}_{\beta\alpha}(\omega)$, 
which implies that in steady-state $J^{(S)}\equiv J_{L}^{(S)} = -J_{R}^{(S)}$. Hence, $\ud E_{M}/\ud t=0$ 
and, as expected, the molecule energy does not change in time.

\subsection{Pumping transport}\label{sec_pumping}

Let us now study the heat current in nanoscopic systems due to a time-dependent external drive, 
as motivated in the introduction. 
As in the stationary case, we employ the NEGF theory, since more standard approaches, like the
Kubo-Greenwood one, are only suitable for bulk systems.

The analysis of heat currents in time-dependent systems is far more involved for bosonic degrees of 
freedom than for the electronic ones. In the latter case, the Fermi energy (and the corresponding 
Fermi velocity) establishes a characteristic time scale for the electronic dynamics. 
In experiments \cite{Switkes1999} the external driving is slow with respect to the electronic dynamics,
which allows to approach the problem using the adiabatic approximation \cite{Buttiker1994, Brouwer1998, 
Vavilov2001, Mucciolo2007, Hernandez2009}.
In the bosonic case there is no internal characteristic time scale and analytical progress has to resort on 
the assumption that the driving force is small to employ perturbation theory. 

As an example of time-dependent transport, we study the case of periodically driven system in time. 
We assume that the coupling between regions depends on time as $\hat{V}(t)=\hat{V} + \varepsilon\,\hat{v}(t)$, where $\varepsilon$ is a dimensionless parameter.
{
The initial state is the fully connected molecule-leads system in equilibrium.
}
Defining an auxiliary matrix $\check{\mathcal{V}}(t)$ as
\begin{align}\label{perturb}
\check{\mathcal{V}}(t)=\varepsilon\,
\begin{pmatrix}
\hat{v}(t) & \hat{0}\\
\hat{0} & \hat{0}
\end{pmatrix},
\end{align}
we can write $\check{\mathcal{K}}(t) = \check{\mathcal{K}} - \check{\mathcal{Q}}\cdot\check{\mathcal{V}}(t)$ or, equivalently, $\check{\mathcal{M}}(t)=\check{\mathcal{M}}+\check{\mathcal{V}}(t)$. 
It follows from Eq.~\eqref{diff} that the Dyson's equation reads
\begin{equation}\label{dyson_II}
\check{\mathcal{G}}(t,t^{\p})=\underline{\check{\mathcal{G}}}(t-t^{\p}) + \int\ud\bar{t}\;\underline{\check{\mathcal{G}}}(t-\bar{t})\cdot\check{\mathcal{V}}(\bar{t})\cdot\check{\mathcal{G}}(\bar{t},t^{\p}),  
\end{equation}
where $\underline{\check{\mathcal{G}}}(t-t^{\p})$ denotes the steady-state Green's function transport, 
given by Eqs.~\eqref{fourier} to \eqref{G_wang}.  
We consider $\varepsilon\ll 1$ and treat the problem using pertubation theory. 
This is an alternative approach to the Floquet analysis used in Refs.~\onlinecite{Arrachea2013,Beraha2016}.
{
We note that the Floquet method is extremely efficient, irrespective of coupling strength, provided the 
ratio between the band width and the driving frequency is not large, a condition that keeps the size of the 
Hilbert space computationally manageable. 
The opposite limit of small $\Omega$ is in general computationally prohibitive for this method.
For electronic systems, however, it has been argued that if the characteristic single particle dwell time 
$\tau_d$ (evaluated at the Fermi energy) in the scattering region is much smaller than $1/\Omega$ only 
few harmonics of the perturbation are coupled. This allows for an effective truncation of the Hilbert space. 
The dwell time $\tau_d$ depends on the spectral density and on the strength of its coupling to the 
leads \cite{Lewenkopf2004}.
Since these quantities typically show a strong energy dependence, one has to verify if $\tau_d\Omega \ll 1$ 
is indeed fulfilled. In general the latter condition rules out the application of the Floquet approach for small 
$\Omega$ to a potentially large number of systems. 
}

The Green's function deviation from steady-state, $\delta \check{\mathcal{G}}(t,t^{\p})
\equiv \check{\mathcal{G}}(t,t^{\p}) - \underline{\check{\mathcal{G}}}(t-t^{\p})$, is conveniently represented by
\begin{align}
\delta \check{\mathcal{G}}(t,t^{\p}) & = \iint\frac{\ud\omega\,\ud\omega^{\p}}{(2\pi)^2}\,\text{e}^{-\ii(\omega t - \omega^{\p} t^{\p})}\;\delta\check{\mathcal{G}}[\omega,\omega^{\p}],
\end{align}
where
\begin{align}
&\delta\check{\mathcal{G}}[\omega,\omega^{\p}]=
\begin{pmatrix}
1 & \ii\,\omega^{\p}\\
-\ii\,\omega & \om\,\om^{\p}
\end{pmatrix}\otimes\sum_{n\geqslant 1}\varepsilon^{n}\,\hat{\Lambda}_{n}[\om,\om^{\p}],
\end{align}
and the set $\left\lbrace\hat{\Lambda}_{n}[\om,\om^{\p}]\right\rbrace$ is defined by the recurrence relation
\begin{subequations}\label{eqs_recurrs}
\begin{equation}\label{eq_recurr}
\hat{\Lambda}_{n}[\om,\om^{\p}]=\hat{G}[\om]\cdot\int\limits_{-\infty}^{\infty}\frac{\ud\nu}{2\pi}\; \hat{v}[\om-\nu]\cdot\hat{\Lambda}_{n-1}[\nu,\omega^{\p}],
\end{equation}
with
\begin{align}\label{set_49}
\hat{\Lambda}_{1}[\omega,\omega^{\p}]=\hat{G}[\omega]\cdot\hat{v}[\om-\om^{\p}]\cdot\hat{G}[\omega^{\p}],
\end{align}
\end{subequations}
where {\small$\hat{G}[\omega]=\big(\omega^2\,\hat{I}-\hat{K}\big)^{-1}$} has been discussed in the previous section and
\begin{equation}\label{fourier_v}
\hat{v}[\om]=\int_{-\infty}^{\infty}\ud t\;\hat{v}(t)\;\text{e}^{\ii \omega t}.
\end{equation}

We model the coupling terms as
\begin{subequations}\label{set_55}
\begin{align}
&v_{\alpha C}(t) =\phi_{\alpha}(t)\,V_{\alpha C},\\
&v_{C \alpha}(t) =\phi_{\alpha}(t)\,V_{C\alpha},\\
&v_{\alpha\alpha}(t) =\phi_{\alpha}(t)\,V_{\alpha\alpha},\\
&v_{CC}(t) = \sum_{\alpha}\phi_{\alpha}(t)\,V_{CC}^{(\alpha)}\label{59d},
\end{align}
\end{subequations}
where $\phi_{\alpha}(t)$ is a dimensionless function that describes the pumping time-dependence 
of the $\alpha$-lead. For a periodic pumping, {\it i.e.}, $\phi_{\alpha}(t+\tau)=\phi_{\alpha}(t)$ the pumping function can be expressed by a Fourier series in harmonic form as 
\begin{equation}\label{pumping}
\phi_{\alpha}(t)=\sum_{n=1}^{\infty}2 a_{n}^{(\alpha)}\cos(\Omega_{n}\,t+\varphi_{n}^{(\alpha)})\;\;\text{for}\;\;\Omega_{n}=n\,\frac{2\pi}{\tau}.
\end{equation}
By construction {\small$\big\langle\phi(t)\big\rangle_{\tau}=0$}, where {\small$\langle\cdots\rangle_{\tau}
\equiv\frac{1}{\tau}\int_{0}^{\tau}\ud t\,(\ldots)$} stands for the time average over a period. We assume that $\vert\phi_{\alpha}(t)\vert_{\text{m{a}x}}=1$.

Expanding the Dyson equation, Eq.~\eqref{dyson_II}, in a power series in $\varepsilon$, 
we write the energy $E_{M}(t)$ of the extended molecule as 
\begin{equation}\label{energy_perturbation}
E_{M}(t) =  E_{M}^{(0)} + \varepsilon \,E_{M}^{(1)}(t) + \varepsilon^2 \,E_{M}^{(2)}(t) + \cdots .
\end{equation}
The explicitly expression for $E_{M}^{(n)}(t)$ are rather lengthy and are given in Appendix~\ref{sec:pertubative}. 
For a periodic pumping we show that $E^{(0)}_{M}$ does not depend on time and  
$E_{M}^{(n)}(t)=E_{M}^{(n)}(t+\tau)$ for $n=1, 2, \ldots$ (see Appendix~\ref{sec:pertubative}).

We express the variation of the extended molecule energy between $t$ and $t+\Delta t$ in the form of a first law of thermodynamics, namely, $\Delta E_{M}^{(\Delta t)} \equiv \sum_{\alpha}Q_{\alpha}^{(\Delta t)} + W^{(\Delta t)}$. 
Note that $-Q_{\alpha}^{(\Delta t)}$ corresponds the heat transferred from the molecule to the $\alpha$-reservoir, 
while $W^{(\Delta t)}$ is the energy transferred to the molecule that does not come from reservoirs, namely,
\begin{align}\label{set_65}
Q_{\alpha}^{(\Delta t)}=\int_{t}^{t+\Delta t}\ud\bar{t}\;J_{\alpha}(\bar{t})\quad\text{and}\quad W^{(\Delta t)} = \int_{t}^{t+\Delta t}\ud\bar{t}\;\Phi(\bar{t}),
\end{align}
where $J_{\alpha}(t)$ and $\Phi(t)$ are, respectively, the thermal current flowing from $\alpha$-reservoir into the molecule and the power developed by the ac sources.

For a pe\-rio\-dic process after a cycle of period $\Delta t=\tau$, we finding 
that $\Delta E_{M}^{(\tau)} = 0$, so that
\begin{equation}\label{eqs_Q+W}
\sum_{\alpha}Q_{\alpha}^{(\tau)}+ W^{(\tau)}=0,
\end{equation}
where we define
\begin{align}\label{eqs_Q_W}
Q_{\alpha}^{(\tau)} = \tau\,\left\langle J_{\alpha}(t)\right\rangle_{\tau}
\qquad\text{and}\qquad 
W^{(\tau)} = \tau\,\left\langle \Phi(t)\right\rangle_{\tau}. 
\end{align}
$\left\langle J_{\alpha}(t)\right\rangle_{\tau}$ and $\left\langle \Phi(t)\right\rangle_{\tau}$ can be evaluated 
by using a perturbative expansion
\begin{subequations}\label{cycles}
\begin{align}
\left\langle J_{\alpha}(t)\right\rangle_{\tau} &= J_{\alpha}^{(S)} + \varepsilon^{2}\,J_{\alpha}^{(P)} + \mathcal{O}(\varepsilon^3),\label{J_cycle}\\
\left\langle \Phi(t)\right\rangle_{\tau} &=  \varepsilon^{2}\,\Phi^{(P)}\; + \;\mathcal{O}(\varepsilon^4)\label{Phi_cycle},
\end{align}
\end{subequations}
where $J_{\alpha}^{(P)}$ and $\Phi^{(P)}$ are discussed in Appendix~\ref{sec:pertubative} and can be 
cast as
\begin{subequations}\label{set_69}
\begin{align}
J^{(P)}_{\alpha}=&\sum_{n=1}^{\infty}\sum_{\beta\gamma}a_{n}^{(\beta)}a_{n}^{(\gamma)}\bigg[\cos\left(\varphi_{n}^{(\beta)}-\varphi_{n}^{(\gamma)}\right)\,A_{\beta\gamma}^{\alpha}(n)\nonumber\\
&-\sin\left(\varphi_{n}^{(\beta)}-\varphi_{n}^{(\gamma)}\right)\,B_{\beta\gamma}^{\alpha}(n)\bigg],\\
\Phi^{(P)}=&\sum_{n=1}^{\infty}\sum_{\beta\gamma}a_{n}^{(\beta)}a_{n}^{(\gamma)}\bigg[\cos\left(\varphi_{n}^{(\beta)}-\varphi_{n}^{(\gamma)}\right)\,D_{\beta\gamma}(n)\nonumber\\
&-\sin\left(\varphi_{n}^{(\beta)}-\varphi_{n}^{(\gamma)}\right)\,E_{\beta\gamma}(n)\bigg],
\end{align}
\end{subequations}
where the quantities $A_{\beta\gamma}^{\alpha}(n)$, $B_{\beta\gamma}^{\alpha}(n)$, 
$D_{\beta\gamma}(n)$, and $E_{\beta\gamma}(n)$ are given by intricate expressions involving combinations of equilibrium Green's functions.
The latter are explicitly given by Eq.~\eqref{set_B14}. 

Note that in Eq.~(\ref{cycles}) the first order contributions in $\varepsilon$ vanish.
The second order terms $J_{\alpha}^{(P)}$ and $\Phi^{(P)}$ depend explicitly on the periodic 
profile $\phi_\alpha(t)$.


The perturbative approach we put forward allows us to write the pumping currents and power order by order in terms 
of products and sums of steady-state Green's functions, which are represented by square matrices of the order of the 
number of degrees of freedom of the system. 
Hence, here the numerical bottleneck for addressing realistic systems is the same as in the steady-state, namely, the 
calculation of the equilibrium Green's functions as a function of the frequency.
Having obtained these objects by any standard method, one needs only to insert the corresponding quantities in the 
expressions given in the App.~\ref{sec:pertubative}. 
We note that the non-perturbative regime requires a calculation of the system Green's functions by directly solving the 
corresponding differential equations, that is in general a very challenging task.

\section{Application: Molecular junction}
\label{sec:application}

\begin{figure}[htb]
\centering
		\includegraphics[width=1.00\columnwidth]{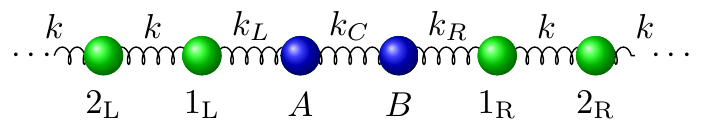}
\caption{(Color online) Sketch of the model system. 
Balls represent the chain sites while springs represent the coupling potential. The central region, formed 
by 2 atoms $A$ and $B$ coupled by a spring with force constant $k_C$, is connected  to left and right 
semi-infinite leads through couplings $k_L$ and $k_R$, respectively. The leads have a constant coupling 
$k$.}
\label{fig:cadeia_linear}
\end{figure}

We investigate the consequences of our findings using the molecular junction model presented 
in Sec.~\ref{sec:model}. We consider a one-dimensional system where a central region with $N$ 
atoms is attached to two semi-infinite linear chains acting as leads, as depicted in Fig.~\ref{fig:cadeia_linear}. 

For the sake of clarity, we consider the simplest non trivial case of a diatomic molecule, namely, $N=2$.
The force constant between the atoms in the leads and its first neighbors is $k$. 
The force constant between the atoms in the central region is $k_C$ while the left (right) atom connects 
to the left (right) lead though a coupling $k_L$ ($k_R$).

In this model the inter-partition and central coupling reduced matrices are
\begin{subequations}\label{set_coupling}
{\small\begin{align}
V_{LL}& = 
\begin{pmatrix}
  k_{L}
\end{pmatrix},&
V_{LC}& = 
\begin{pmatrix}
 -\,k_{L} &  0                
\end{pmatrix},&
V_{LR}& = 
\begin{pmatrix}
  0
\end{pmatrix}&\nonumber\\
V_{CL}& = 
\begin{pmatrix}
 -k_{L} \\
   0
\end{pmatrix},&
V_{CC}& = 
\begin{pmatrix}
k_{L} & 0 \\
  0  & k_{R}               
\end{pmatrix},&
V_{CR}& = 
\begin{pmatrix}
 0  \\
-k_{R}                
\end{pmatrix},&\nonumber\\
V_{RL}& = 
\begin{pmatrix}
  0
\end{pmatrix}, &
V_{RC}& = 
\begin{pmatrix}
 0 & -k_{R} 
\end{pmatrix},&
V_{RR}& = 
\begin{pmatrix}
k_{R}
\end{pmatrix},\label{vcllc}
\end{align}}
and
{\small\begin{align}
K_{CC}^{0}=\begin{pmatrix}
k_{C} & -k_{C}\\
-k_{C} & k_{C}
\end{pmatrix}.
\end{align}}
Here the matrices $V_{CC}^{(L)}$ and $V_{CC}^{(R)}$ introduced in \eqref{59d} read 
\begin{align}
&V_{CC}^{(L)}=\begin{pmatrix}
k_{L} & 0\\
0 & 0
\end{pmatrix}, &
V_{CC}^{(R)}=\begin{pmatrix}
0 & 0\\
0 & k_{R}
\end{pmatrix}, 
\end{align}
and satisfy $V_{CC} = V_{CC}^{(L)} + V_{CC}^{(R)}$.
\end{subequations}

The retarded and advanced components of the modified Green's functions are
\begin{equation}\tilde{g}_{\alpha}^{r,a}[\om]=
 \begin{cases} 
      \dfrac{1}{2}\,\dfrac{\om^2-2k_{\alpha}\mp\ii\,\om\sqrt{4k-\omega^2}}{(k-k_{\alpha})\,\om^2 + k_{\alpha}^{2}},& \vert\omega\vert\leqslant\sqrt{4k}\\
      \dfrac{1}{2}\,\dfrac{\om^2-2k_{\alpha} -\sqrt{\om^{2}\left(\omega^2-4k\right)}}{(k-k_{\alpha})\,\om^2 + k_{\alpha}^{2}},& \vert\omega\vert >\sqrt{4k},  
   \end{cases}\label{gfree}
\end{equation}
for $\alpha=L,R$. Note that the property $\tilde{g}_{\alpha}^{r}[-\om]=\tilde{g}_{\alpha}^{a}[\om]$ is 
satisfied according to the Eq.~\eqref{15a}. The derivation of Eqs.~\eqref{gfree} is presented in 
App.~\ref{sec:pertubative}. 

\subsection{Steady-state}\label{sec_application_ss}

Equations \eqref{set_coupling} and \eqref{gfree}, allow us to calculate the retarded and 
advanced self-energies $\tilde{\Sigma}_{L(R)}^{r,a}[\om]$ defined in Eq.~\eqref{eq:self-energy_VgV}, 
the level-width functions $\tilde{\Gamma}_{L(R)}[\om]$ given by Eqs.~\eqref{width} and \eqref{width_app},
 and the central region Green's functions $G_{CC}^{r,a}[\om]$. The local density of states (LDOS) at
the site $j=A,B$ in the central region reads
\begin{align}\label{DOS}
	\text{DOS}_{j}(\omega) = -\frac{2\omega}{\pi}\,\text{Im}\Big[ G^{r}_{CC}[\om]\Big] _{jj}.
\end{align}
The factor $2\omega$ is present to convert the value co\-ming 
directly from the imaginary part of $G^r_{CC}[\om]$ into the DOS per unit of $\omega$, 
ensuring that $\int\text{DOS}(\omega)\,\ud\omega$ equals the number of propagating 
channels in the system.

For the equal force constant case we can calculate the LDOS and the transmission analytically, namely
\begin{subequations}
\begin{align}
& \text{DOS}_{j}(\om)=\frac{2}{\pi\sqrt{4k-\om^2}}\;\Theta(4k-\om^2),\quad\forall\,j\\
& \mathcal{T}(\om)=\Theta(4k-\om^2).
\end{align}
\end{subequations}

\begin{figure}[htbp]
	\centering
	\includegraphics[width=1.00\columnwidth]{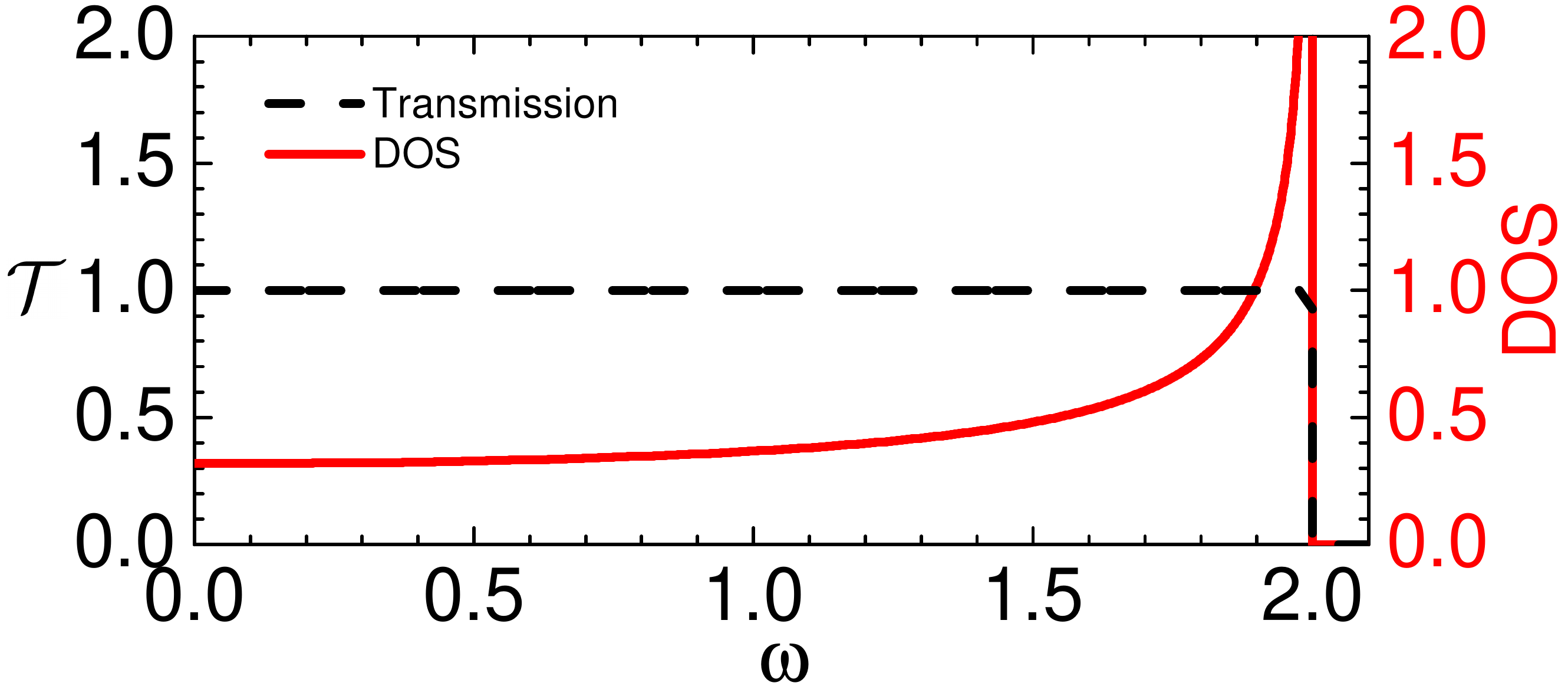}
	\caption{(Color online) DOS and transmission ${\cal T}$ as functions of the frequency $\omega$ in units of $\sqrt{k}$ for $k_L=k_C=k_R=k$.}
	\label{fig:TDOS}
\end{figure}

Figure~\ref{fig:TDOS} shows the DOS at one of the sites in the central region for $k_L=k_R=k_C=k$. 
Our formalism recovers the standard DOS for a linear chain.
The singularity at $\omega=\sqrt{4k}$ agrees with the frequency in which the dispersion relation of a 
linear chain $\omega = \sqrt{4k}\sin(k_xa/2)$ becomes flat, {\it i.e.}, at the edge of the first Brillouin 
zone. Here $k_x$ is the longitudinal momentum and $a$ is the lattice parameter. 
Also, the transmission coefficient $\mathcal{T}(\omega)$ corresponds to a perfect transmission 
inside the frequency band of the leads $\vert\omega\vert<\sqrt{4k}$ and it is zero otherwise.
%

In the limit of small temperatures and small temperature differences, namely, 
$T_{L/R} = T \pm \Delta T/2$ with $\Delta T\ll T$ for $T\to 0$, the thermal current 
for steady-state can be written as $ J^{(S)}_{L,R}= \pm\,\sigma(T)\;\Delta T$, where $\sigma(T)$ 
corresponds to the thermal conductance defined by
{
\begin{align}
\sigma(T)= \frac{2k_B^2T}{h}\int_{0}^{\frac{\hbar\omega_c}{2k_BT}} \ud x\frac{x^2}{\sinh^{2}x}\, \mathcal{T}\left(\frac{2k_BT}{\hbar} x\right)
\end{align}
where $\omega_{c}\equiv \sqrt{4 k}$.
From Eq.~\eqref{transmission} it is possible to verify that $\mathcal{T}(T\rightarrow 0)=1$.
}
The low temperature limit of $\sigma(T)$ is 
\begin{equation}\label{quantum_conductance}
\sigma_0 = \frac{\pi^2 k_{B}^{2}}{3\,h}\,T
\end{equation} 
as theoretically predicted \cite{Pendry1983,Rego1998} and experimentally observed \cite{Schwab2000}.
{
Thus, at low temperatures the thermal conductance $\sigma(T) \propto T$ 
} 
vanishes for $T\to 0$, as 
required by the third law of thermodynamics.

\begin{figure}[htbp]
	\centering
		\includegraphics[width=1.00\columnwidth]{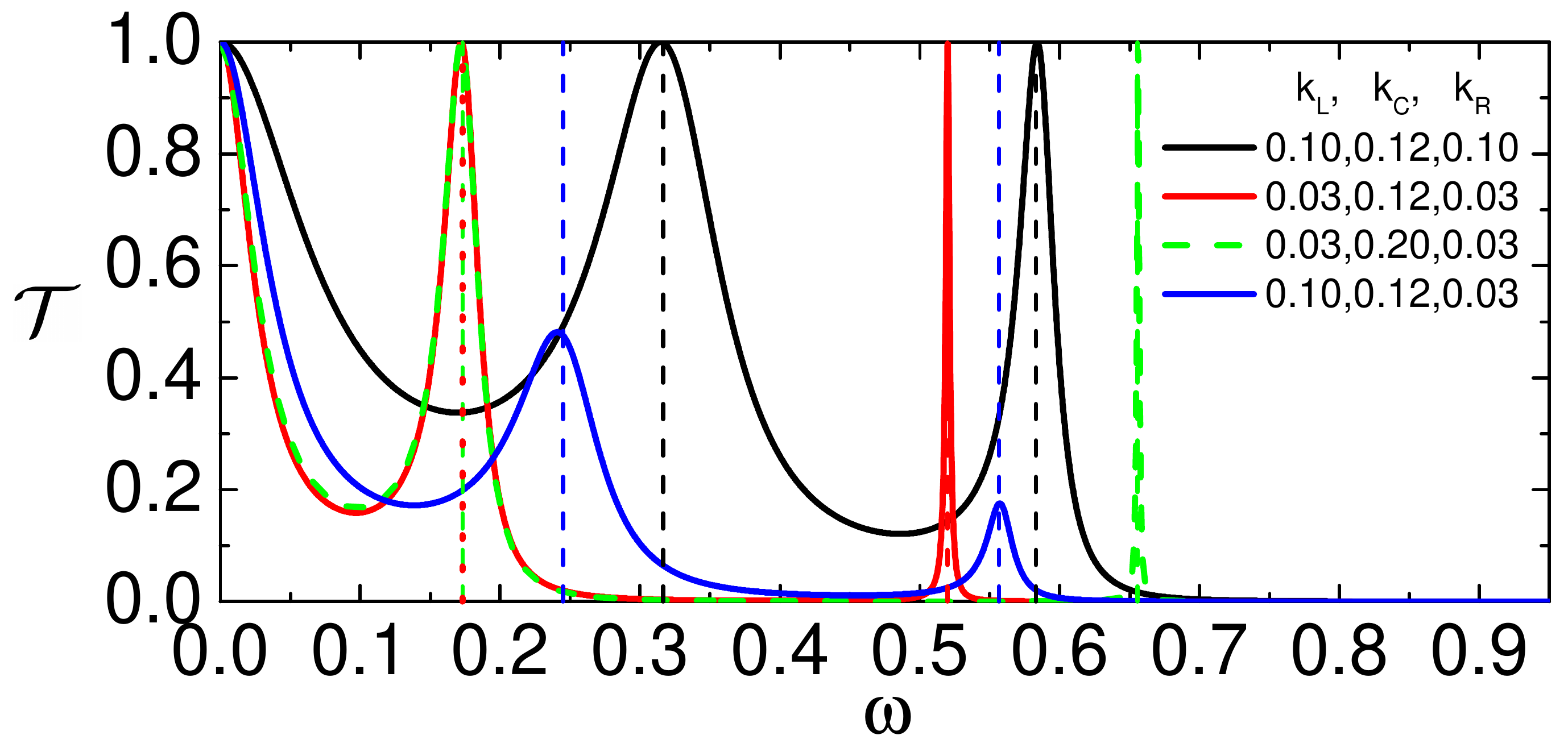}
	\caption{(Color online) Transmission ${\cal T}$ as a function of the frequency $\omega$ in units of $\sqrt{k}$ in the weak coupling regime. 
	The values of $k_L$, $k_R$ and $k_C$ are indicated in the picture in units of $k$. 
	The vertical dashed lines are the frequencies given by Eq.~(\ref{classicalomega}).} 
	\label{fig:tsmallkl}
\end{figure}

Let us now study situations where the force cons\-tants are different.
In the weak coupling limit, $k_R,k_L \ll k,k_C$, the central region is nearly disconnected from 
the outside world having only one resonant level at $\omega_C=\sqrt{2k_C}$. 
Thus, the conductance is only expected to be significant at the vicinity of $\omega_C$. Instead, 
Fig.~\ref{fig:tsmallkl} shows one peak at $\omega \approx \omega_C$ and two additional strong 
peaks, one at zero frequency and another intermediate peak at $0<\omega<\omega_C$. 

\begin{figure}[htbp]
	\centering
		\includegraphics[width=1.00\columnwidth]{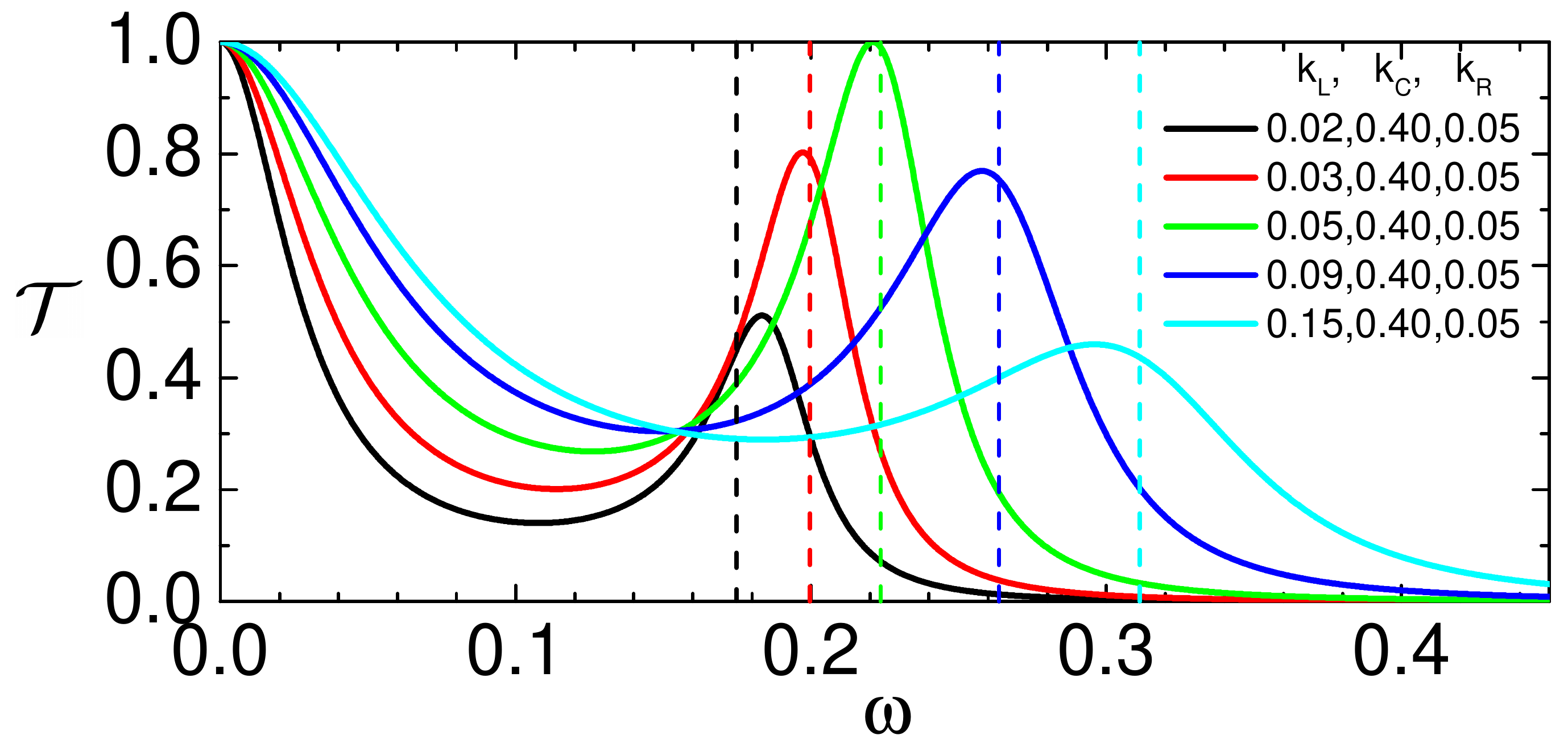}
	\caption{(Color online) Transmission ${\cal T}$ as a function of the frequency $\omega$ in units of $\sqrt{k}$  
	for different values of $k_R$ (indicated in the figure) with $k_C=0.4$ and $k_L=0.05$.
	All constants are in units of $k$. The maximum transmission occurs when $k_L=k_R$.
         The vertical dashed lines are the frequencies given by Eq.~(\ref{classicalomega}).}
	\label{fig:tklkr}
\end{figure}

The first peak at $\omega=0$ corresponds to the acoustic mode that has an infinite long 
wavelength so that the short ranged ``defects" introduced by $k_L, k_R, k_C\neq k$ do not 
affect the transport across the system. This picture is reenforced by noticing that the zero frequency 
peak is robust against changes in the value of $k_L$, $k_C$ and $k_R$ in the weak coupling regime, 
see Fig.~\ref{fig:tklkr}.

By coupling the diatomic molecule to leads, the resonance level at $\omega_{C}$ is shifted and acquires broadening, as described by the self-energy $\tilde{\Sigma}^{r}[\om]$.
Hence the peak near $\omega_C$ is very sensitive to variations in $k_L$ and $k_R$. 
These features, are illustrated in Fig.~\ref{fig:tsmallkl}, by inspecting a set of transmission curves where we keep $k_C$ constant and increase $k_L=k_R$.

On the other hand, the remaining peak at $0<\omega<\omega_C$ depends only on the values of 
$k_L$ and $k_R$. In the weak coupling regime, a semi-classical picture explains this additional 
transmission peak. 
The natural interfaces frequencies $\omega_\alpha \propto \sqrt{k_\alpha}$, with $\alpha=L,R$, are much 
smaller then $\omega_{C}$. The large separation in frequencies suggest that the resonance close to 
$\omega_{C}$ is dominated by the isolated molecule mode, while the other corresponds to an oscillation 
of a frozen  central region. The Green's functions of such a system gives resonances at the frequencies
\begin{align}
	\omega_{1,2} = \sqrt{k_C + \left(\frac{k_L+k_R}{2}\right) \pm \sqrt{  \left(\frac{k_L-k_R}{2}\right)^2 + k_C^2} } .
	\label{classicalomega}
\end{align}
For $k_L=k_R$, $\omega_{1} = \sqrt{2k_C+k_L}$ and $\omega_{2} = \sqrt{k_L}$ that are plotted in Fig.~\ref{fig:tsmallkl} as vertical dotted lines matching the peaks positions.
For $k_{L}\neq k_{R}$, the symmetry is broken and the maximum transmission at all the peaks, except for the one with zero frequency, is no longer perfect.

{
We note that our results are qualitatively similar to those in Ref.~\cite{Wang2007}, that analyze 
the steady-state transport through a benzene ring.
There is an important difference though: Taking into account $V_{aa}$ in $\tilde g_\alpha$ guarantees 
that ${\cal T} (\omega \rightarrow 0) \rightarrow 1$, which is a necessary condition to obtain the quantum 
of thermal conductance for $T\rightarrow 0$. In distinction, by using $g_\alpha$ as the surface Green's 
function, as done in Refs.~\cite{Wang2007,Wang2008,Wang2014}, 
one obtains ${\cal T} (\omega \rightarrow 0) \rightarrow 0$.
}

\subsection{Pumping}

For simplicity, let us analyze a pumping process between reservoirs at the same temperature. In this case, 
the steady-state current from the $\alpha$-reservoir is $J_{\alpha}^{(S)}=0$. 
Hence, $J_\alpha^{(P)}$ gives the leading contribution to the heat flow.

We consider the case of pumping functions with a phase difference $\varphi$, namely, 
$\phi_{L}(t)=\phi_{R}(t-\varphi/\Omega)$, which implies that $\varphi_{n}^{(L)}-\varphi_{n}^{(R)}=
n\,\varphi$ and $a_{n}^{(L)}=a_{n}^{(R)}\equiv a_{n}$ for $n\geqslant 1$. According to 
Eq.~\eqref{set_69}, we can express the $\alpha$-thermal pumped current as
\begin{align}
J^{(P)}_{\alpha}(\Omega)=&\sum_{n=1}^{\infty}a_{n}^2\bigg[\mathcal{A}_{\text{homo}}^{\alpha}(n\Omega) + \cos(n\varphi)\,\mathcal{A}_{\text{hete}}^{\alpha}(n\Omega) \nonumber\\
&-\sin\left(n\varphi\right)\,\mathcal{B}^{\alpha}(n\Omega)\bigg],\label{JP}
\end{align}
where $\mathcal{A}^{\alpha}_{\text{homo}}(n\Omega)\equiv A^{\alpha}_{LL}(n)+A^{\alpha}_{RR}(n)$, $\mathcal{A}^{\alpha}_{\text{hete}}(n\Omega)\equiv A^{\alpha}_{LR}(n)+A^{\alpha}_{RL}(n)$, $\mathcal{B}^{(\alpha)}(n\Omega)\equiv B^{\alpha}_{LR}(n)-B^{\alpha}_{RL}(n)$ for $\alpha=L,R$. 
For the symmetric coupling case, {\it i.e.}, $k_{L}=k_{R}\neq k_{C}$, we can show that $\mathcal{A}^{L}_{\text{homo/hete}}(n\Omega)=\mathcal{A}^{R}_{\text{homo/hete}}(n\Omega)$ and $\mathcal{B}^{L}(n\Omega)=-\mathcal{B}^{R}(n\Omega)$.

\begin{figure*}[!htbp]
 \centering
  {\includegraphics[width=0.32\textwidth]{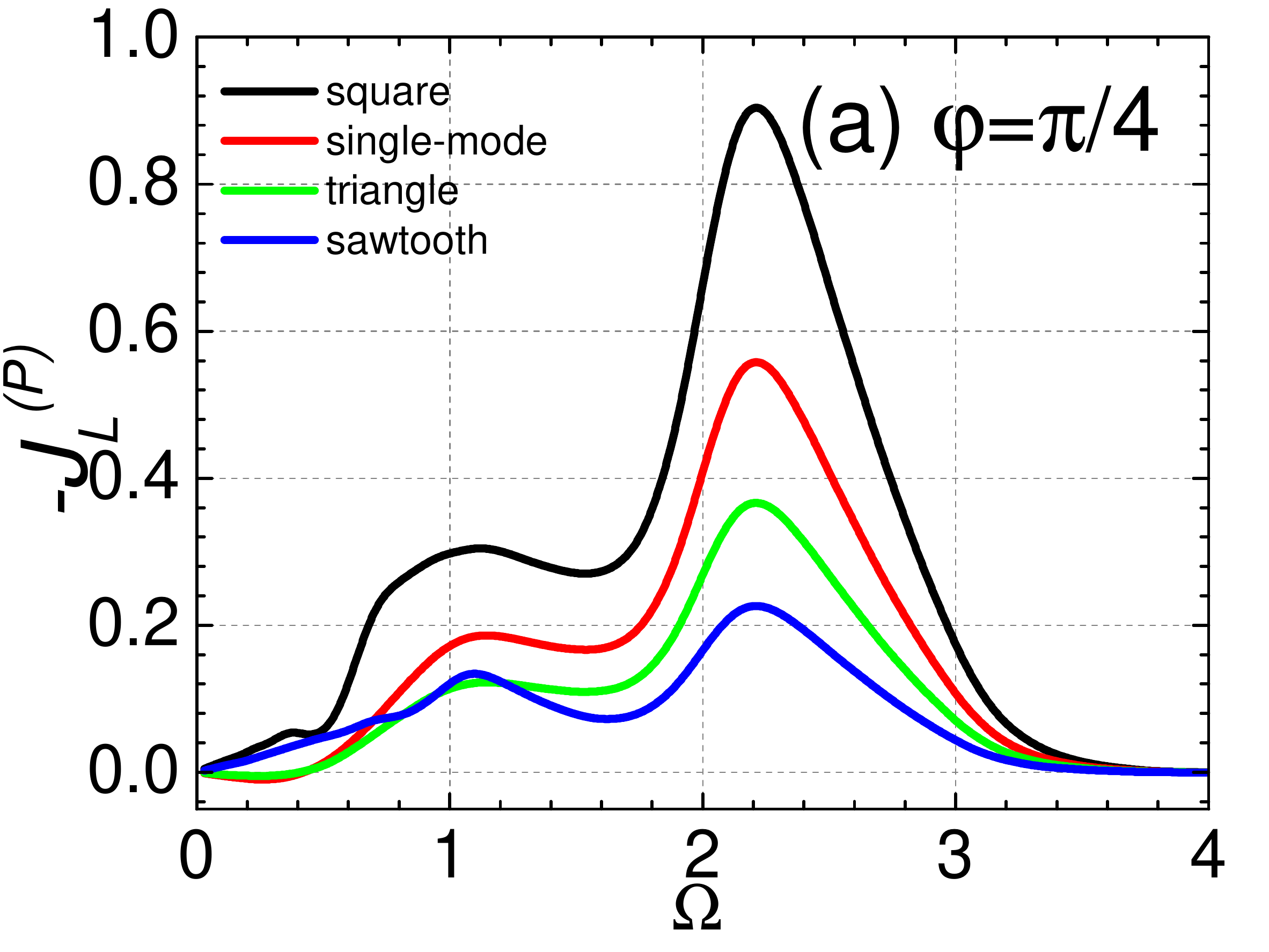}}
  {\includegraphics[width=0.32\textwidth]{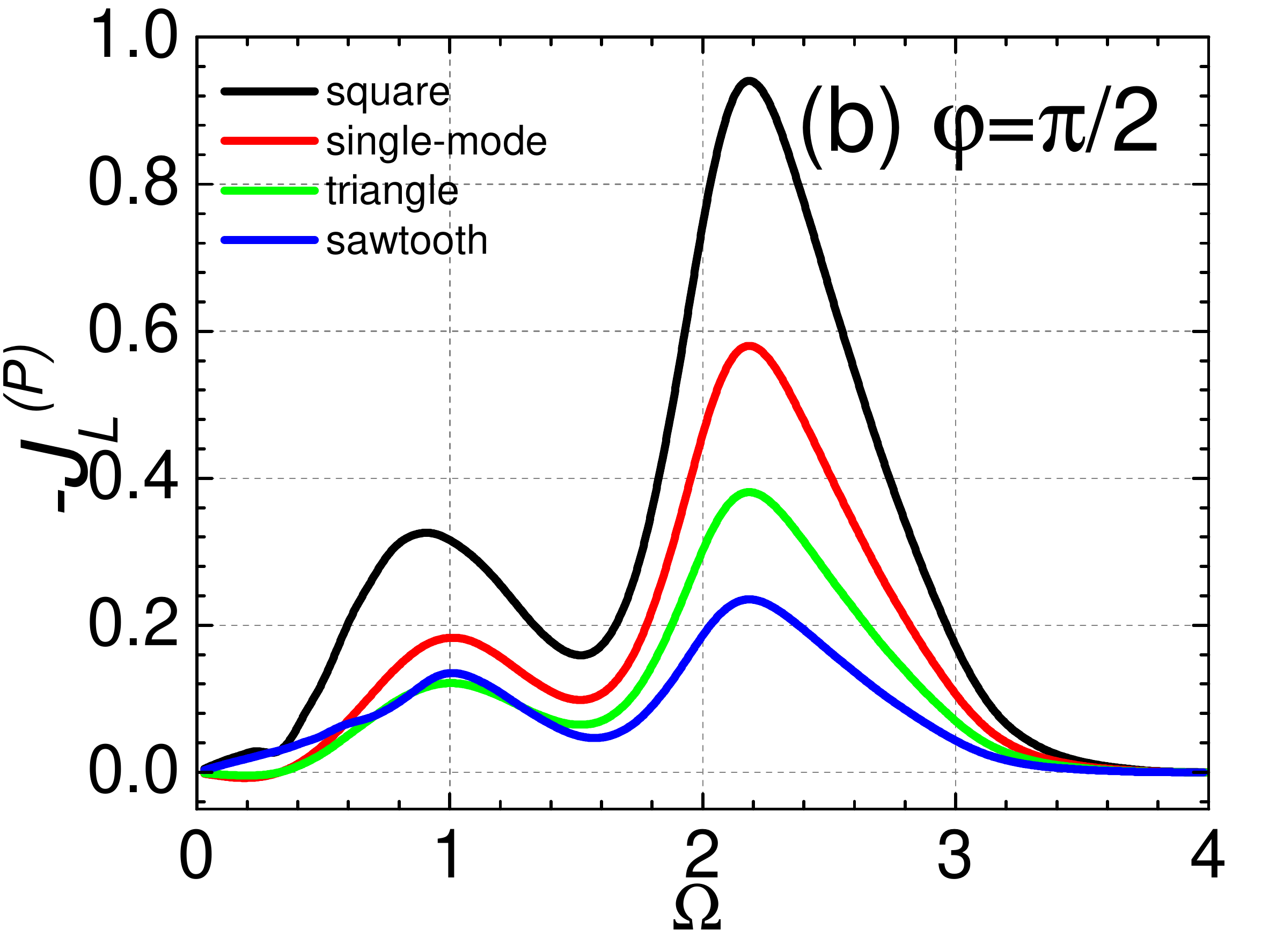}}
  {\includegraphics[width=0.32\textwidth]{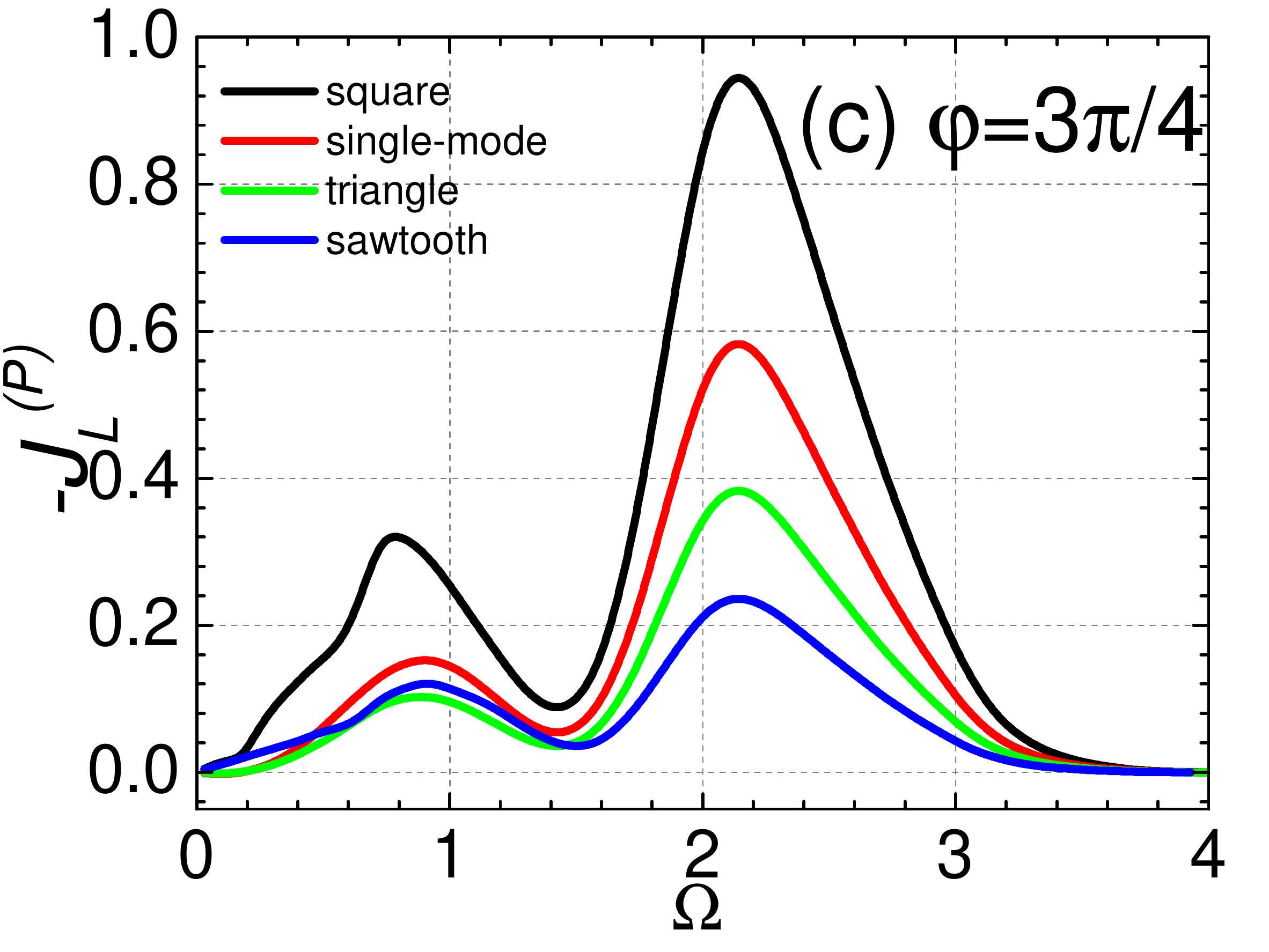}}
  {\includegraphics[width=0.32\textwidth]{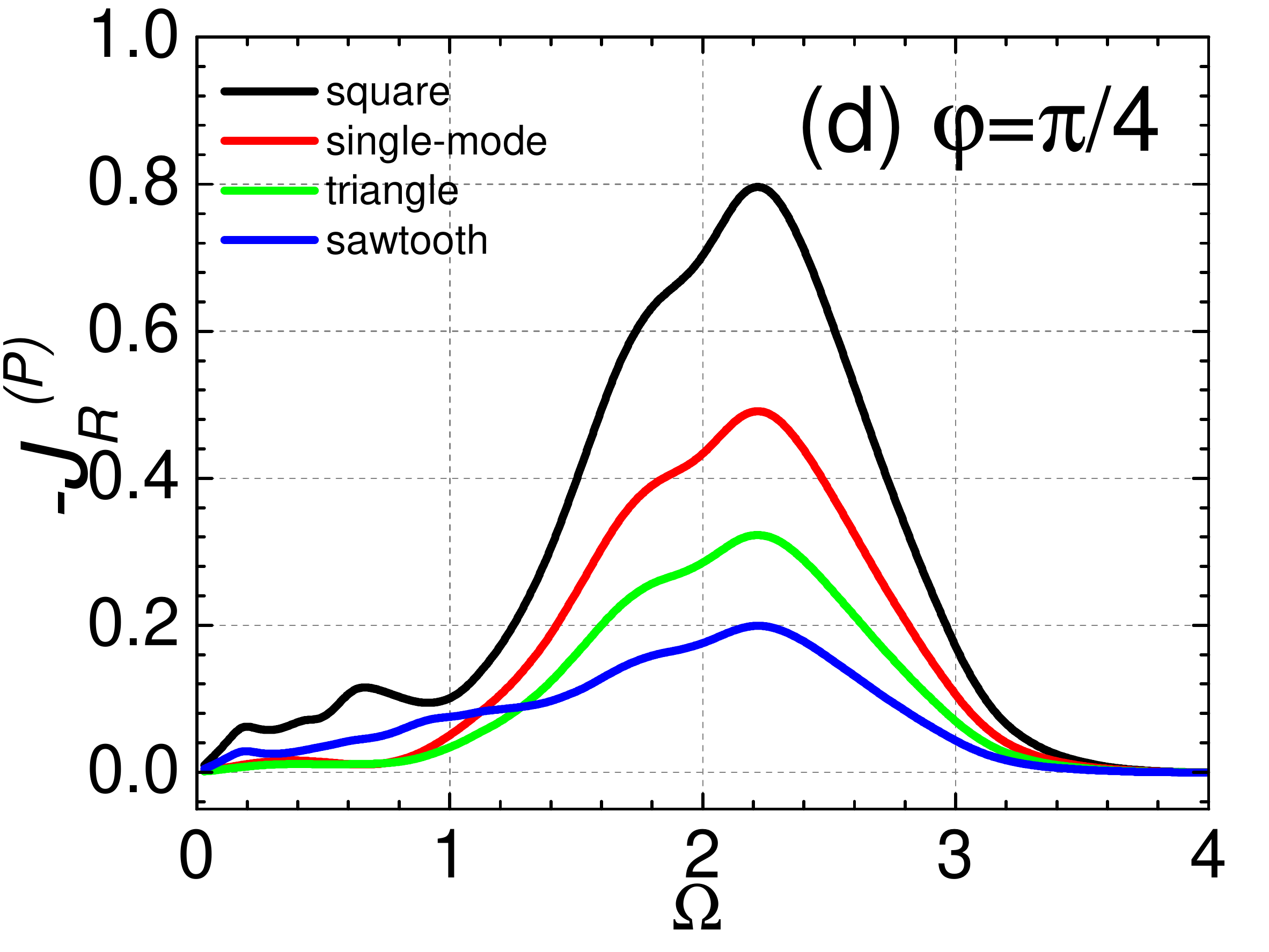}}
  {\includegraphics[width=0.32\textwidth]{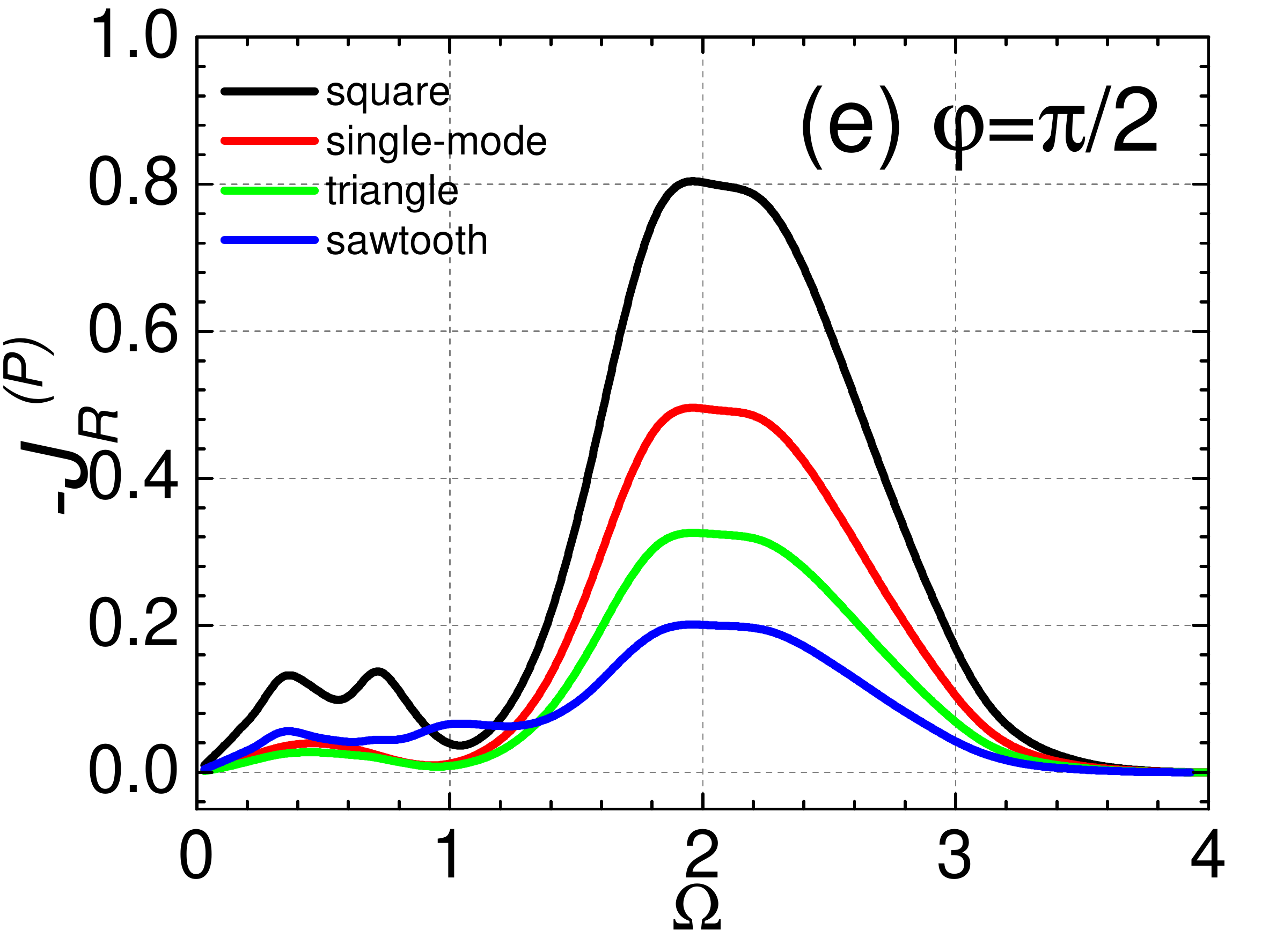}}
  {\includegraphics[width=0.32\textwidth]{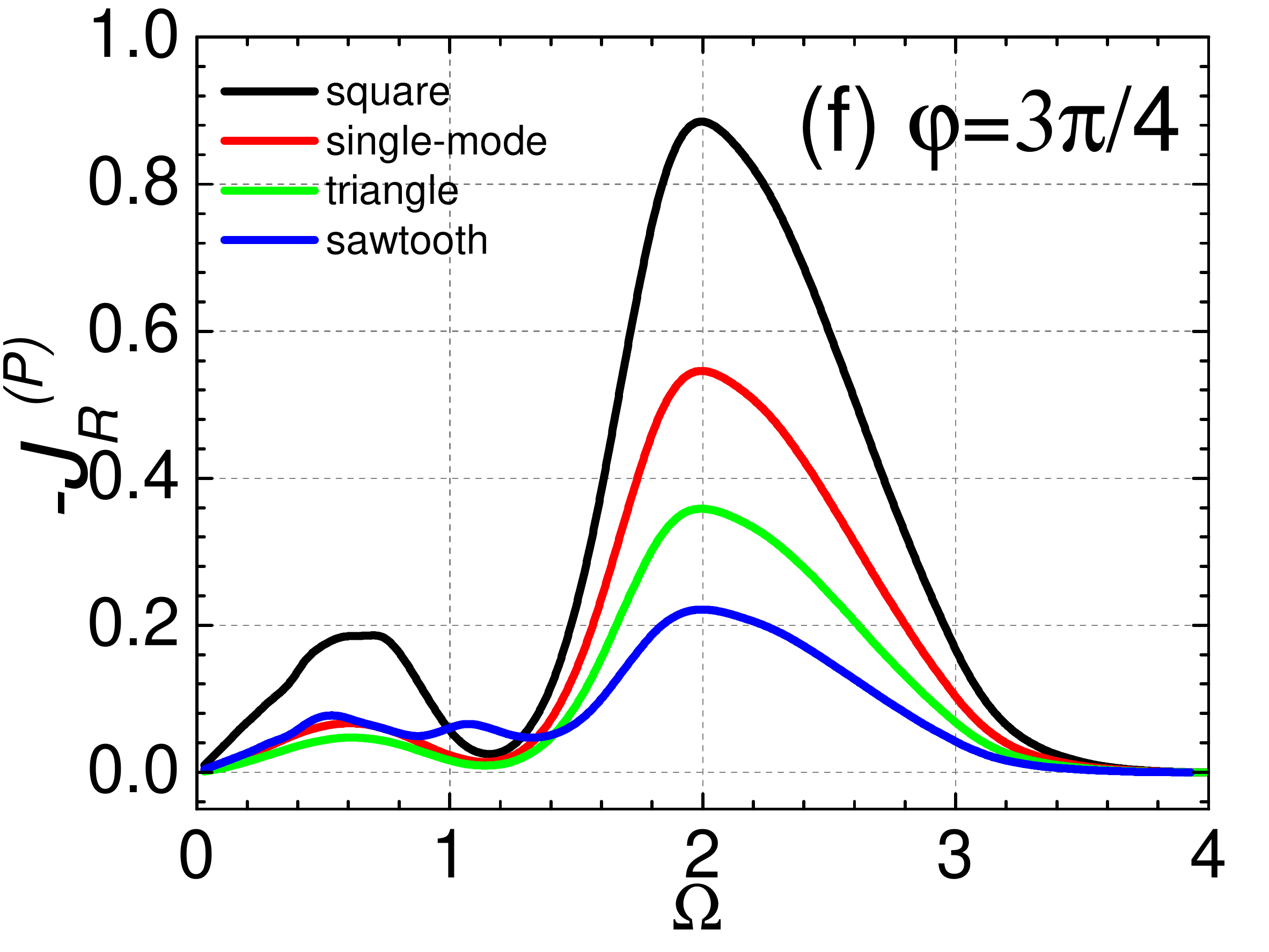}}
	{\includegraphics[width=0.32\textwidth]{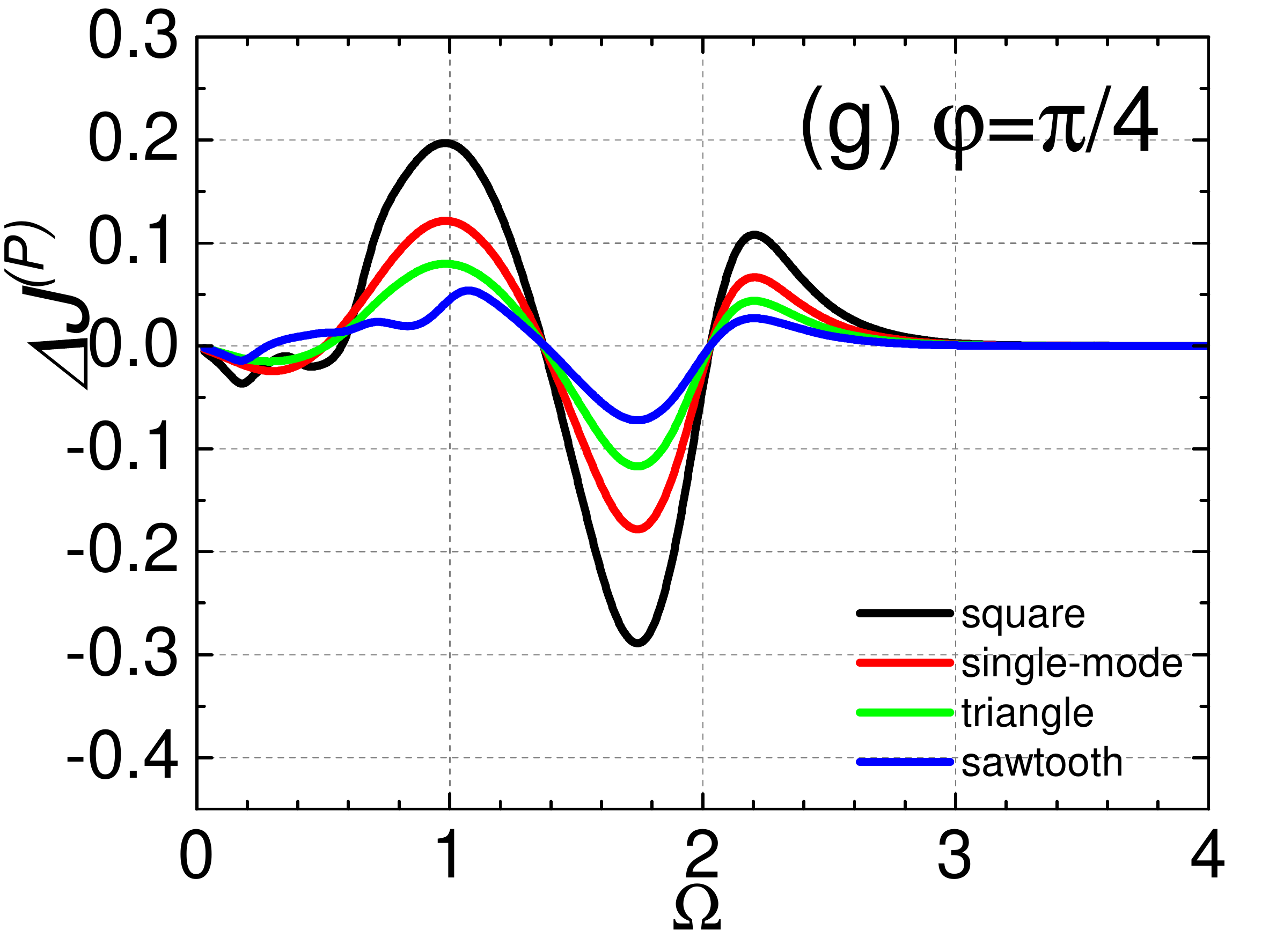}}
  {\includegraphics[width=0.32\textwidth]{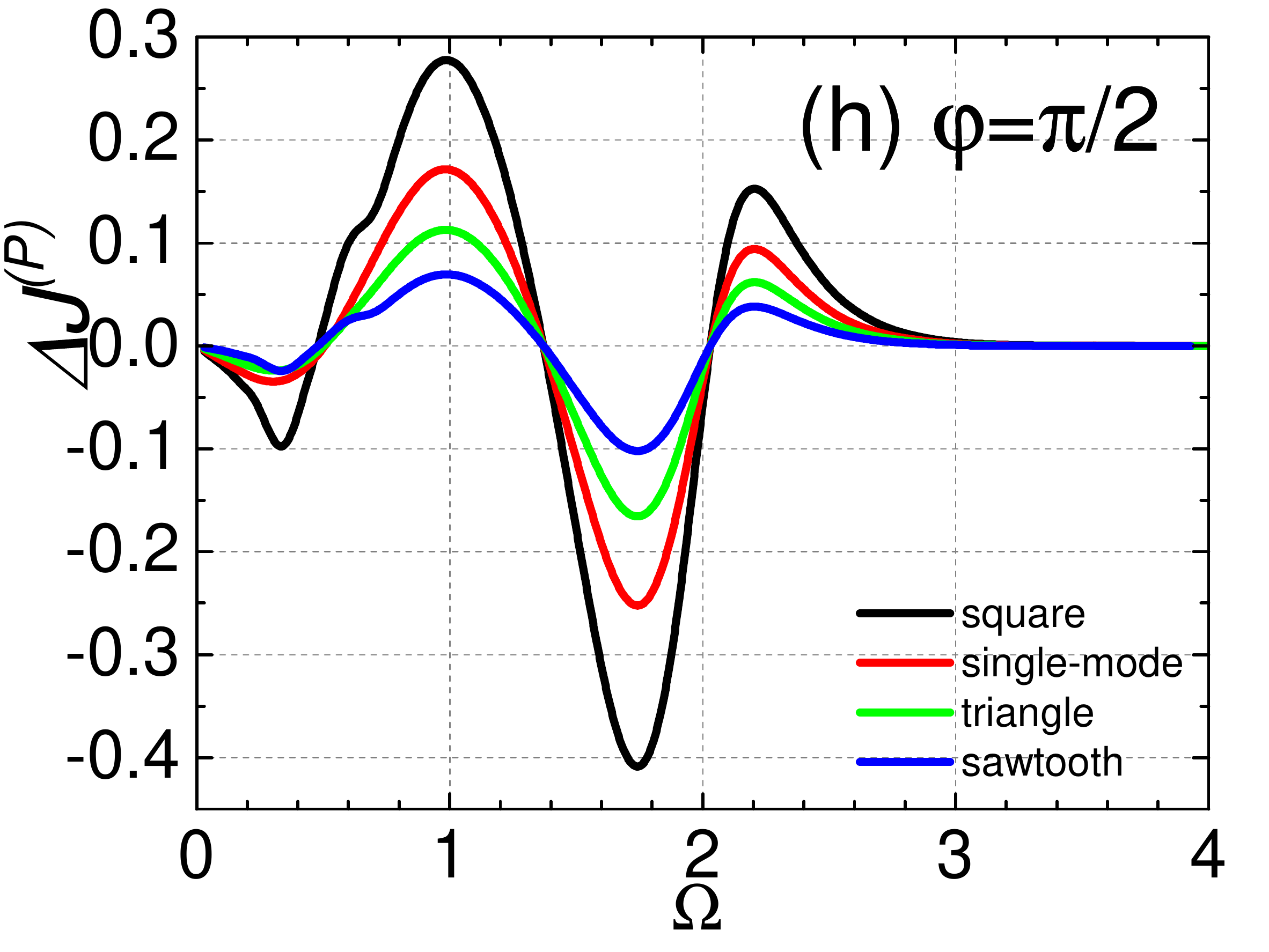}}
  {\includegraphics[width=0.32\textwidth]{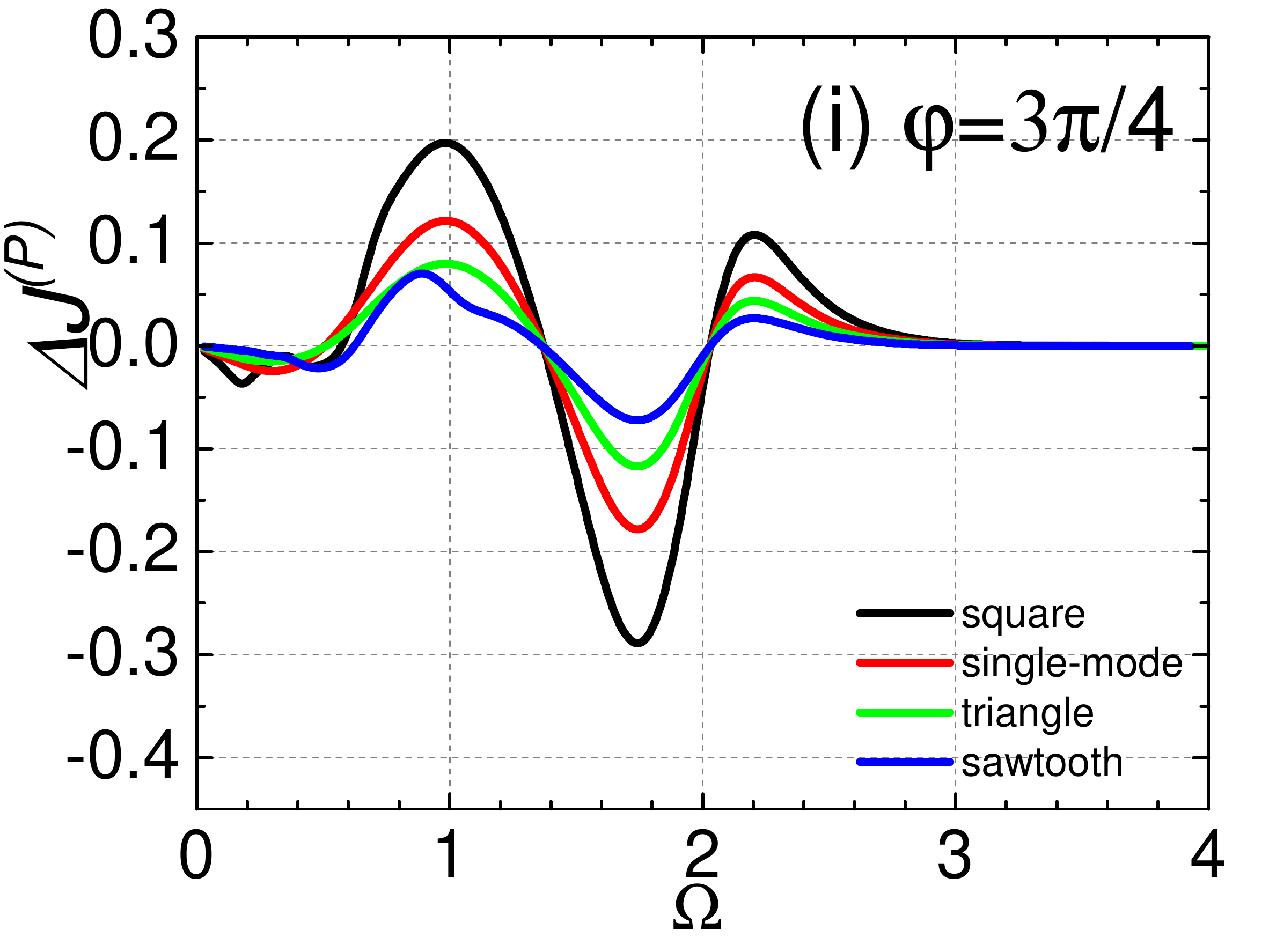}}
	{\includegraphics[width=0.32\textwidth]{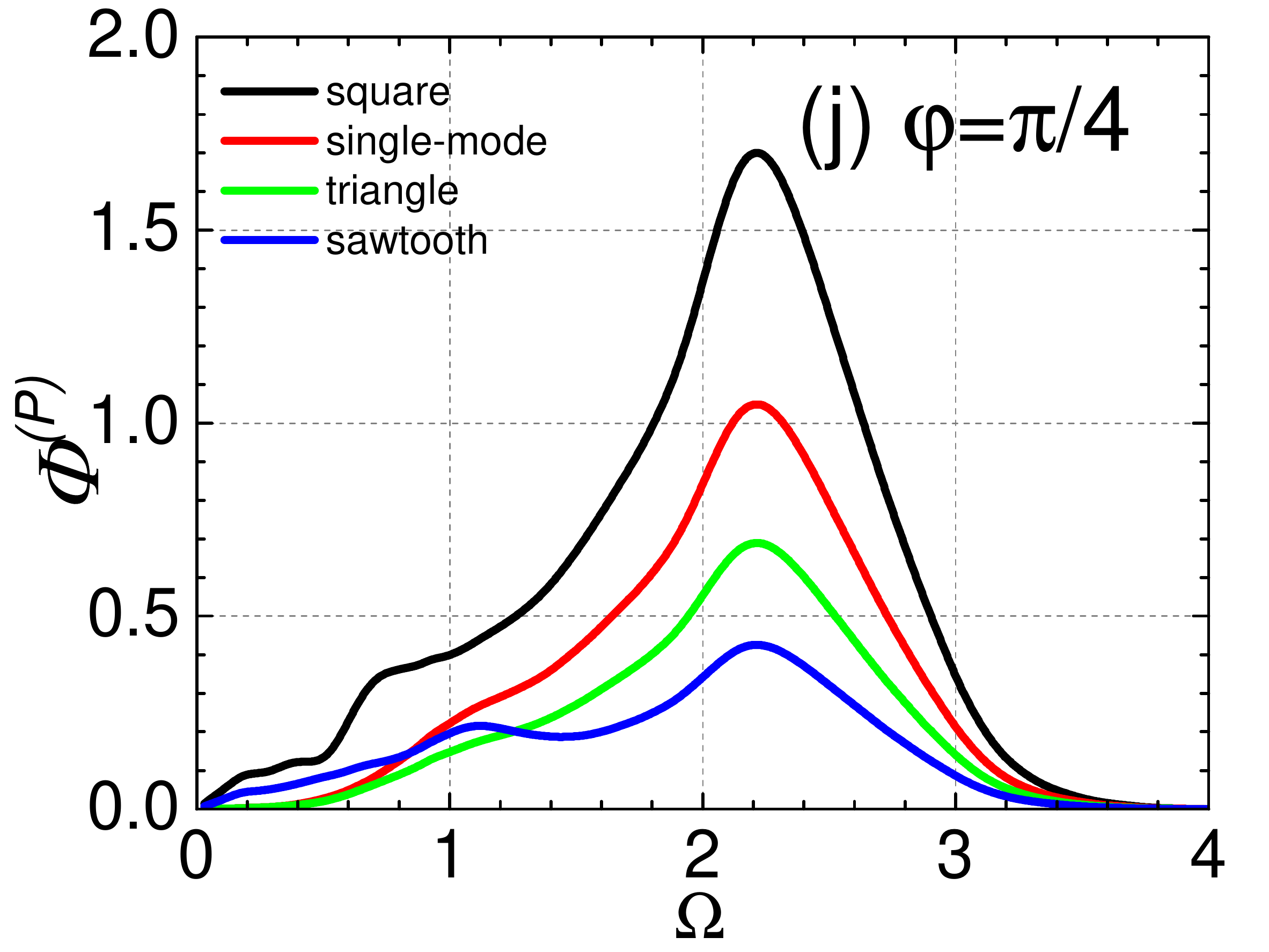}}
  {\includegraphics[width=0.32\textwidth]{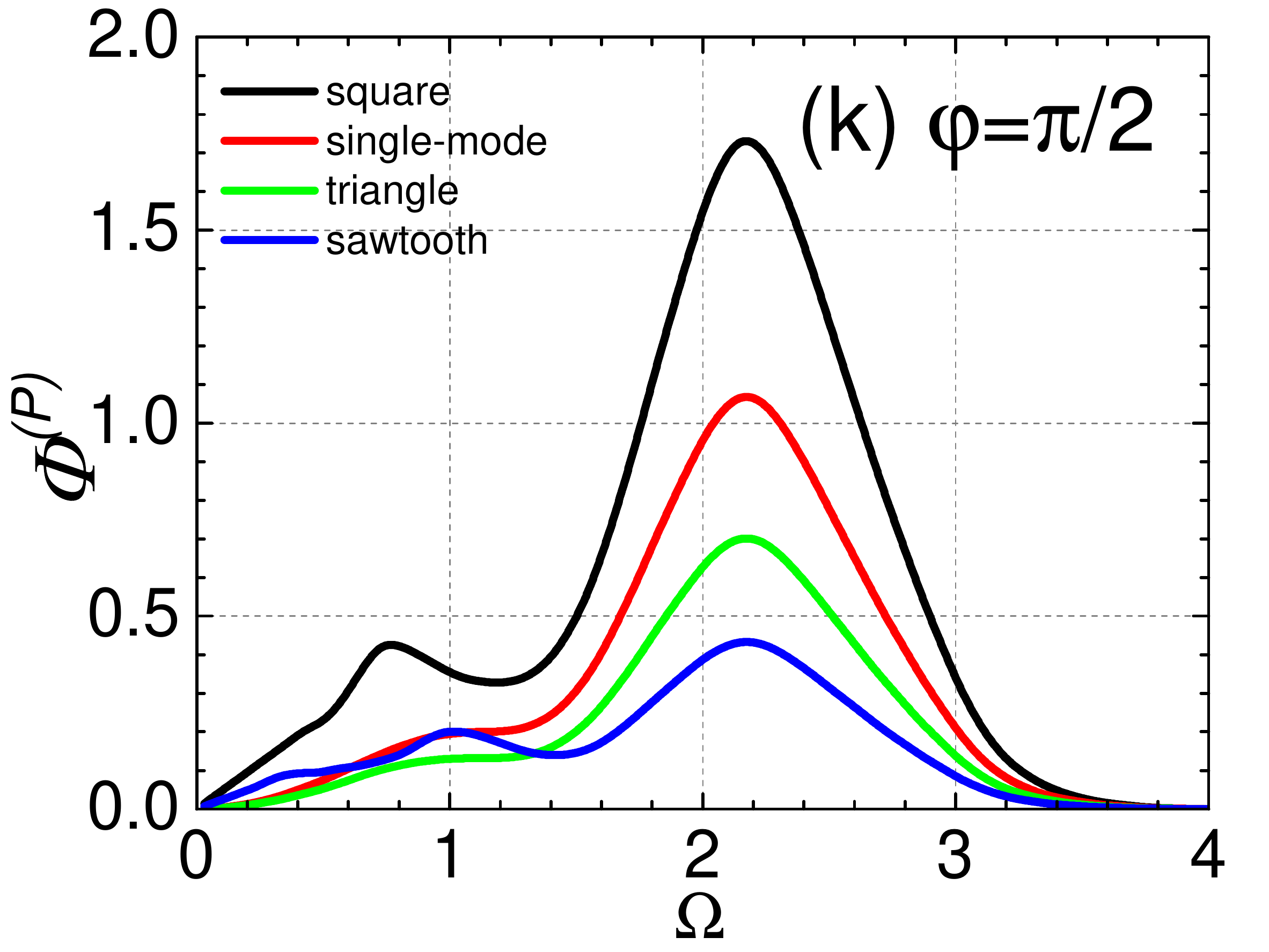}}
  {\includegraphics[width=0.32\textwidth]{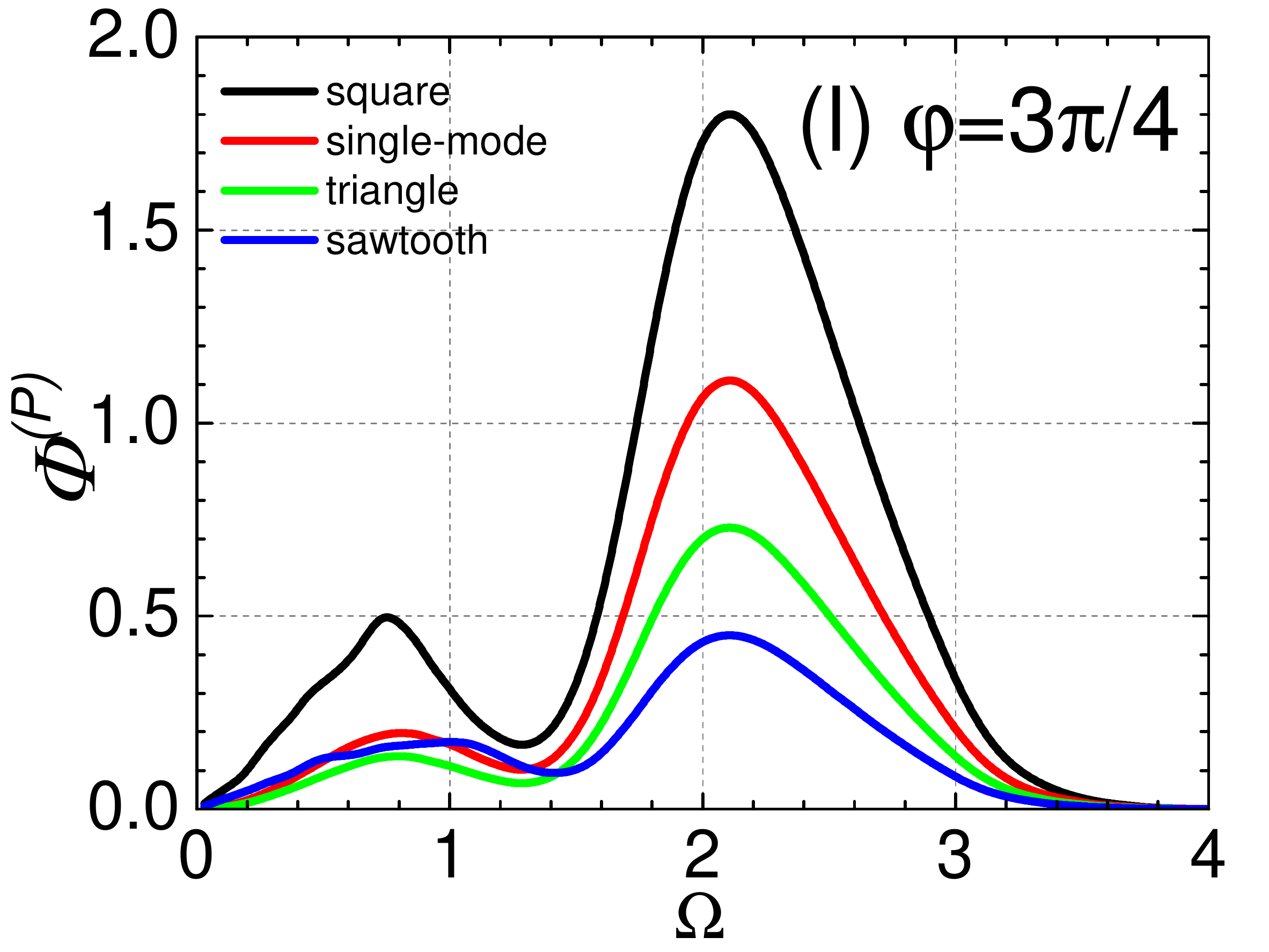}}
 \caption{(Color online) Thermal current absorbed by the \textit{left/right}-reservoir $-J_{L/R}^{(P)}$ (in units of $\hbar k/\varepsilon$), effective thermal current transmitted from \textit{left}- to \textit{right}-reservoir $\Delta J^{(P)}\equiv J_{R}^{(P)}-J_{L}^{(P)}$ (in units of $\hbar k/\varepsilon$) and power injected $\Phi^{(P)}$ (in units of $\hbar k/\varepsilon$) into the system as a function of the pumping frequency $\Omega$ (in units of $\sqrt{k}$) using $k_L = k_R = 0.5 k$ and $k_{C}=0.25 k$ for equal temperatures with $k_{B}T_{L}=k_{B}T_{R}=1$ (in units of $\hbar\sqrt{k}$) and for distinct pumping functions. }
\label{figs_pumping}
\end{figure*}

Similarly, the pumped power reads
\begin{align}
\Phi^{(P)}(\Omega)=&\sum_{n=1}^{\infty}a_{n}^2\bigg[\mathcal{D}_{\text{homo}}(n\Omega) + \cos\left(n\varphi\right)\,\mathcal{D}_{\text{hete}}(n\Omega)\nonumber\\
&-\sin\left(n\varphi\right)\,\mathcal{E}(n\Omega)\bigg]\label{Phi},
\end{align}
where $\mathcal{D}_{\text{homo}}(n\Omega)\equiv D_{LL}(n)+D_{RR}(n)$, $\mathcal{D}_{\text{hete}}(n\Omega)\equiv D_{LR}(n)+D_{RL}(n)$ and $\mathcal{E}(n\Omega)\equiv E_{LR}(n)-E_{RL}(n)$ defined for $0< n\,\Omega<4\sqrt{k}$ and zero otherwise. For further details see Appendix~\ref{sec:pertubative}. For a symmetric setup (i.e. $k_{L}=k_{R}\neq k_{C}$), we can show that $\mathcal{E}(n\Omega)\equiv 0$.   Note that $\Phi^{(P)}$ 
satisfies the condition $\Phi^{(P)}>0$, 
{
as e\-xem\-pli\-fied for the diatomic molecule in the Fig.~\ref{figs_pumping}, which corres\-ponds to positive rate of work performed on the system. Therefore, for reservoirs at the same temperature, the entropy production per cycle cast as $(\Delta S)_{\text{cycle}}/\tau=\varepsilon^2\, \dot{\mathcal{S}}^{(P)}+\mathcal{O}(\varepsilon^{4})$ satisfies
\begin{equation}
\dot{\mathcal{S}}^{(P)} = \frac{-J_{L}^{(P)}}{T} + \frac{-J_{R}^{(P)}}{T} = \frac{\Phi^{(P)}}{T} > 0,
\end{equation}
as expected from the second law of thermodynamics.
Note that the overall partition scheme, the definitions of the heat currents, and the pumped power are consistent with general thermodynamic properties.
}

The pumping function is determined by the choice of the parameter set $\lbrace a_{n}\rbrace$ and phase difference $\varphi$. We study four examples of pumping functions:  \textit{single-mode} represented by $a_{n}=\delta_{n,1}/2$; \textit{square} oscillation represented by $a_{n}=2/(n\pi)$ for $n$ odd (and zero otherwise);  \textit{triangle} oscillation by $a_{n}=\frac{4}{n^{2}\pi^{2}}(-1)^{\frac{n-1}{2}}$ for $n$ odd (and  zero otherwise);  \textit{sawtooth} oscillation by $a_{n}=-1/(\pi n)$ for all $n$.

The thermal current absorbed by the $\alpha$-reservoir, $-J_{\alpha}^{(P)}>0$, as a function of pumping frequency $\Omega$ for different pumping profiles and phase difference is shown in Fig.~\ref{figs_pumping}. We find a suppression of thermal current according to the type of pumping in the following order: \textit{square}, \textit{single-mode}, \textit{triangle} and \textit{sawtooth}. The pumping peak occurs in the frequency window $ 2\sqrt{k}<\Omega<3\sqrt{k}$ and a sub-peak within $0<\Omega<\sqrt{k}$, accompanied by a weak suppression in the domain $\sqrt{k} <\Omega <2\sqrt{k}$ and the strong suppression for $\Omega\gtrsim 4\sqrt{k}$. As the phase difference is increased, the suppression in $\sqrt{k} <\Omega <2\sqrt{k}$ is intensified. 
Note that the unperturbed proposed setup is symmetric ($k_{L}=k_{R}\neq k_{C}$).
The phase shift $\varphi$ causes the difference between $J_{L}^{(P)}$ and $J_{R}^{(P)}$, see Fig.~\ref{figs_pumping}. 

The effective heat flux between the two reservoirs $\Delta J^{(P)}(\Omega)\equiv J^{(P)}_{R}(\Omega)-J^{(P)}_{L}(\Omega)$ for the symmetric coupling case (\textit{i.e.} $k_{L}=k_{R}\neq k_{C}$) reads
\begin{equation}\label{DJ}
\Delta J^{(P)}(\Omega)= \sum_{n=1}^{\infty}2a_{n}^{2}\,\sin(n\varphi)\,\mathcal{B}^{L}(n\Omega),
\end{equation}
where $\Delta J^{(P)}>0$ corresponds to the heat flux from \textit{right}-reservoir to \textit{left}-reservoir and $\Delta J^{(P)}<0$ to reverse direction. Note that for $\varphi = 0, \pm 2\pi, \pm 4\pi, \pm 6\pi, \ldots$,  we obtain $\Delta J^{(P)} =0$, as expected. For the \textit{single-mode} case we verify that the maximum value of $\vert\Delta J^{(P)}\vert $ occurs for $\varphi = \pm\pi/2, \pm 3\pi/2, \pm 5\pi/2, \ldots$. 
$\Delta J^{(P)}$ versus the pum\-ping frequency $\Omega$ is represented in the Figs.~\ref{figs_pumping}(g)-(i). 

The time-dependent drive breaks the system symmetry. 
Hence, one can engineer configurations of $\Omega$ and $\phi_\alpha(t)$ that direct the pumped heat either to the left or to the right. Equation~(\ref{DJ}) shows that by re\-pla\-cing $\varphi \rightarrow -\varphi$ the directionality is reversed, namely, $\Delta J^{(P)}(\varphi) = -\Delta J^{(P)}(-\varphi)$. 

For all considered pumping profiles and phase diffe\-ren\-ces, we find $\Delta J^{(P)}=0$ at $\Omega=\Omega_{1} \approx 1.371\sqrt{k}$  and $\Omega=\Omega_{2}\approx 2.029\sqrt{k}$, see Fig.~\ref{figs_pumping}. 
The largest negative peak of $\Delta J^{(P)}$ occurs between $\Omega_{1}$ and $\Omega_{2}$. 
Other two positive peaks occur on $\Omega<\Omega_{1}$ and $\Omega>\Omega_{2}$.  

The external drives contribute to the e\-ner\-gy transfer by exciting the original unperturbed pro\-pa\-ga\-ting energy $\hbar\omega$ to $\hbar(\omega + n\Omega)$ where $n=1,2,3,\ldots $.
In our model, the energy transfer between leads is only possible if the injected energy $\hbar\omega$ and the excited energy $\hbar(\omega + n\Omega)$ satisfy the conditions $\vert\omega\vert \le \omega_c$ and $|\omega + n\Omega| \le \omega_c$, res\-pec\-ti\-ve\-ly, where $\omega_c \equiv 2\sqrt{k}$.
Outside this frequency window the leads linewidths vanish and no energy transfer is allowed.
Thus, only modes with $n \le 2\omega_c/\Omega$ will contribute to the energy transfer between reservoirs.
As $\Omega$ increases a smaller number of modes contribute and the overall energy transport decreases, as we see in Fig.~\ref{figs_pumping}. 
For $\Omega > 2\omega_c = 4\sqrt{k}$ the external drives do not induce energy transport to the reservoirs as no positive integer can satisfy $n<1$.
For small $\Omega$, many modes $n$ with $n \le 2\omega_c/\Omega$ compete and contribute to the energy transfer.

\begin{figure}[!htbp]
 \centering
 \includegraphics[width=0.85\columnwidth]{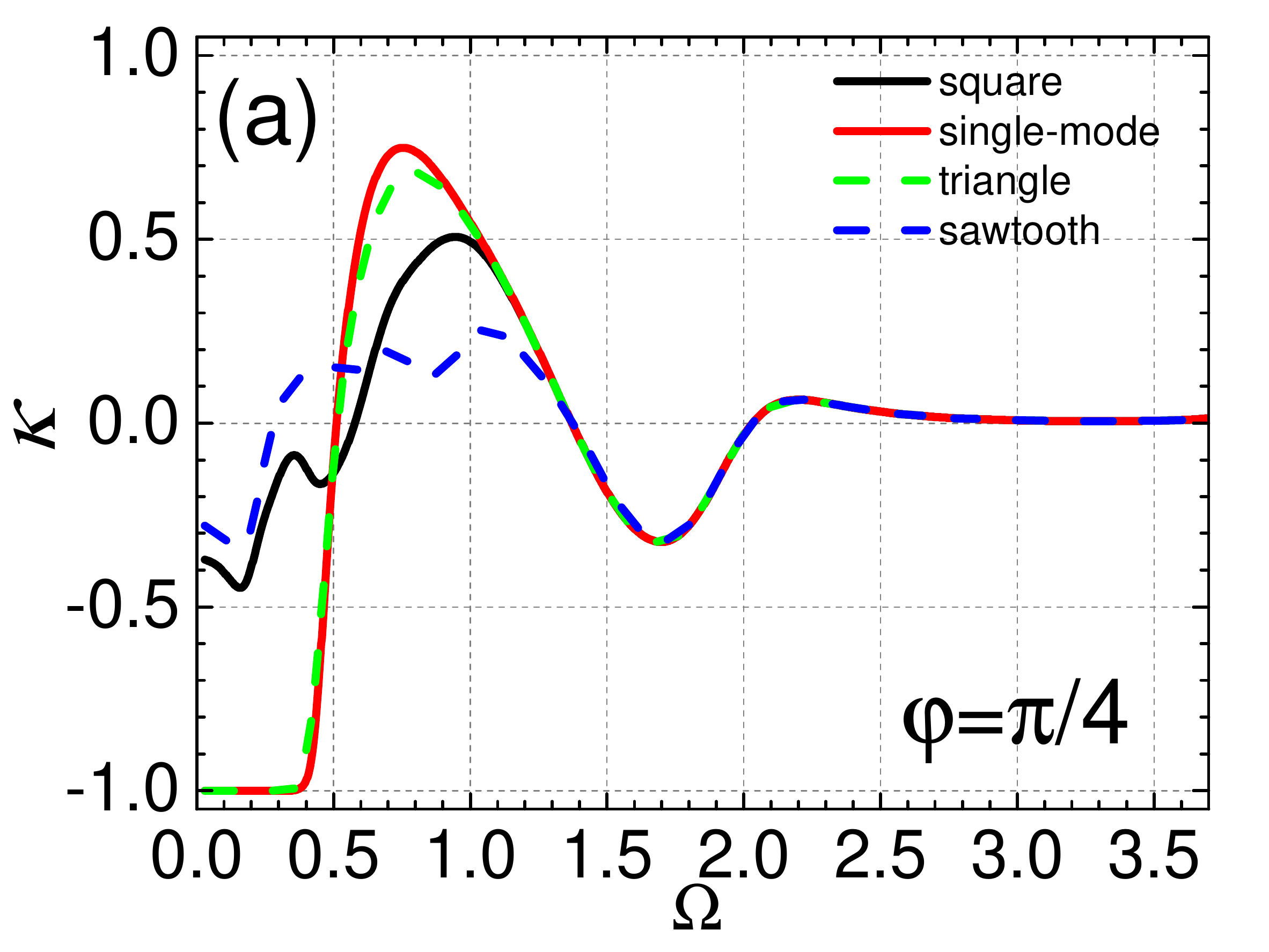}
 \includegraphics[width=0.85\columnwidth]{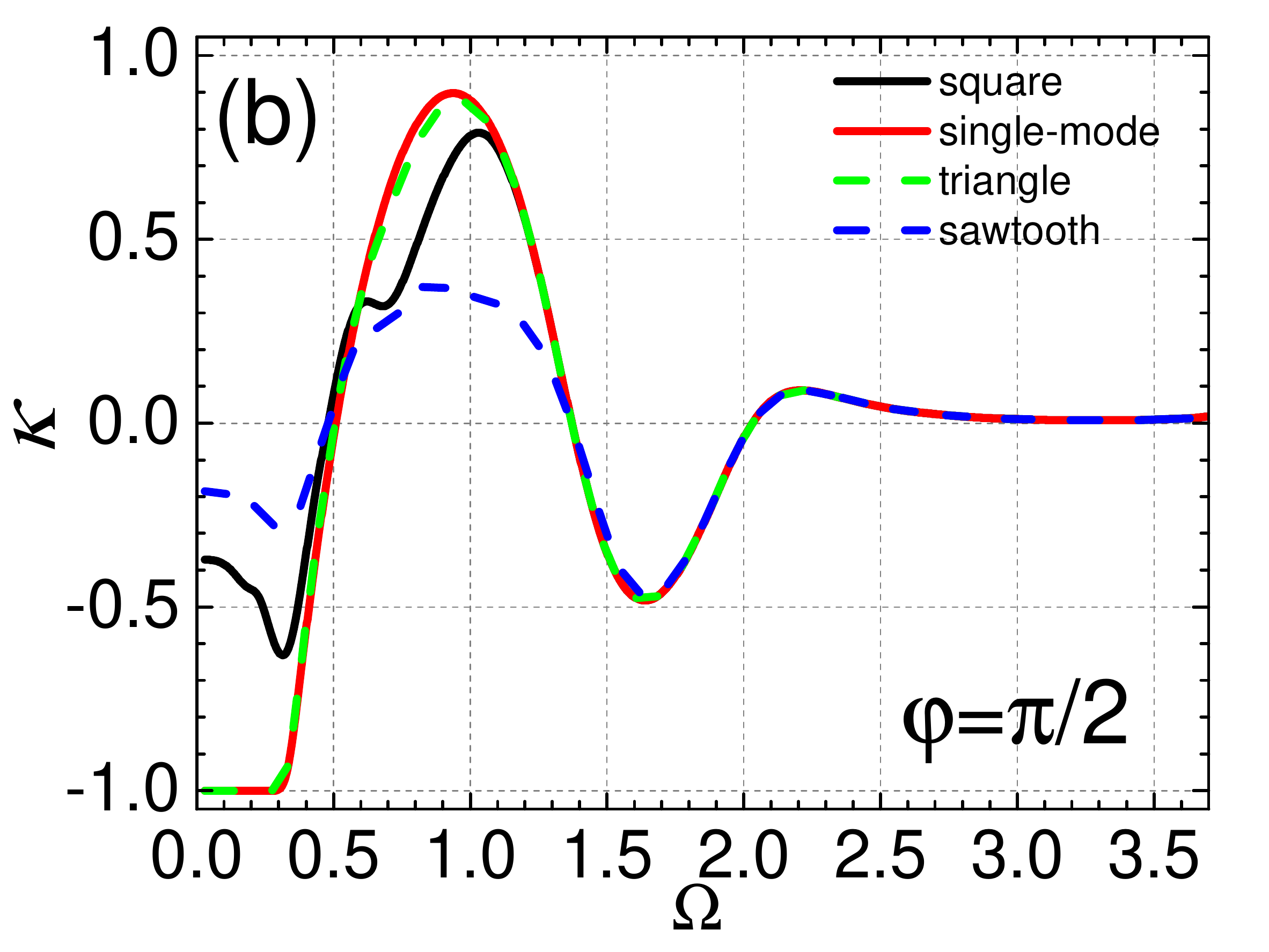}
 \includegraphics[width=0.85\columnwidth]{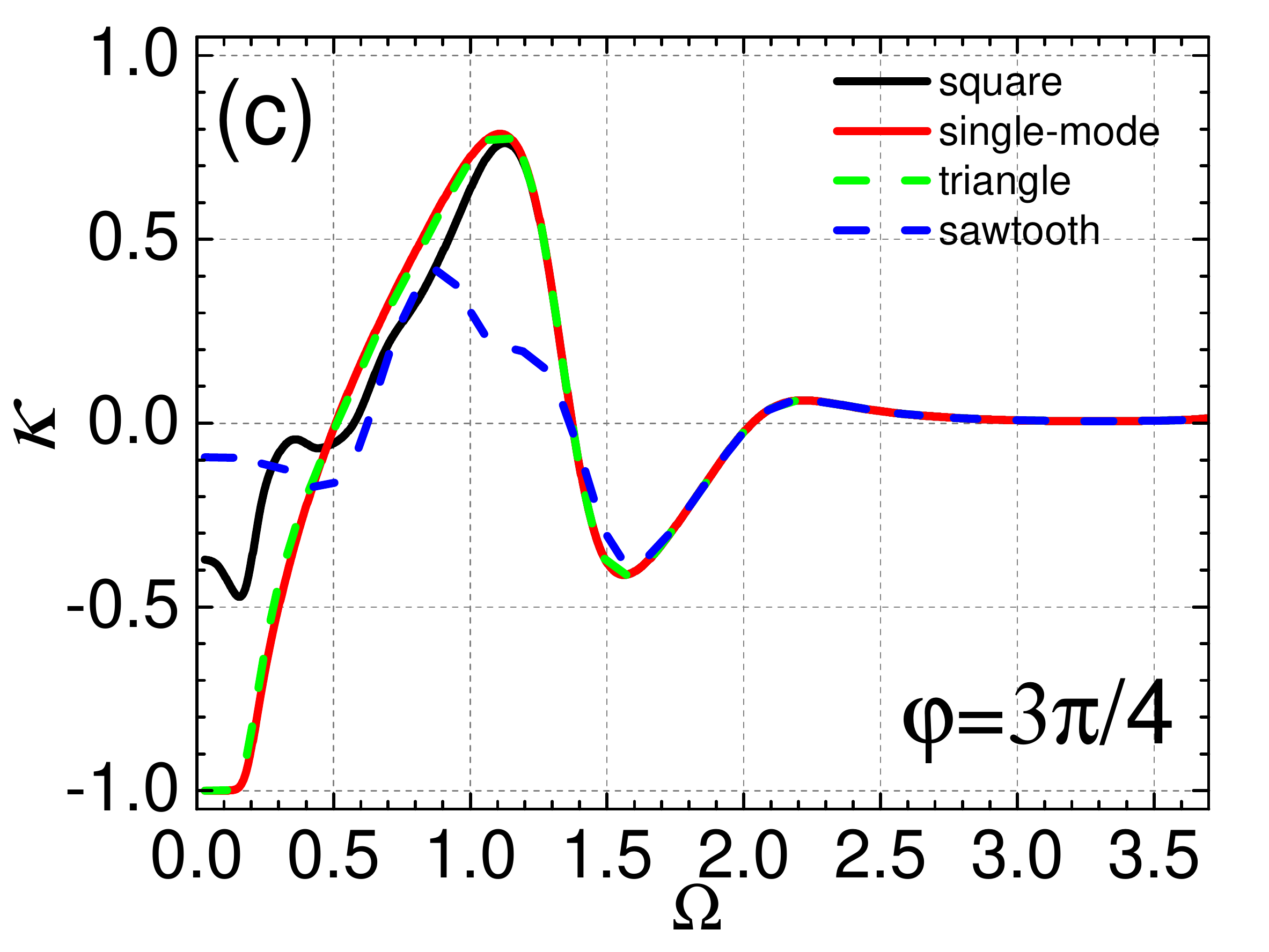}
 \caption{
(Color online) Efficiency $\kappa$ of the system as a function of the pumping frequency 
 $\Omega$ (in units of $\sqrt{k}$) using $k_L = k_R = 0.5 k$ and $k_{C}=0.25 k$ for equal temperatures
  with $k_{B}T_{L}=k_{B}T_{R}=1$ (in units of $\hbar\sqrt{k}$) and for distinct pumping functions.
  }
\label{kappa}
\end{figure}

From Eqs.~\eqref{eqs_Q_W} and \eqref{cycles}, we define the (cooling) effi\-ci\-ency $\kappa$ of the heat pump that operates between two reservoirs \textit{left} ($L$) and \textit{right} ($R$) at equal temperatures, considering a full period as
\begin{align}\label{eqq_kappa}
\kappa \equiv\frac{Q_{R}^{(\tau)}-Q_{L}^{(\tau)}}{|Q_{R}^{(\tau)}|+|Q_{L}^{(\tau)}|} = \frac{J_{R}^{(P)}-J_{L}^{(P)}}{|J_{R}^{(P)}|+|J_{L}^{(P)}|} + \mathcal{O}(\epsilon).
\end{align}
$\vert\kappa \vert$ is the ratio between the net current and the total heat current driven by the external ac source per cycle.
Fi\-gu\-re~\ref{kappa} shows that $\kappa$ can be positive or negative depending on $\Omega$ and $\phi_{L,R}(t)$.
$\kappa=\pm 1$ correspond to situations where the heat currents have opposite signs, independent of their magnitudes.

For $\Omega \gtrsim 0.5\sqrt{k}$, the thermal energy is always absorbed by the reservoirs for all the studied pum\-ping profiles (see Fig.~\ref{figs_pumping}).
Thus, the denominator of Eq.~\eqref{eqq_kappa}, $\vert Q_{R}^{(\tau)} \vert + \vert Q_{L}^{(\tau)}\vert$, corresponds to realized work $W>0$.
In contrast, when $\Omega \lesssim 0.5\sqrt{k}$ for the single-mode and the triangle profiles, we find a plateau $\kappa=-1$, indicated in the Fig.\ref{kappa}. 
In this cases, the external drive pumps heat from the left ($-Q_{L}^{(\tau)}<0$) to the right reservoir ($-Q_{R}^{(\tau)}>0$) and $\kappa=-1$ indicates all the thermal energy extracted from the left reservoir by the external drive is transferred to the right one.
Unfortunatly, the corresponding heat current is rather small.
Outside of this regime, we find that in order to optimize the pumped heat $\Delta J^{(P)}$ one should: (i) tune the pumping frequency to $\Omega \approx \sqrt{k}$  or $\Omega \approx 1.75\sqrt{k}$ producing positive ($\Delta J^{(P)}>0$) or negative ($\Delta J^{(P)}<0$), respectively, heat transfer (see Fig.~\ref{figs_pumping}) and (ii) tune the phase shift to $\pi/2$ (see Eq.~(\ref{DJ})).

The high frequency, $\Omega\gg\sqrt{k}$, asymptotic behavior of $\kappa$ (see Fig.~\ref{kappa}) is trivial 
due to the contribution of a smaller number of modes $n$ as discussed before.
We also note a surprisingly similar behavior of the efficiency curves for \textit{single-mode} and 
\textit{triangle} pumping profiles.
This can be explained by inspecting Eq.~\eqref{DJ} and recalling the rapid decrease of $a_{n}^2 \propto 1/n^4$ 
with $n$ for the latter case.
In other words, the pumped current, Eq.~(\ref{JP}), for the \textit{triangle} profile is dominated by the first mode $n=1$, 
resulting in a frequency dependence similar to the \textit{single-mode} profile.

\section{Conclusions}
\label{sec:conclusion}

In this paper we have presented a rigorous description of quantum thermal transport properties due 
to phonons in molecular and nanomechanical systems using the non-equilibrium Green's function 
theory. We approached the problem using a phase-space representation on the quantum correlations 
functions in the Keldysh contour, a convenient generalization of the standard Green's functions 
technique \cite{Rammer1986,Kamenev2011}. 

We have shown that in the stationary regime our approach recovers a Landauer-like transmission formula, as expected. 
Our derivation solves some inconsistencies of previous theoretical works based on NEGF \cite{Wang2007, Wang2008, Wang2014}. 
For instance, the use of phase-space correlation functions avoids the necessity of taking the Fourier 
transform of $\vec{u}$ and canonically conjugate $\vec{p}$ operators, that is troublesome 
for the commutation relations $[\vec{u},\vec{p}]$. The partition we put forward in Sec.~\ref{sec:model}, 
avoids conceptual difficulties with the adiabatic switch-on picture on picture on which the 
formalism is based. 
{
Finally, starting with a symmetrized Hamiltonian, our formalism avoids the necessity of imposing the \textit{ad hoc} symetrization 
$J=\left(J_{L}+J^{\ast}_{L}-J_{R}-J^{\ast}_{R}\right)/4$ used in Refs.~\cite{Wang2007, Wang2008, Wang2014}.
}

{We} extend the formalism to study the heat transport in  
systems subjected to a time-dependent external drive which opens the possibility of addressing situations of interest for applications in phononics. 
In distinction to the electronic case where the 
Fermi velocity and the system size give a characteristic time scale for the dynamics, the absence of such time 
scale in bosonic systems leads us to develop a {new} perturbation theory {scheme} assuming that the external drive is weak.

We apply our results to a model of a diatomic molecule coupled to semi-infinite linear chains in equilibrium with thermal reservoirs. The simplicity of the model allows for an amenable computation and to understand its main physical features using simple analytical considerations. This gives us confidence on the method  and {we expect it to be used} to treat more realistic systems.

\begin{acknowledgments}
This work is supported by the Brazilian funding agencies CAPES, CNPq, FAPERJ and FAPEAL.
The authors thank the hospitality of the International Institute of Physics (IIP) in Natal (Brazil), where
this work was concluded.
\end{acknowledgments}

\appendix

\section{Canonical commutation relations and Fourier transform in frequency space}\label{commutation}

The canonical quantization procedure of a classical Hamiltonian expressed in terms of the set of independent variables represented by the displacements $\lbrace u_{i}\rbrace$ and by canonically conjugated momenta $\lbrace p_{j}\rbrace$, renders the commutation relations $\left[u_i(t), p_{j}(t)\right]=\ii\hbar\,\delta_{i,j}$ and $\left[u_i(t), u_{j}(t)\right]=0=\left[p_i(t), p_{j}(t)\right]$.  

The standard approach \cite{Wang2007} is not consistent with the above relations, as we show below.

Let us consider the Fourier transform of $u_i(t)$ and $p_{i}(t)\equiv \dot{u}_i(t)$ as \cite{Wang2007, Wang2008, Wang2014}
\begin{subequations}
	\begin{align}
		u_{i} (t) &= \int_{-\infty}^{\infty}\frac{\ud\om}{2\pi}\,e^{-\ii \om t}\,u_{i}[\om], \\
		p_i (t) &= \dot{u}_{i}(t) = \int_{-\infty}^{\infty}\frac{\ud\om}{2\pi}\,e^{-\ii\om t}\,\left(-\ii\,\om\,u_{i}[\om]\right),
		\label{transformation}
	\end{align}
\end{subequations}
where the condition $(u_{i}[-\om])^{\dagger}=u_{i}[\om]$ must be sa\-tis\-fied as a result of  $\left(u_{i}(t)\right)^{\dagger}=u_{i}(t)$ (and reciprocally $\left(p_{i}(t)\right)^{\dagger}=p_{i}(t)$).

Hence, the canonical commutation relations $\left[u_i(t), p_{j}(t)\right]=\ii\hbar\,\delta_{i,j}$ and $\left[u_i(t), u_{j}(t)\right]=0$ can be written as
\begin{subequations}
	\begin{align}
			&\int_{-\infty}^{\infty}\frac{\ud\om}{2\pi}\,\int_{-\infty}^{\infty}\frac{\ud\om^{\prime}}{2\pi}\,e^{-\ii (\om + \om^\prime) t}
			(-\ii\,\om^{\prime})\,\Big[u_{i}[\om], \,u_j[\om^{\prime}]\Big] =
			\nonumber\\
			& \hspace{6cm}=\ii\hbar\,\delta_{ij} \label{eq2a}\\
			&\int_{-\infty}^{\infty}\frac{\ud\om}{2\pi}\int_{-\infty}^{\infty}\frac{\ud\om^{\prime}}{2\pi}\,e^{-\ii (\om + \om^\prime) t}
			\Big[u_{i}[\om], u_j[\om^{\prime}]\Big] = 0 \label{eq2b}
	\end{align}
\end{subequations}

Note that it is not possible to obtain a consistent result for $\left[u_{i}[\om], u_j[\om^{\prime}]\right]$ with the \eqref{eq2a} e \eqref{eq2b}, simultaneously. 
This results from the fact that canonicity relations involve operators at equal times, which is not compatible with the transformation (\ref{transformation}).
Thus, the frequency Fourier transform (\ref{transformation}) is not a canonical transformation.

In our construction we obtain the equations of motion for the NEGFs. 
The Fourier transform to frequency space is performed in the Green's functions arguments and not in the displacement and momentum operators, as standard. 
Hence, circumventing potential problems with the commutation relations.


\section{Surface Green's functions for semi-infinite lattices}\label{sec:freegf}

{
In this Appendix we present a {novel} direct analytical calculation of $\tilde{g}_{\alpha}^{r,a}[\om]$ for the 
non-ideal coupling case, namely, $k_\alpha\neq k$. (The results of Refs.~\cite{Wang2007, Wang2008} are recovered by taking $k_{\alpha}=k$.)
Next, we discuss the importance of the term $V_{aa}$ which originates the difference between ${g}_{\alpha}^{r,a}[\om]$ and $\tilde{g}_{\alpha}^{r,a}[\om]$.
}

According to \eqref{set_32} and \eqref{set_33_c}, we write
\begin{equation}\label{hessian-ring}
\tilde{g}_{\alpha}^{r,a}[\om]= 
  \langle e_{1}\vert\left(\begin{array}{cccccc}
\om^{2}_{\pm}-k_{\alpha}-k &  \;\; k               &        \\ \cline{2-4}
  k             &\temp{}   &                       &        \\
                &\temp{}   & \mathcal{D}_{n}^{\pm} &        \\
                &\temp{}   &                       &        \\ 
 \end{array}\right)^{-1}\vert e_{1}\rangle ,
\end{equation}
where $\vert e_{1}\rangle = (1, 0, \cdots, 0)^{\text{T}}$ represents the surface site and
\begin{equation}
\mathcal{D}_{n}^{\pm}=\left(\begin{array}{ccccc}
\om_{\pm}^2-2k &    k           &                &              &\\ 
  k            & \om_{\pm}^2-2k &      k         &              &\\ 
               &     k          & \om_{\pm}^2-2k &     k        &\\ 
               & \qquad\ddots   & \qquad\ddots   & \qquad\ddots &  
\end{array}\right)_{n\times n}, 
\end{equation}
with $\om_{\pm}\equiv\om \pm\ii\,0^{+}$ and $n\to\infty$. 

Applying the method of co-factors in \eqref{hessian-ring} we write  
\begin{equation}\label{16}
\tilde{g}_{\alpha}^{r,a}[\om]=\left(\,k-k_{\alpha} - \lim_{n\to\infty}\frac{d^{\pm}_{n+1}}{d^{\pm}_{n}}\,\right)^{-1},
\end{equation}
where $d_{n}^{\pm}\equiv (-1)^{n}\det\mathcal{D}_{n}^{\pm}$. The Laplace's method gives the following
recurrence equation 
\begin{equation}\label{17a}
d^{\pm}_{n+1}+ (\om^{2}_{\pm}-2k)\, d^{\pm}_{n} + k^2\,d^{\pm}_{n-1} =0.
\end{equation}

The discriminant $\Delta = \om_{\pm}^2\,\left(\om_{\pm}^2 -4k\right)$ of the associated characteristic 
e\-qua\-tion has non trivial roots $\vert\om\vert =\sqrt{4k}$. 
Hence, we split the solution of the recurrence equation in two frequency domains: 
(\emph{i}) $\vert\om\vert\leqslant\sqrt{4k}\,$ and (\emph{ii}) $\vert\om\vert > \sqrt{4k}$.

(\emph{i}) For  $\vert\om\vert\leqslant\sqrt{4k}$ we introduce the  parametrization 
$\om_{\pm}\equiv\om\pm\ii\,0^{+}=\sqrt{4k}\,\sin\left(\theta_{\pm}/2\right)$ with 
$\theta_{\pm}=\theta \pm \ii  \eta$ and $\eta=0^{+}$ for $\theta\in [-\pi,\pi]$. Substituting 
the latter in \eqref{17a}, we find
\begin{equation}
d_{n}^{\pm}=k^{n}\,\frac{\sin[(n+1)\,\theta_{\pm}]}{\sin\theta_{\pm}}.
\end{equation}
Since $\tan\left[m(\theta\pm\ii\,\eta)\right]\sim\pm\ii$ for $m\gg 1$ and $\eta>0$, 
%
\begin{equation}\label{A6}
\lim_{\eta=0^{+}}\lim_{n\to\infty}\frac{d^{\pm}_{n+1}}{d^{\pm}_{n}}= k \,\text{e}^{\mp\ii\,\theta}.
\end{equation}
that, with the help of the identities $2k\cos\theta=2k-\om^2$ and $2k\sin\theta=\om\sqrt{4k-\om^2}$, 
leads to
\begin{equation}\label{A7}
\tilde{g}^{r,a}_{\alpha}[\om]=\frac{1}{2}\,\frac{\om^2-2k_{\alpha}\mp\ii\,\om\sqrt{4k-\omega^2}}{(k-k_{\alpha})\,\om^2 + k_{\alpha}^{2}}.
\end{equation}
Note that \eqref{A7} satisfies the property $\tilde{g}_{\alpha}^{r}[-\om]=\tilde{g}_{\alpha}^{a}[\om]$ 
in line with \eqref{15a}. 

(\emph{ii}) For $\vert\om\vert>\sqrt{4k}$, we parametrize 
$\om_{\pm}\equiv\om\pm\ii\,0^{+}=\sqrt{4k}\,\cosh\left(\theta_{\pm}/2\right)\,\text{sgn}(\theta)$ 
with $\theta_{\pm}=\theta\pm\ii \eta$ and $\eta=0^{+}$ for $\theta\in\mathbb{R}^{\ast}$, 
where $\text{sgn}(\theta)$ is the sign function of $\theta$. 
Substituting this parametrization 
in \eqref{17a} we find
\begin{equation}
d_{n}^{\pm}=(-k)^{n}\,\frac{\sinh[(n+1)\,\theta_{\pm}]}{\sinh\theta_{\pm}}.
\end{equation}
Since $\coth\left[m(\theta\pm\ii\,\eta)\right]\sim\text{sgn}(\theta)$ for $m\gg 1$ and $\eta>0$, 
%
\begin{equation}\label{A6_1}
\lim_{\eta=0^{+}}\lim_{n\to\infty}\frac{d^{\pm}_{n+1}}{d^{\pm}_{n}} =- k \,\text{e}^{\vert\theta\vert}.
\end{equation}
Using $2k\,\cosh\theta = \om^2 - 2k$ and $2k\,\text{sgn}(\theta)\,\sin\theta=\sqrt{\om^{2}(\om^2-4k)}$, we obtain
\begin{equation}\label{A11}
\tilde{g}^{r,a}_{\alpha}[\om]=\frac{1}{2}\,\frac{\om^2-2k_{\alpha} -\sqrt{\om^{2}\left(\omega^2-4k\right)}}{(k-k_{\alpha})\,\om^2 + k_{\alpha}^{2}}.
\end{equation}
Note that $\tilde{g}^{r,a}_{\alpha}[\om]\sim 1/\om^{2}\to 0$ for $\vert\om\vert\gg\sqrt{4 k}$, which guarantees  
convergence in the integrations. 

We can write $\tilde{g}^{r,a}_{\alpha}[\om]$ in a convenient form as
\begin{equation}\label{eq_B11}
\tilde{g}^{r,a}_{\alpha}[\om]=\frac{1}{2}\left(\tilde{\mu}_{\alpha}[\om]\mp\ii\,\tilde{\gamma}_{\alpha}[\om]\right),
\end{equation}
where the real au\-xi\-liary functions $\tilde{\gamma}_{\alpha}[\om]$ and $\tilde{\mu}_{\alpha}[\om]$ are 
\begin{subequations}
\begin{align}
\tilde{\gamma}_{\alpha}[\om] &= \frac{\om\sqrt{4k-\om^2}}{(k-k_{\alpha})\,\om^{2}+k_{\alpha}^{2}}\,\Theta(4k-\om^{2}),\label{gamma_alpha}\\
\tilde{\mu}_{\alpha}[\om] &= \frac{\om^2-2k_{\alpha}-\sqrt{\om^2(\om^2-4k)}\,\Theta(\om^2-4k)}{(k-k_{\alpha})\,\om^{2}+k_{\alpha}^{2}}\label{mu_alpha}.
\end{align}\label{eq_B12}
\end{subequations}
{It is straightforward to verify that $\tilde{g}_{\alpha}^{r}[\om]$ satisfies the Kramers-Kronig relations, 
as it should \cite{Tuovinen2016,Tuovinen2016PhD}.}

The $\alpha$-contact line width function, $\tilde{\Gamma}_{\alpha}[\om]$, Eq.~\eqref{width}, 
becomes
\begin{equation}\label{width_app}
\tilde{\Gamma}_{\alpha}[\om] = V_{C\alpha}\cdot\tilde{\gamma}_{\alpha}[\om]\cdot V_{\alpha C}.
\end{equation}
  
Let us now analyze the role of term $V_{aa}$ in the transmission $\mathcal{T}(\om\rightarrow 0)$, given by Eq.~(\ref{transmission}).
We consider a system where the central region is composed by a dimer as shown in Fig.~\ref{fig:cadeia_linear}.
According to Eqs.~(\ref{eq:self-energy_VgV}), (\ref{set_coupling}) and (\ref{A7}) the low frequency limit of $\tilde{\Gamma}_{\alpha}[\om]$ and $\tilde{\Sigma}^{r,a}[\om]$ are
\begin{align}
\tilde{\Gamma}_{\alpha}[\om] &\approx 2\,\sqrt{k}\,\om\,P_{\alpha},\label{Gamma}\\
\tilde{\Sigma}^{r,a}[\om] &\approx -V_{CC} \mp\ii \sqrt{k}\,\om\,(P_{L}+P_{R}),\label{embed}
\end{align}
where 
$P_{L}=\begin{pmatrix}
1 \;  &\; 0\\
0 \;  &\;  0
\end{pmatrix}$,
$P_{R}=\begin{pmatrix}
0 \;  &\; 0\\
0 \;  &\;  1
\end{pmatrix}$,
$V_{CC} = \sum_{\alpha}k_{\alpha}\,P_{\alpha}$ 
and $\alpha=L,R$.
The central region Green's function is 
\begin{align}
	G^{r,a}_{CC}[\omega]=\left( \omega_{\pm} ^2\,\text{I}_{2} - K_{CC}^0 - V_{CC} - \tilde{\Sigma}^{r,a}[\om] \right)^{-1}, 
	\label{gcentral}
\end{align}
where $I_2$ is a $2\times 2$ identity matrix. 

Using Eq.~\eqref{embed} in Eq.~(\ref{gcentral}), we obtain $G^{r,a}_{CC}[\om]=[-K_{CC}^{0}\,\pm\,\ii\,\sqrt{k}\,\om\,(P_{L}+P_{R}) + \mathcal{O}(\om^2)]^{-1}$, where the spring-constant matrix of the decoupled central region $K_{CC}^{0}$ is singular and gives rise to the expansion 
\begin{align}
\label{GGr}
G_{CC}^{r,a}[\om] = \frac{\mp\ii}{2\sqrt{k}\,\om} 
\begin{pmatrix}
1    \;\, & \,\; 1\\
1    \;\, & \,\; 1
\end{pmatrix}
+ \mathcal{O}(\om^{0}).
\end{align} 
Substituting Eqs.~\eqref{Gamma} and \eqref{GGr} into Eq.\eqref{transmission}, we obtain $\mathcal{T}(\om)=1+\mathcal{O}(\om)$, that is, $\mathcal{T}(\omega\rightarrow 0)=1$.

{
In summary, the term $V_{aa}$ in Eq.~(\ref{set_coupling}) leads to  $\text{Re}\lbrace\tilde{\Sigma}^{r,a}
[0]\rbrace=-V_{CC}$.
The latter cancels out the term $-V_{CC}$ in Eq.~(\ref{gcentral}), leading to Eq.~(\ref{GGr}), that results in 
unit transmission for $\omega\rightarrow 0$. 
We conclude that for the general non-ideal coupling case, the use of $\tilde{g}^{r,a}_{\alpha}[\om]$ is 
 key to obtain the correct transmission low-frequency behavior.
 }

\section{Perturbative weak pumping regime }\label{sec:pertubative}

In this Appendix, we derive the perturbation expansion for the energy $E_{M}(t)$ of the extended molecule 
 and analyze the periodic behavior for pumped-induced heat transport. Next, we obtain 
the perturbation expansion in $\varepsilon$ for thermal current $J_{\alpha}(t)$ and the power 
 $\Phi(t)$.

\subsection{Energy $E_{M}(t)$}

We expand the Dyson equation, Eq.~\eqref{dyson_II}, in a power series in $\check{\mathcal{V}}(t)$, to obtain
(after a lengthy but straightforward calculation) the energy $E_{M}(t)$ as
\begin{widetext}
\begin{subequations}
\begin{align}
E_{M}^{(0)}=&\,\frac{\ii\hbar}{2}\,\int_{-\infty}^{\infty}\frac{\ud\om}{2\pi}\,\text{Tr}\Big\lbrace\left(G^{<}[\om]\cdot\underline{K}\right)_{CC} + \underline{K}_{CC}\cdot G^{<}_{CC}[\om]+ \sum_{\alpha}\left(V_{C\alpha}\cdot G_{\alpha C}^{<}[\om] + G_{C\alpha}^{<}[\om]\cdot V_{\alpha C}\right)\Big\rbrace,\label{EM_0}\\
E_{M}^{(1)}(t)=&\,\frac{\ii\hbar}{2}\sum_{\alpha}\phi_{\alpha}(t)\int_{-\infty}^{\infty}\frac{\ud\om}{2\pi}\,\text{Tr}\Big\lbrace V_{CC}^{(\alpha)}\cdot G_{CC}^{<}[\om]+ V_{C\alpha}\cdot G_{\alpha C}^{<}[\om] + G_{C\alpha}^{<}[\om]\cdot V_{\alpha C}\Big\rbrace
+\frac{\ii\hbar}{2}\sum_{\beta}\iint\frac{\ud\om\,\ud\om^{\p}}{(2\pi)^2}\,\text{e}^{-\ii(\om-\om^{\p})t}\nonumber\\
&\times\,\phi_{\beta}[\om-\om^{\p}]\,\text{Tr}\Big\lbrace \left(I_{C}\,\om\om^{\p}+\underline{K}_{CC}\right)\cdot\Xi_{CC,\beta}^{<}[\om,\om^{\p}] +\sum_{\alpha}\left(V_{C\alpha}\cdot\Xi_{\alpha C,\beta}^{<}[\om,\om^{\p}]+ \Xi_{C\alpha,\beta}^{<}[\om,\om^{\p}]\cdot V_{\alpha C} \right)\Big\rbrace,\label{EM_1}
\end{align}
\begin{align}
E_{M}^{(2)}(t) =&\,\frac{\ii\hbar}{2}\,\sum_{\alpha,\beta}\phi_{\alpha}(t)\,\iint\frac{\ud\om\,\ud\om^{\p}}{(2\pi)^{2}}\text{e}^{-\ii(\om-\om^{\p})t}\,\phi_{\beta}[\om-\om^{\p}]\,\text{Tr}\Big\lbrace V_{C\alpha}^{(\alpha)}\cdot\Xi_{\alpha C,\beta}^{<}[\om,\om^{\p}] + \Xi_{C\alpha,\beta}^{<}[\om,\om^{\p}]\cdot V_{\alpha C} \nonumber\\
&+ V_{CC}^{(\alpha)}\cdot\Xi_{CC,\beta}^{<}[\om,\om^{\p}] \Big\rbrace\; +\;\frac{\ii\hbar}{2}\sum_{\beta,\gamma}\iiint\frac{\ud\om\,\ud\nu\,\ud\om^{\p}}{(2\pi)^3}\,\text{e}^{-\ii(\om-\om^{\p})t}\,\phi_{\beta}[\om-\nu]\,\phi_{\gamma}[\nu-\om^{\p}]\,\nonumber\\
&\times \text{Tr}\Big\lbrace{\left(\om\,\om^{\p} I_{C} + \underline{K}_{CC}\right)}\cdot\Xi_{CC,\beta\gamma}^{<}[\om,\nu,\om^{\p}]\;+\;\sum_{\alpha}\left(V_{C\alpha}\cdot\Xi^{<}_{\alpha C,\beta\gamma}[\om,\om^{\p}] + \Xi_{C\alpha,\beta\gamma}^{<}[\om,\om^{\p}]\cdot V_{\alpha C}\right) \Big\rbrace,\label{EM_2}
\end{align}
\end{subequations}
where $\underline{\hat{K}}\equiv \hat{K}^{0} + \hat{V}$ and $\big(\hat{G}[\om]\cdot\underline{\hat{K}}\big)_{CC}=\sum_{\alpha}\big( G_{C\alpha}^{<}[\om]\cdot V_{\alpha C} + G_{CC}^{<}[\om]\cdot V_{CC}^{(\alpha)}\big)$.
\end{widetext}
Here, $\phi_{\alpha}[\om]$ is the Fourier's transform of the pum\-ping function, Eqs.~\eqref{fourier_v} and \eqref{pumping}, given by

\begin{align}\label{fourier_phi}
\phi_{\alpha}[\om]=& \int_{-\infty}^{\infty}\ud t\,\text{e}^{\ii\om t}\,\phi_{\alpha}(t)
\nonumber \\ 
= & \sum_{n=1}^{\infty}\sum_{\sigma=\pm 1} a_{n}^{(\alpha)}
2\pi\,\delta(\omega+\sigma\Omega_{n})\,\text{e}^{\ii\sigma\varphi_{n}^{(\alpha)}}.
\end{align} 

$\Xi_{a b,\beta}^{<}[\omega,\omega^{\p}]$ and $\Xi_{ab,\beta\gamma}^{<}[\om,\nu,\om^{\prime}]$ are \emph{lesser} components of
\begin{subequations}\label{set_Xi}
\begin{align}
&\Xi_{a b,\beta}[\omega,\omega^{\p}] = G_{a\beta}[\omega]\cdot V_{\beta\beta}\cdot G_{\beta b}[\omega^{\p}]\nonumber\\
& + G_{a C}[\omega]\cdot V_{C C}^{(\beta)}\cdot G_{C b}[\omega^{\p}] + G_{a\beta}[\omega]\cdot V_{\beta C}\cdot G_{C b}[\omega^{\p}]\nonumber\\
& + G_{a C}[\omega]\cdot V_{C \beta}\cdot G_{\beta b}[\omega^{\p}],
\end{align}
and
\begin{align}
&\Xi_{ab,\beta\gamma}[\om,\nu,\om^{\p}]= G_{a\beta}[\om]\cdot V_{\beta\beta}\cdot \Xi_{\beta b,\gamma}[\nu,\om^{\prime}]\nonumber\\
& + G_{aC}[\om]\cdot V_{CC}^{(\beta)}\cdot\Xi_{Cb,\gamma}[\nu,\om^{\prime}] + G_{a\beta}[\om]\cdot V_{\beta C}\cdot\Xi_{Cb,\gamma}[\nu,\om^{\prime}]\nonumber\\
&  + G_{aC}[\om]\cdot V_{C\beta}\cdot \Xi_{\beta b, \gamma}[\nu,\om^{\prime}],
\end{align}
\end{subequations}
respectively, with latin letters corresponding to reservoirs or $C$ and greek letters corresponding to reservoirs only.

Note that $\Lambda_{1}^{<}$ and $\Lambda_{2}^{<}$ of Eq.~\eqref{eqs_recurrs} are related to $\Xi_{a b,\beta}^{<}$ and $\Xi_{ab,\beta\gamma}^{<}$ by
\begin{subequations}
\begin{align}
\left(\Lambda_{1}^{<}[\omega,\omega^{\p}]\right)_{ab} &= \sum_{\beta} \Xi_{a b,\beta}^{<}[\omega,\omega^{\p}]\;\phi_{\beta}[\omega-\omega^{\p}],\\
\left(\Lambda_{2}^{<}[\om,\om^{\p}]\right)_{ab} &= \sum_{\beta\gamma}\int\frac{\ud\nu}{2\pi}\;\Xi_{ab,\beta\gamma}^{<}[\om,\nu,\om^{\prime}]\nonumber\\
&\qquad\qquad\times\phi_{\beta}[\om-\nu]\,\phi_{\gamma}[\nu-\om^{\prime}].
\end{align}
\end{subequations}

Note that $E_{M}^{(0)}$ is constant. Substituting \eqref{fourier_phi} in Eqs. \eqref{EM_1} and \eqref{EM_2}, 
we can see that $E_{M}^{(n)}(t)=E_{M}^{(n)}(t+\tau)$ for $n=1, 2, \ldots$.  

\subsection{Current $J_{\alpha}(t)$ and power developed by the ac sources $\Phi(t)$}

Substituting the results of Eqs.~\eqref{dyson_II}-\eqref{set_55} into \eqref{eq_ILL}-\eqref{eq_Phii}, we obtain the current $J_{\alpha}(t)$ from $\alpha$-lead and the power developed by the ac sources 
$\Phi(t)$ in the form of a perturbative series in $\varepsilon$ as
\begin{subequations}\label{series_app}
\begin{align}
& J_{\alpha}(t)=J_{\alpha}^{(S)}+\varepsilon\,J^{(1)}_{\alpha}(t) +\varepsilon^{2}\,J^{(2)}_{\alpha}(t) + \cdots\\
& \Phi(t)=\varepsilon\;\Phi^{(1)}(t) +\varepsilon^{2}\;\Phi^{(2)}(t) + \cdots
\end{align}
\end{subequations}
where $J^{(n)}_{\alpha}(t)$ and $\Phi^{(n)}(t)$ are $n$-order contribution of the series of $J_{\alpha}(t)$ and $\Phi(t)$, respectively. 
Below we give the explicit expressions for the first and second-order contributions.

\subsubsection{First-order contribution}

The coefficients $J_{\alpha}^{(1)}(t)$ and $\Phi^{(1)}(t)$ read
\begin{subequations}
\begin{align}
& J^{(1)}_{\alpha}(t)=\,\text{Re}\bigg[\phi_{\alpha}(t)\,\int\limits_{-\infty}^{\infty}\frac{\ud\omega}{2\pi}\,\hbar\omega\,\text{Tr}\left\lbrace V_{C\alpha}\cdot G^{<}_{\alpha C}[\omega]\right\rbrace\nonumber\\
&\;+ \sum_{\beta}\iint \frac{\ud\omega\,\ud\omega^{\p}}{(2\pi)^2}\,\text{e}^{-\ii\,(\omega-\omega^{\p})t}\,\phi_{\beta}[\om-\om^{\p}]\nonumber\\
&\;\times \text{Tr}\left\lbrace\hbar\omega\,V_{C\alpha}\cdot \Xi_{\alpha C,\beta}^{<}[\omega,\omega^{\p}]\right\rbrace\bigg],\\
&\text{and}\nonumber\\
&\Phi^{(1)}(t)=\text{Re}\bigg[\sum_{\alpha}\ii\hbar\,\dot{\phi}_{\alpha}(t)\int\limits_{-\infty}^{\infty}\frac{\ud\om}{2\pi}\,\text{Tr}\bigg\lbrace\frac{1}{2}\,V_{CC}^{(\alpha)}\cdot G_{CC}^{<}[\om]\nonumber\\
&\; + V_{C\alpha}\cdot G_{\alpha C}^{<}[\om]\bigg\rbrace\bigg].
\end{align}
\end{subequations}

Using Eqs.~(\ref{fourier_v}) and (\ref{fourier_phi}), we find
\begin{subequations}
\begin{align}
&(i)\quad & &\big\langle\phi_{\alpha}(t)\big\rangle_{\tau}=0,\\
&(ii)\quad & &\big\langle\dot{\phi}_{\alpha}(t)\big\rangle_{\tau}=0,\\
&(iii)\quad & &\big\langle\text{e}^{-\ii(\om-\om^{\p})t}\,\phi_{\beta}[\om-\om^{\p}]\big\rangle_{\tau} =0,
\end{align}
\end{subequations}
where $\langle\cdots\rangle_{\tau}=\frac{1}{\tau}\int_{0}^{\tau}\ud t\,(\cdots)$.
Thus,
\begin{equation}
\big\langle J_{\alpha}^{(1)}(t)\big\rangle_{\tau} = 0 = \big\langle \Phi^{(1)}(t)\big\rangle_{\tau}.
\end{equation}

\subsubsection{Second-order contribution}

The coefficients $J_{\alpha}^{(2)}(t)$ and $\Phi^{(2)}(t)$ read
\begin{subequations}\label{set_B9}
\begin{align}
&J^{(2)}_{\alpha}(t)=\text{Re}\bigg[\sum_{\beta\gamma}\iiint\frac{\ud\om\,\ud\nu\,\ud\om^{\p}}{(2\pi)^3}\,\hbar\omega\,\text{e}^{-\ii(\om-\om^{\p})t}\nonumber\\
&\times\phi_{\beta}[\om-\nu]\,\phi_{\gamma}[\nu-\om^{\p}]\,\text{Tr}\left\lbrace V_{C\alpha}\cdot\Xi_{\alpha C,\beta\gamma}^{<}[\om,\nu,\om^{\p}] \right\rbrace\nonumber\\
& + \sum_{\beta}\iint\frac{\ud\om\,\ud\om^{\p}}{(2\pi)^{2}}\,\hbar\om\,\text{e}^{-\ii(\om-\om^{\p})t}\,\phi_{\alpha}(t)\,\phi_{\beta}[\om-\om^{\p}]\nonumber\\
&\times \text{Tr}\left\lbrace V_{C\alpha}\cdot\Xi^{<}_{\alpha C,\beta}[\om,\om^{\p}]\right\rbrace\bigg],\\
&\text{and}\nonumber\\
&\Phi^{(2)}(t)=\text{Re}\bigg[\sum_{\alpha,\beta}\iint\frac{\ud\om\,\ud\om^{\p}}{(2\pi)^{2}}\,\text{e}^{-\ii(\om-\om^{\p})t}\,\ii\hbar\,\dot{\phi}_{\alpha}(t)\nonumber\\
&\times\phi_{\beta}[\om-\om^{\p}]\;\text{Tr}\bigg\lbrace\frac{1}{2}\,V_{CC}^{(\alpha)}\cdot\Xi_{CC,\beta}^{<}[\om,\om^{\p}]\nonumber\\
& + V_{C\alpha}\cdot\Xi_{\alpha C,\beta}^{<}[\om,\om^{\p}]\bigg\rbrace\bigg].
\end{align}
\end{subequations}

Using the Eqs. \eqref{pumping} and \eqref{fourier_phi}, we obtain
\begin{subequations}\label{set_B10}
\begin{align}
&\emph{(i)}\quad\big\langle\text{e}^{-\ii(\om-\om^{\p})t}\big\rangle_{\tau}\;\phi_{\beta}[\om-\nu]\,\phi_{\gamma}[\nu-\om^{\p}] =\nonumber\\
&\quad = 2\pi\,\delta(\om-\om^{\prime})\sum_{n=1}^{\infty} a_{n}^{(\beta)}a_{n}^{(\gamma)}\nonumber\\
&\quad\times\sum_{\sigma=\pm 1} 2\pi\,\delta(\om^{\p}-\nu+\sigma\,\Omega_{n})\,\text{e}^{\ii\sigma\left(\varphi_{n}^{(\beta)}-\varphi_{n}^{(\gamma)}\right)},\\
&\emph{(ii)}\quad\big\langle\text{e}^{-\ii(\om-\om^{\p})t}\,\phi_{\alpha}(t)\big\rangle_{\tau}\;\phi_{\beta}[\om-\om^{\p}] =\sum_{n=1}^{\infty}a_{n}^{(\alpha)}\,a_{n}^{(\beta)}\nonumber\\
&\quad \times\sum_{\sigma=\pm 1}2\pi\,\delta(\omega-\omega^{\p}+\sigma\,\Omega_{n})\,\text{e}^{\ii\sigma \left(\varphi_{n}^{(\beta)}-\varphi_{n}^{(\alpha)}\right)},\\
&\emph{(iii)}\quad\big\langle\text{e}^{-\ii(\om-\om^{\p})t}\,\dot{\phi}_{\alpha}(t)\big\rangle_{\tau}\,\phi_{\beta}[\om-\om^{\p}] =-\sum_{n=1}^{\infty} a_{n}^{(\alpha)}\,a_{n}^{(\beta)}\nonumber\\
&\times\sum_{\sigma=\pm 1}2\pi\Omega_{n}\,\ii\sigma\,\delta(\omega-\omega^{\p}+\sigma\Omega_{n})\,\text{e}^{\ii\sigma \left(\varphi_{n}^{(\beta)}-\varphi_{n}^{(\alpha)}\right)}.
\end{align}
\end{subequations}
Hence,
\begin{subequations}
\begin{multline}
\big\langle J_{\alpha}^{(2)}(t)\big\rangle_{\tau}=\sum_{n=1}^{\infty}\sum_{\beta,\gamma} a_{n}^{(\beta)}\,a_{n}^{(\gamma)}\\
\times\sum_{\sigma=\pm 1}\,\text{Re}\left[\text{e}^{\ii\sigma\left(\varphi_{n}^{(\beta)}-\varphi_{n}^{(\gamma)}\right)}\,\mathcal{J}_{\beta\gamma}^{(\alpha)}(\sigma\Omega_{n})\right]
\end{multline}
and
\begin{multline}
\big\langle \Phi^{(2)}(t)\big\rangle_{\tau}=\sum_{n=1}^{\infty}\sum_{\beta,\gamma} a_{n}^{(\beta)}\,a_{n}^{(\gamma)}\\
\times\sum_{\sigma=\pm 1}\,\text{Re}\left[\text{e}^{\ii\sigma\left(\varphi_{n}^{(\beta)}-\varphi_{n}^{(\gamma)}\right)}\,\mathcal{F}_{\beta\gamma}(\sigma\Omega_{n})\right],
\end{multline}
\end{subequations}
where we introduced the following integrals
\begin{subequations}\label{set_12}
\begin{align}
\mathcal{J}_{\beta\gamma}^{(\alpha)}(\sigma\Omega_{n})&=\int\limits_{-\infty}^{\infty}\frac{\ud\om}{2\pi}\,\hbar\om\,\text{Tr}\bigg\lbrace V_{C\alpha}\cdot\Xi_{\alpha C,\beta\gamma}^{<}[\om,\om+\sigma\Omega_{n},\om]\nonumber\\
&+\delta_{\alpha\gamma}\,V_{C\alpha}\cdot\Xi_{\alpha C,\beta}^{<}[\om,\om+\sigma\Omega_{n}]\bigg\rbrace,\\
\mathcal{F}_{\beta\gamma}(\sigma\Omega_{n})&=\int\limits_{-\infty}^{\infty}\frac{\ud\om}{2\pi}\,\hbar\sigma\Omega_{n}\,\text{Tr}\bigg\lbrace \frac{1}{2}\, V_{CC}^{(\gamma)}\cdot\Xi_{CC,\beta}^{<}[\om,\om+\sigma\Omega_{n}]\nonumber\\
&+\,V_{C\gamma}\cdot\Xi_{\gamma C,\beta}^{<}[\om,\om+\sigma\Omega_{n}]\bigg\rbrace.
\end{align}
\end{subequations}

Defining $J^{(P)}_{\alpha}\equiv\big\langle J_{\alpha}^{(2)}(t)\big\rangle_{\tau}$ and $\Phi^{(P)}\equiv\big\langle\Phi^{(2)}(t)\big\rangle_{T}$, we get
\begin{subequations}\label{set_B13}
\begin{align}
J^{(P)}_{\alpha}=&\sum_{n=1}^{\infty}\sum_{\beta\gamma}a_{n}^{(\beta)}a_{n}^{(\gamma)}\bigg[\cos\left(\varphi_{n}^{(\beta)}-\varphi_{n}^{(\gamma)}\right)\,A_{\beta\gamma}^{\alpha}(n)\nonumber\\
&-\sin\left(\varphi_{n}^{(\beta)}-\varphi_{n}^{(\gamma)}\right)\,B_{\beta\gamma}^{\alpha}(n)\bigg],\\
\Phi^{(P)}=&\sum_{n=1}^{\infty}\sum_{\beta\gamma}a_{n}^{(\beta)}a_{n}^{(\gamma)}\bigg[\cos\left(\varphi_{n}^{(\beta)}-\varphi_{n}^{(\gamma)}\right)\,D_{\beta\gamma}(n)\nonumber\\
&-\sin\left(\varphi_{n}^{(\beta)}-\varphi_{n}^{(\gamma)}\right)\,E_{\beta\gamma}(n)\bigg],
\end{align}
\end{subequations}
where 
\begin{subequations}\label{set_B14}
\begin{align}
& A_{\beta\gamma}^{\alpha}(n)=\text{Re}\left[\sum_{\sigma=\pm 1}\mathcal{J}_{\beta\gamma}^{(\alpha)}(\sigma\Omega_{n})\right],\\
& B_{\beta\gamma}^{\alpha}(n)=\text{Im}\left[\sum_{\sigma=\pm 1}\sigma\mathcal{J}_{\beta\gamma}^{(\alpha)}(\sigma\Omega_{n})\right],\\
& D_{\beta\gamma}(n)=\text{Re}\left[\sum_{\sigma=\pm 1}\mathcal{F}_{\beta\gamma}(\sigma\Omega_{n})\right],\\
& E_{\beta\gamma}(n)=\text{Im}\left[\sum_{\sigma=\pm 1}\sigma\mathcal{F}_{\beta\gamma}(\sigma\Omega_{n})\right].
\end{align}
\end{subequations}

Equations \eqref{set_B13} and \eqref{set_B14} and the energy conservation $\sum_{\alpha}J_{\alpha}^{(P)} + \Phi^{(P)}=0$ (according Eqs.~(\ref{eqs_Q+W})-(\ref{cycles})), lead to the following conditions 
\begin{subequations}
\begin{align}
&\sum_{\alpha}A_{(\beta\gamma)}^{\alpha}(n) + D_{(\beta\gamma)}(n) = 0,\\
&\sum_{\alpha}B_{[\beta\gamma]}^{\alpha}(n) + E_{[\beta\gamma]}(n) = 0,
\end{align}
\end{subequations}
where we introduced symmetrization $O_{(\beta\gamma)}\equiv\frac{1}{2}\left(O_{\beta\gamma}+O_{\gamma\beta}\right)$ and anti-symmetrization $O_{[\beta\gamma]}\equiv\frac{1}{2}\left(O_{\beta\gamma}-O_{\gamma\beta}\right)$ shorthand notations.

For the calculation of $\mathcal{J}_{\beta\gamma}^{(\alpha)}(\sigma\Omega_{n})$ and $\mathcal{F}_{\alpha\beta}(\sigma\Omega_{n})$, we use $V_{C\alpha}\cdot\Xi_{\alpha C,\beta}[\om,\om^{\p}]$, $\Xi_{C C,\beta}[\om,\om^{\p}]$ and $V_{C\alpha}\cdot\Xi_{\alpha C,\beta\gamma}[\om,\om^{\p},\om]$ of \eqref{43} and \eqref{set_Xi}, as
\begin{subequations}\label{set_B16}
\begin{align}
&V_{C\alpha}\cdot\Xi_{\alpha C,\beta}[\om,\om^{\p}] = \,\delta_{\alpha\beta}\,\Pi_{\alpha}^{(1)}[\om,\om^{\p}]\cdot G_{CC}[\om^{\p}]\nonumber\\
&\;\; + \tilde{\Sigma}_{\alpha}[\om]\cdot G_{CC}[\om]\cdot \Pi_{\beta}^{(2)}[\om,\om^{\p}]\cdot G_{CC}[\om^{\p}],\\
&\nonumber\\
&\Xi_{C C,\beta}[\om,\om^{\p}] = G_{CC}[\om]\cdot\Pi_{\beta}^{(2)}[\om,\om^{\p}]\cdot G_{CC}[\om^{\p}],
\end{align}
\begin{align}
&V_{C\alpha}\cdot\Xi_{\alpha C,\beta\gamma}[\om,\om^{\p},\om] =\, \delta_{\alpha\beta}\,\delta_{\alpha\gamma}\,\Pi_{\alpha}^{(3)}[\om,\om^{\p}]\cdot G_{CC}[\om]\nonumber\\
&\;\; +\delta_{\alpha\beta}\,\Pi_{\alpha}^{(1)}[\om,\om^{\p}]\cdot G_{CC}[\om^{\p}]\cdot\Pi_{\gamma}^{(2)}[\om^{\p},\om]\cdot G_{CC}[\om]\nonumber\\
&\;\; + \delta_{\beta\gamma}\,\tilde{\Sigma}_{\alpha}[\om]\cdot G_{CC}[\om]\cdot\Pi_{\beta}^{(4)}[\om,\om^{\p}]\cdot G_{CC}[\om]\nonumber\\
&\;\; +\tilde{\Sigma}_{\alpha}[\om]\cdot G_{CC}[\om]\cdot \Pi_{\beta}^{(2)}[\om,\om^{\p}]\cdot G_{CC}[\om^{\p}]\cdot\Pi_{\gamma}^{(2)}[\om^{\p},\om]\nonumber\\
&\;\;\cdot G_{CC}[\om],
\end{align}
\end{subequations}
where we define
\begin{subequations}\label{set_Pi}
\begin{align}
&\Pi_{\theta}^{(1)}[\om,\om^{\p}]\equiv\tilde{\Sigma}_{\theta}[\om,\om^{\p}]+\tilde{\Sigma}_{\theta}[\om],\\
&\Pi_{\theta}^{(2)}[\om,\om^{\p}]\equiv\tilde{\Sigma}_{\theta}[\om,\om^{\p}]+\tilde{\Sigma}_{\theta}[\om]+\tilde{\Sigma}_{\theta}[\om^{\p}]+V_{CC}^{(\theta)},\\
&\Pi_{\theta}^{(3)}[\om,\om^{\p}]\equiv\tilde{\Sigma}_{\theta}[\om,\om^{\p},\om]+\tilde{\Sigma}_{\theta}[\om,\om^{\p}],\\
&\Pi_{\theta}^{(4)}[\om,\om^{\p}]\equiv\Pi_{\theta}^{(3)}[\om,\om^{\p}]+\Pi_{\theta}^{(1)}[\om^{\p},\om]
\end{align}
\end{subequations}
where 
\begin{equation}\label{SIGMA}
\tilde{\Sigma}_{\theta}[\om_{1}, \ldots, \om_{n}]\equiv V_{C\theta}\cdot\tilde{g}_{\theta}[\om_{1}]\cdot V_{\theta\theta}\cdot\ldots \cdot\tilde{g}_{\theta}[\om_{n}]\cdot V_{\theta C}.
\end{equation}

\begin{subequations}
Hence, we obtain the \emph{lesser} components of \eqref{set_B16} as 
\begin{widetext}
\begin{align}
& V_{C\alpha}\cdot\Xi_{\alpha C, \beta}^{<}[\om,\om^{\p}]=\,\delta_{\alpha\beta}\;\Big[\left(\Pi_{\alpha}^{(1)}[\om,\om^{\p}]\right)^{r}\cdot G^{<}_{CC}[\om] + \left(\Pi_{\alpha}^{(1)}[\om,\om^{\p}]\right)^{<}\cdot G^{a}_{CC}[\om]\;\Big]\nonumber\\
&\;\; +\tilde{\Sigma}_{\alpha}^{r}[\om]\cdot G^{r}_{CC}[\om]\cdot\left(\Pi_{\beta}^{(2)}[\om,\om^{\p}]\right)^{r}\cdot G_{CC}^{<}[\om^{\p}] +\tilde{\Sigma}_{\alpha}^{r}[\om]\cdot G^{r}_{CC}[\om]\cdot\left(\Pi_{\beta}^{(2)}[\om,\om^{\p}]\right)^{<}\cdot G_{CC}^{a}[\om^{\p}]\nonumber\\
&\;\; +\tilde{\Sigma}_{\alpha}^{r}[\om]\cdot G^{<}_{CC}[\om]\cdot\left(\Pi_{\beta}^{(2)}[\om,\om^{\p}]\right)^{a}\cdot G_{CC}^{a}[\om^{\p}] +\tilde{\Sigma}_{\alpha}^{<}[\om]\cdot G^{a}_{CC}[\om]\cdot\left(\Pi_{\beta}^{(2)}[\om,\om^{\p}]\right)^{a}\cdot G_{CC}^{a}[\om^{\p}],\\
&\nonumber\\
&\Xi_{C C,\beta}^{<}[\om,\om^{\p}] = G_{CC}^{r}[\om]\cdot\left(\Pi_{\beta}^{(2)}[\om,\om^{\p}]\right)^{r}\cdot G_{CC}^{<}[\om^{\p}] + G_{CC}^{r}[\om]\cdot\left(\Pi_{\beta}^{(2)}[\om,\om^{\p}]\right)^{<}\cdot G_{CC}^{a}[\om^{\p}] \nonumber\\
&\;\; + G_{CC}^{<}[\om]\cdot\left(\Pi_{\beta}^{(2)}[\om,\om^{\p}]\right)^{a}\cdot G_{CC}^{a}[\om^{\p}],
\end{align}
\begin{align}
&V_{C\alpha}\cdot\Xi_{\alpha C,\beta\gamma}^{<}[\om,\om^{\p},\om] = \delta_{\alpha\beta}\,\delta_{\alpha\gamma}\,\Big[\,\left(\Pi_{\alpha}^{(3)}[\om,\om^{\p}]\right)^{r}\cdot G_{CC}^{<}[\om] +\left(\Pi_{\alpha}^{(3)}[\om,\om^{\p}]\right)^{<}\cdot G_{CC}^{a}[\om]\,\Big]\nonumber\\
&\;\;+\delta_{\alpha\beta}\Big[\left(\Pi_{\alpha}^{(1)}[\om,\om^{\p}]\right)^{r}\cdot G_{CC}^{r}[\om^{\p}]\cdot\left(\Pi_{\gamma}^{(2)}[\om^{\p},\om]\right)^{r}\cdot G_{CC}^{<}[\om]+\left(\Pi_{\alpha}^{(1)}[\om,\om^{\p}]\right)^{r}\cdot G_{CC}^{r}[\om^{\p}]\cdot\left(\Pi_{\gamma}^{(2)}[\om^{\p},\om]\right)^{<}\cdot G_{CC}^{a}[\om]\nonumber\\
&\;\;+\left(\Pi_{\alpha}^{(1)}[\om,\om^{\p}]\right)^{r}\cdot G_{CC}^{<}[\om^{\p}]\cdot\left(\Pi_{\gamma}^{(2)}[\om^{\p},\om]\right)^{a}\cdot G_{CC}^{a}[\om]+\left(\Pi_{\alpha}^{(1)}[\om,\om^{\p}]\right)^{<}\cdot G_{CC}^{a}[\om^{\p}]\cdot\left(\Pi_{\gamma}^{(2)}[\om^{\p},\om]\right)^{a}\cdot G_{CC}^{a}[\om]\,\Big]\nonumber\\
&\;\;+\delta_{\beta\gamma}\,\Big[\,\tilde{\Sigma}_{\alpha}^{r}[\om]\cdot G_{CC}^{r}[\om]\cdot\left(\Pi_{\beta}^{(4)}[\om,\om^{\p}]\right)^{r}\cdot G_{CC}^{<}[\om] +\tilde{\Sigma}_{\alpha}^{r}[\om]\cdot G_{CC}^{r}[\om]\cdot\left(\Pi_{\beta}^{(4)}[\om,\om^{\p}]\right)^{<}\cdot G_{CC}^{a}[\om]\nonumber\\
&\;\;+\tilde{\Sigma}_{\alpha}^{r}[\om]\cdot G_{CC}^{<}[\om]\cdot\left(\Pi_{\beta}^{(4)}[\om,\om^{\p}]\right)^{a}\cdot G_{CC}^{a}[\om]+\tilde{\Sigma}_{\alpha}^{<}[\om]\cdot G_{CC}^{a}[\om]\cdot\left(\Pi_{\beta}^{(4)}[\om,\om^{\p}]\right)^{a}\cdot G_{CC}^{a}[\om]\,\Big]\nonumber\\
&\;\; +\tilde{\Sigma}_{\alpha}^{r}[\om]\cdot G_{CC}^{r}[\om]\cdot\left(\Pi_{\beta}^{(2)}[\om,\om^{\p}]\right)^{r}\cdot G_{CC}^{r}[\om^{\p}]\cdot\left[\,\left(\Pi_{\gamma}^{(2)}[\om^{\p},\om]\right)^{r}\cdot G_{CC}^{<}[\om] + \left(\Pi_{\gamma}^{(2)}[\om^{\p},\om]\right)^{<}\cdot G_{CC}^{a}[\om]\,\right]\nonumber\\
&\;\; + \tilde{\Sigma}_{\alpha}^{r}[\om]\cdot G_{CC}^{r}[\om]\cdot\left[\,\left(\Pi_{\beta}^{(2)}[\om,\om^{\p}]\right)^{r}\cdot G_{CC}^{<}[\om^{\p}] + \left(\Pi_{\beta}^{(2)}[\om,\om^{\p}]\right)^{<}\cdot G_{CC}^{a}[\om^{\p}]\,\right]\cdot \left(\Pi_{\gamma}^{(2)}[\om^{\p},\om]\right)^{a}\cdot G_{CC}^{a}[\om]\nonumber\\
&\;\; + \left[\,\tilde{\Sigma}_{\alpha}^{r}[\om]\cdot G_{CC}^{<}[\om] + \tilde{\Sigma}_{\alpha}^{<}[\om]\cdot G_{CC}^{a}[\om]\,\right]\cdot\left(\Pi_{\beta}^{(2)}[\om,\om^{\p}]\right)^{a}\cdot G_{CC}^{a}[\om^{\p}]\cdot \left(\Pi_{\gamma}^{(2)}[\om^{\p},\om]\right)^{a}\cdot G_{CC}^{a}[\om],
\end{align}
\end{widetext}
\end{subequations}
where we define the \emph{lesser} component of the set \eqref{set_Pi} as
\begin{align}
\left(\Pi_{\theta}^{(1)}[\om,\om^{\p}]\right)^{<}&\equiv \;\tilde{\Sigma}_{\theta}^{r,<}[\om,\om^{\p}]+\tilde{\Sigma}_{\theta}^{<,a}[\om,\om^{\p}]\nonumber\\
&\;\;\;+\tilde{\Sigma}_{\theta}^{<}[\om],\\
\left(\Pi_{\theta}^{(2)}[\om,\om^{\p}]\right)^{<}&\equiv \;\tilde{\Sigma}_{\theta}^{r,<}[\om,\om^{\p}]+\tilde{\Sigma}_{\theta}^{<,a}[\om,\om^{\p}]\nonumber\\
&\;\;\;+\tilde{\Sigma}_{\theta}^{<}[\om]+\tilde{\Sigma}_{\theta}^{<}[\om^{\p}],
\end{align}
\begin{align}
&\left(\Pi_{\theta}^{(3)}[\om,\om^{\p}]\right)^{<}\equiv \;\tilde{\Sigma}_{\theta}^{r,r,<}[\om,\om^{\p},\om]+\tilde{\Sigma}_{\theta}^{r,<,a}[\om,\om^{\p},\om]\nonumber\\
&\;\;\;+\tilde{\Sigma}_{\theta}^{<,a,a}[\om,\om^{\p},\om]+\tilde{\Sigma}_{\theta}^{r,<}[\om,\om^{\p}]+\tilde{\Sigma}_{\theta}^{<,a}[\om,\om^{\p}],
\end{align}
and the \emph{retarded}-($r$) and \emph{advanced}-($a$) component of \eqref{set_Pi} as
\begin{subequations}
\begin{align}
\left(\Pi_{\theta}^{(1)}[\om,\om^{\p}]\right)^{x}\equiv &\;\tilde{\Sigma}_{\theta}^{x,x}[\om,\om^{\p}]+\tilde{\Sigma}_{\theta}^{x}[\om],\\
\left(\Pi_{\theta}^{(2)}[\om,\om^{\p}]\right)^{x}\equiv &\;\tilde{\Sigma}_{\theta}^{x,x}[\om,\om^{\p}]+\tilde{\Sigma}_{\theta}^{x}[\om]\nonumber\\
&+\tilde{\Sigma}_{\theta}^{x}[\om^{\p}]+V_{CC}^{(\theta)},\\
\left(\Pi_{\theta}^{(3)}[\om,\om^{\p}]\right)^{x}\equiv &\;\tilde{\Sigma}_{\theta}^{x,x,x}[\om,\om^{\p},\om]+\tilde{\Sigma}_{\theta}^{x,x}[\om,\om^{\p}],
\end{align}
\end{subequations}
with $x= r,a$ and where we denote the components of generalized function as
\begin{multline}\label{74d}
\tilde{\Sigma}_{\alpha}^{x_{1},\ldots , x_{n}}[\om_1, \ldots\om_{n}] \equiv\\
 V_{C\alpha}\cdot\tilde{g}_{\alpha}^{\,x_{1}}[\om_{1}]\cdot V_{\alpha\alpha}\cdot\ldots \cdot\tilde{g}_{\alpha}^{\,x_{n}}[\om_{n}]\cdot V_{\alpha C}.
\end{multline}

\bibliography{quantum_thermal_transport}

\end{document}